# Conjugate-dual clusters

Silei Wang[1], Jing Tian[1], Jiayu Li[1], Tian Gong[2], Xing Yan[3], Jijun Zhao[4] and Xin-Gao Gong[5], Xiao Gu[1*]

[1] School of Physical Science and Technology, Ningbo University, Ningbo 315211, China

[2] School of Mathematics and Statistics, Ningbo University, Ningbo 315211, China

[3] School of Civil & Environmental Engineering and Geography Science, Ningbo University, Ningbo 315211, China

[4] Guangdong Provincial Key Laboratory of Quantum Engineering and Quantum Materials, School of Physics, South China Normal University, Guangzhou 510006, China

[5] Department of Physics, Key Laboratory for Computational Science (MOE), State Key Laboratory of Surface Physics, Fudan University, Shanghai 200433, China

*Email: guxiao@nbu.edu.cn*

# Abstract

Discovery of clusters with high symmetrical geometry, such as $C_{60}$ fullerene, always attract lots of interest because of their diverse nature. However, most of such interesting cluster were sporadically discovered, is there any systematic method to explore all possible high symmetrical clusters? Herein, we propose an idea to systematically construct high symmetrical structures based on novel conjugate-dual (co-dual) concept. A co-dual structure would be constructed by conjugately combining one high symmetrical polyhedron and its corresponding dual. In such co-dual structures, the symmetry of the original polyhedron could be kept or even promoted in some special cases. This provides a way to explore high symmetrical polyhedra and leads to a new cluster family, co-dual clusters. In this paper, we have systematically studied the spherical co-dual clusters with one- or two-element shells, and found a series of stable cage-like and core-shell clusters. The co-dual structures would be a new treasure box for hunting novel clusters.

# 1. Introduction

Over the years, extensive research has been conducted on homo- and heteroatom clusters of noble gases, simple, transition and rare earth metals, semiconductor elements, and molecular species. Clusters have sustained interest due to their immense potential across numerous applications such as catalysis, nanoelectronics, and even life science. Among various types of clusters, highly symmetric structures are considered particularly special because of their complexity and aesthetic appeal. Specifically, metal clusters that exhibit the characteristics of well-known Platonic (regular polyhedra) or Archimedean polyhedra (semi-regular polyhedra), as shown in Figure S1, present visually appealing structures and intriguing physical properties.[1-4] However, predicting and synthesizing highly symmetric clusters present new challenges for chemists and materials scientists, as these outcomes are occasional discoveries. Thus, envisioning orderly arrangements of elements based on highly symmetric geometric structures offers a reliable and novel approach to effectively predict stable unknown clusters.

The ancient Greeks identified Platonic polyhedral, belonging to highly symmetric point groups: the tetrahedron ($T_d$), cube ($O_h$), octahedron ($O_h$), dodecahedron ($I_h$), and icosahedron ($I_h$). As depicted in Figure S1(a), these regular polyhedra consist of identical regular polygons on all faces, with all vertices and edges being equal. Another class of highly symmetric polyhedra is the Archimedean polyhedra,[5] which are composed of two or more different regular polygons sharing identical vertices. Figure S1(b) shows all the 13 Archimedean polyhedra. Numerous clusters featuring highly symmetric structures have been extensively investigated. The octahedral structure {$M_6O_{19}$} (M = V, Mo, W, Ta, Nb, etc.) exhibits $O_h$ symmetry.[6] Comparative analyses of experimental vibrational spectra against theoretical predictions have established that the neutral $(MgO)_4$ cluster adopts a cubic structure.[7] The regular dodecahedral hollow cage clusters $Sn_{12}^{2-}$ and $Pb_{12}^{2-}$ demonstrate high stability,[8] while W@$Au_{12}$,[9] Mo@$Ag_{12}$,[10] and Cr@$Cu_{12}$[11] also possess similar cage-like architectures. A $Si_{20}$ regular dodecahedron was synthesized through a chlorine-induced disproportionation reaction, which contains an endohedral chloride ion that imparts a net negative charge, resulting in a stable structure.[12] The $C_{20}$ cluster, which encapsulates a Cf atom, is capable of stabilizing into a highly symmetric icosahedral structure.[13] The truncated tetrahedral structure of $Mn_{12}$ possesses distinct attributes that influence its magnetic properties and electrocatalytic $H_2O$ oxidation characteristics.[14] Computed at the PW91/QZ4P level of theory, the neutral Mn@$Au_{12}$ cluster favors the $O_h$ structure rather than the $I_h$ one, with a significant magnetic moment of 5 $\mu_B$ and a moderate HOMO−LUMO gap of 0.36 eV.[15] Certain coordination cages based on metal clusters have a truncated cubic structure, with their stiffness and permanent porosity confirmed through gas adsorption studies.[16] The hollow cage cluster $Sn_{12}Se_{12}$ has been successfully assembled in a truncated octahedral structure, demonstrating excellent thermoelectric performance and making it a promising candidate for thermoelectric materials.[17] The cluster $B_{24}N_{24}$, with $O_h$ symmetry (truncated cuboctahedral structure), can provide a confined space for stable polynitrogen clusters, representing a novel type of nanoscale high-energy energetic material.[18] Additionally, cryo-electron microscopy has revealed that the cowpea chlorotic mottle virus consists of a truncated icosahedral structure formed by 12 pentamers and 20 hexamers.[19]

# 2. Results and Discussion

By the duality principle in projective geometry, for every polyhedron, there exists another dual polyhedron, which would be constructed by replacing each face with a vertex, each vertex with a face. Figure 1 shows the dual pairs of the 5 Platonic solids: tetrahedron is self-dual, cube and octahedron are dual structures to each other, and dodecahedron and icosahedron are dual structures to each other. Since the dual pairs would share the same symmetry, they could be carefully aligned to each other to construct a new structure, whose symmetry could be kept or even been promoted in some special cases. To achieve this, the dual pairs should have the same center, and their face-vertex pair should be aligned in the conjugate way. Herein, we propose that a conjugate-dual (co-dual) structure is the one by conjugately combining a polyhedron and its dual structure.

Figure 1 also shows the co-dual structure of the three dual pairs of the 5 Platonic solids, $Z_1$, $Z_2$, and $Z_3$ respectively. The co-dual structures are also polyhedra, and they surely would have their own dual polyhedra, and co-dual structure. Figure S3 shows the hierarchy of co-dual structures of tetrahedron. In geometry the co-dual could happen to infinite into a sphere. Inspired by the 5 Platonic solids, efforts were made to explore highly symmetrical structures for clusters by including all the 13 Archimedean solids. By applying each vertex-plane pair to the Archimedean solids, the corresponding co-dual structures, the fully conjugated dual structures, are depicted in Figure S4, labeled as f-$A_1$ to f-$A_{13}$.

All these co-dual structures could be used to generate a new family of clusters. Assuming atoms could occupy the vertex of the co-dual polyhedra, then edges represent the chemical bonds. When imposing chemical elements, the vertex and the edge length should be also considered in the reasonable region. By choosing proper vertex and the size of polyhedra, co-dual structures of Platonic and Archimedean could result in spherical clusters with high symmetry. Actually, in recent decades, lots of clusters with co-dual structures based on Platonic and Archimedean polyhedra had been serendipitously discovered.

Co-dual clusters by Tetrahedron: The tetrahedron is its own dual polyhedron. The co-dual cluster is denoted as $Q_4$ (tetrahedron)@$T_4$ (tetrahedron), where Q and T denote identical atoms. Complex metal-containing clusters often feature skeletal frameworks composed of basic units such as metal atoms $M^{0/n+}$, metal-oxo $MO_x^{m-}$. As early as 1984, co-dual clusters were constructed by incorporating organic composites. An octa-nuclear heterometal cluster $[V^{III}ZnO(O_2CC_6H_5)_3(THF)]_4$ was successfully synthesized, featuring two concentric tetrahedra: $V_4$@$Zn_4$. The remarkable ideal tetrahedral ($T_d$) symmetry of this structure was noted.[20] This unique co-dual structure contributes to the stability of many metal clusters. For instance, in $Ag_8\{Ti(Sal)_3\}_4$, tetrahedral $Ag_4$ units are encapsulated by four metal-ligand motifs $Ti(Sal/5\text{-}FSal)_3$, remaining stable in air for several months.[21] In 2009，the Cu-H cluster with phosphorochalcogenides as the supporting ligand was first reported: the octanuclear copper complex $Cu_4(H)(\mu_3\text{-}Cu)_4[S_2P(O^iPr)_2]_6(PF_6)$ features a $Cu_4$@$Cu_4$ core for hydrogen capture, serving as efficient precatalysts for Ulmann-type coupling reactions.[22]

Co-dual clusters by Octahedron and Cube**:** The octahedron and cube are dual structures of each other, with their co-dual cluster represented as $Q_6$ (octahedron)@$T_8$ (cube). In 2021, the homoleptic alkynyl-protected AgCu superatomic nanocluster $[Ag_9Cu_6(^tBuC{\equiv}C)_{12}]^+$ comprises an Ag@$Cu_6$@$Ag_8$ metal core, with Ag atoms occupying each triangle center of the $Cu_6$ octahedron, demonstrating excellent stability at room temperature and elevated temperatures.[23] Another highly stable silver nanoclusters example is $[Ag_{14}(C_2B_{10}H_{10}S_2)_6$(pyridine/*p*-methyl-pyridine)$_8]$

($C_2B_{10}H_{10}S_2$=1,2-dithiolate-*o*-carborane), with $Ag_6@Ag_8$ superatoms shielded by ligands, remaining stable up to 150 °C in air.[24] Here, the nanocluster $[Ag_{62}S_{12}(SBu^t)_{32}]^{2+}$ with an $Ag_6@Ag_8$ core exhibits unique optical properties.[25] Various $M^I_6Ag^I_8$ (M = Cu, Ag, and Au) clusters protected by different ligands have been continuously reported, e.g., $[Cl@M^I_6Ag^I_8(C{\equiv}CFc)_{12}]^+$ ($\equiv CFc^-$ = ferrocenylacetylide).[26] These molecular clusters serve as precursors for functional materials. Their cores consist of nested spheres of octahedra and cubes: $M_6@E_8$ (M = metal; E = chalcogen).[27] $Mo_6@S_8$ clusters catalyze $CO_2$ hydrogenation to methanol, benefiting from their cage-like geometry.[28] Clusters composed of $[Co_6Se_8(PEt_3)_6][C_{60}]_2$ and $[Cr_6Te_8(PEt_3)_6][C_{60}]_2$, with the $Co_6@Se_8$ and $Cr_6@Te_8$ cores, exhibit magnetic ordering at low temperatures.[29] The nanocluster $[Au@Cu_{14}(SPh^tBu)_{12}(PPh(C_2H_4CN)_2)_6]^+$ with an $Au@Cu_6@Cu_8$ framework, shows reduced sensitivity to air regarding phosphorescence emission, valuable for lighting and biomedical sciences.[30] Redox- and photo-active coordination cages, such as the Metal-Organic Cage (MOC) $Pd_6(RuL_3)_8$, aid in preserving photosensitive matrices and controlling photocatalytic reactions.[31] The composite hybrid cluster includes a metal halide cluster (MHC)$[I_6@Cu_8]^{2+}$ linked to Anderson-type anionic polyoxometalates (POMs) $[HCrMo_6O_{18}(OH)_6]^{2-}$, creating crystalline materials with novel properties.[32] Other high-nuclearity metal-containing clusters with $Q_6$(octahedron)@$T_8$(cube) structure include Si@$(AlCp^*)_6$@Al,[33] Ni@$Ni_8$ (cube)@$Te_6$ (octahedron)@$P_8$ (cube) with a double-dual core-shell structures[29] and $[Cl_6Ag_8@Ag_{30}(^tBuC{\equiv}C)_{20}(ClO_4)_{12}]\cdot Et_2O$ featuring a $Cl_6Ag_8$ core.[34]

Co-dual clusters by Icosahedron and Dodecahedron: The icosahedron and dodecahedron are dual structures to each other, with their co-dual cluster represented as $Q_{12}$ (icosahedron)@$T_{20}$ (dodecahedron). Utilizing the ultra-soft pseudopotential method (VASP), relativistic ADF, and Dmol3 codes confirm that the one-shell $Au_{32}$ hollow cluster exhibits shell closure and spherical aromaticity. An overlap occurs between the icosahedron and dodecahedron in a concentric circular arrangement of comparable sizes.[35] The bimetallic 18-electron superatomic nanocluster, $Au_{24}Ag_{20}$(2-SPy)$_4$(PhC≡C)$_{20}$Cl$_2$ (abbreviated as $Au_{24}Ag_{20}$, 2-SPy = 2-pyridylthiolate), incorporates three ligands (2-SPy$^-$, PhC≡C$^-$, Cl$^-$) on the $Au_{12}@Ag_{20}$ surface, with an $Au_{12}$ core within the metal-ligand shell. Moreover, ligand exchange occurs easily, resulting in various derivatives with the same metal core and novel functionalities.[36] Recently, an all-metal fullerene: $[K@Au_{12}Sb_{20}]^{5-}$ was synthesized by a wet-chemistry method,[37] with $Au_{12}Sb_{20}$ forming an $I_h$ co-dual structure.

Co-dual clusters by Truncated Tetrahedron**:** The truncated tetrahedron consists of four triangles and four hexagons, forms a compound structure denoted as $Q_4$ (tetrahedron)@$T_{12}$ (truncated tetrahedron). Metal clusters like $Au_4@Au_{12}$ have been extensively studied using trapped ion electron diffraction experiments and DFT calculations. $Au_{16}$ displays a cage-like spherical structure derived from dualizing the centers of four hexagons within the truncated tetrahedron.[38] Similarly, $Li_{16}$ was discovered through genetic algorithm structural searches, with its lowest energy among all isomers when forming a tetrahedral $T_d$ structure with a $Q_4@T_{12}$ shell of 16 lithium atoms and one enclosed heavy atom M@$Li_{16}$ (M=Ca, Sr, Ba, Ti, Zr, and Hf).[39]

Particularly, $Au_{20}$ features a stable pyramidal structure with dual formed at the vertices of the truncated tetrahedron.[40] Similar structures were found in both $A_4B_{16}$ (A = P, As)[41] and $Ti_8C_{12}$.[42] Following the discovery of $C_{60}$, fullerene-like silicon structures were sought. Ab initio pseudopotential plane wave calculations revealed that $Ti@Si_4@Si_{12}$ exhibits a significant HOMO-LUMO gap (2.35 eV).[43] In 1993, a 16-metal polyoxomolybdates, $[(Mo^{VI}O_3)_4Mo^{V}{}_{12}O_{28}(OH)_{12}]^{8-}$ ($Mo_{16}$), was reported. The truncated tetrahedral structure $Mo^{V}_{12}O_{28}(OH)_{12}$ with tetrahedral arrangements of four $\{Mo^{VI}O_3\}$ motifs constitutes the $Mo_{16}$ with complete tetrahedral $T_d$ symmetry.[44] This was followed by the ε-Keggin polyoxocation $[\varepsilon\text{-}PMo^{V}_{8}Mo^{VI}_{4}O_{36}(OH)_4\{La(H_2O)_4\}_4]^{5+}$ (ε-$La_4PMo_{12}$), where ε-$PMo_{12}O_{40}$ is stabilized by tetrahedral $\{La(H_2O)_4\}^{3+}$ groups, forming the $La_4@Mo_{12}$ framework, widely applicable in heterogeneous catalysis.[45] Complex iron clusters also exhibit similar architectures. Controlled hydrolysis of $Fe^{III}$ yielded two clusters: $Fe^{III}$-oxo/hydroxy cluster $[Fe_{17}O_{16}(\mu_4\text{-}OH)_{12}(py)_{12}Cl_4]^{3+}$ ($Fe_{17}$) formed from truncated tetrahedra with tetrahedrally arranged motifs,[46] and $[Fe_{34}O_{38}(\mu_2\text{-}OH)_{12}Br_{12}(py)_{18}]$ ($Fe_{34}$), which has hexagons covered by $\{Fe^{III}(\mu_3\text{-}O)_3Br\}_3$ motifs, resulting in an external $(Fe_3)_4@Fe_{12}$ framework.[47]

Co-dual clusters by Cuboctahedron: The co-dual cluster of a cuboctahedron, characterized by six squares as dual faces, is designated as $Q_6$ (octahedron)@$T_{12}$. Investigating into gold's tendency to form planar and cage-like clusters, it was discovered that $Au_6$ (octahedral) and $Au_{12}$ (cuboctahedron) constitute the cage structure of $Au_{18}$.[48] A similar structural is seen in $Mn_{12}Cl_6$, constructed through supramolecular methods, where a $\{Cl_6\}$ octahedral stabilizes the six squares of the cuboctahedron.[49] For high-nuclearity metal clusters, organic composite in small polyoxovanadates (POVs) yield V-O clusters with co-dual structure. For instance, $[H_8(V^{V}O_4)V^{IV}_{18}O_{42}]^{7-}$ exhibits a polyhedral arrangement of $V@V_6$ (octahedron)@$V_{12}$ (cuboctahedron).[50] Similarly, $\{PNb_{12}O_{40}(V^{IV}O)_6\}^{3-}$ and $[Ce^{IV}_{12}(V^{V}O)_6O_{24}(H_2O)_{24}(SO_4)_8]^{2+}$ exhibit $V_6$ (octahedron)@$Nb_{12}$[51] and $V_6$ (octahedron)@$Ce_{12}$[52] arrangements, respectively. Their functionalization with organic ligands achieves adjustable catalytic reaction activity. Polyoxoniobates (PONbs), developed later, integrate various metals; for example, the $[N(CH_3)_4]_{10}[Ti_{12}Nb_6O_{44}]$ framework features $Nb_6@Ti_{12}$, where a cuboctahedral $Ti_{12}$-shell $[Ti_{12}O_{14}]^{20+}$ is capped by six $\{NbO_5\}$ units in an octahedral fashion.[53] Notable high-nuclearity iron clusters include the $Bi_6$-coated $Fe_{13}$-oxo/hydroxy cluster, $Bi_6[FeO_4Fe_{12}O_{12}(OH)_{12}(O_2C(CCl_3)_{12}]^{+}$ ($Bi_6Fe_{13}$). This cluster features a central $\{Fe^{III}O_4\}$ tetrahedron surrounded by a cuboctahedral shell $\{Fe_{12}\}$, with six Bi atoms capping the structure in an octahedral arrangement.[54] The $[HFe_{19}O_{14}(OEt)_{30}]$ ($Fe_{19}$) cluster also possesses an $[Fe(\mu_6\text{-}O)_6]$ octahedral core akin to ferritin protein, with a similar framework structure $Fe@Fe_6$ (octahedral)@$Fe_{12}$ (cuboctahedron).[55] The largest $Fe_{42}$ cyanide-bridged polynuclear cluster comprises 18 $Fe^{III\text{-}HS}$ ions forming a $Fe_6@Fe_{12}$ framework, with 24 $Fe^{II\text{-}LS}$ ions bridging them via cyanide, exhibiting a rare hollow structure.[56] Other complexes with an $Q_6@T_{12}$ structure encompass transition metal cage complexes like $Eu_6@Cu_{12}$[57] and the MOC $\{Pd_6@Pd_{12}\}$.[58]

Co-dual clusters by Truncated Octahedron: Taking dual vertices from the eight hexagonal faces of a truncated octahedron yields the co-dual cluster $Q_8$ (cube) @$T_{24}$.

The compound $[Co^{II}_{24}Co^{III}_{8}O_{24}(TCA)_6(H_2O)_{24}]$ features a metal framework composed of a soladite-type $Co^{II}_{24}$ fragment and an encapsulated $Co^{III}_{8}$ cube, illustrating a $Co_8$ (cube) @$Co_{24}$ structure.[59] The core structure of the high-nuclearity POM cage $[(V^{V}O)_{16}(V^{IV}O)_8O_{24}(O_3AsC_6H_5)_8]$ is $As_8$@$V_{24}$. Such supramolecular coordination cages are useful for molecular recognition and catalysis.[60] By utilizing Bi-C bond cleavage reactions enables the synthesis of $[Bi_{38}(\mu_3\text{-}O)_{22}(\mu_4\text{-}O)_{22}(\mu_6\text{-}O)\{3,5\text{-}(NO_2)_2C_6H_3COO\}_{20}(OH)_4(DMSO)_{16}]\cdot4DMSO\cdot11H_2O$ (DMSO = dimethyl sulfoxide). Characterization of the $Bi_{38}$ oxocarboxylate cage revealed a structure where a $\{Bi_8\}$ cube is encapsulated by a truncated octahedral $\{Bi_{24}\}$ cage, containing a $\{Bi_6\}$ octahedral core.[61] Another example is the high-nuclearity POV cage $\{Cl_6@[(V^{V}O)_{16}(V^{IV}O)_8O_{24}(O_3AsC_6H_5)_8]\}_6$ ($Cl_6$@$As_8$@$V_{24}$), where As atoms cover the hexagonal faces of the V-based truncated octahedron, encapsulating a $\{Cl_6\}$ octahedron.[60] Extensive studies on clusters with actinide elements include the plutonium(IV) oxide nanocluster $[Pu^{IV}_{38}O_{56}Cl_{54}(H_2O)_8]^{14-}$ ($Pu_{38}$), which features a framework of $Pu_6$ (octahedron)@$Pu_8$ (cube)@$Pu_{24}$ (truncated octahedron), with a $\mu_6$-$O^{2-}$ centered octahedron $\{Pu_6\}$.[62] Transition metal cations with high-nuclearity coordination clusters have garnered increasing attention due to their intriguing electronic and magnetic properties. For instance, the bi-metallic $\{Co^{II}_{24}Mo_8\}$ nanospheres, $[Co^{II}_{24}(TC_4A)_6(MoO_4)_8Cl_6]^{2+}$, exhibit a core structure of $Cl_6$ (octahedron)@$Mo_8$ (cube)@$Co_{24}$ (truncated octahedron).[63] Furthermore, complex clusters like $\{Co^{II}_{24}\text{-}Co^{III}_{8}(\mu_3\text{-}O)_{24}(H_2O)_{24}(TC_4A)_6\}$ ($Co_{32}$) exemplify the $Co_8$@$Co_{24}$ cage architecture.[59] Most high-nuclearity metal carbonyl clusters containing interstitial Ni or Pt atoms exhibit multivalence encompassing several redox changes, such as $[Pt_{14}Ni_{24}(CO)_{44}]^{4-}$ ($Pt_6$@$Pt_8$@$Ni_{24}$), characterized by a distinct HOMO-LUMO gap.[64]

Co-dual clusters by Icosidodecahedron: The icosidodecahedron is composed of 30 triangles and 12 pentagons. The dual vertices corresponding to its pentagonal faces form the co-dual cluster represented as $Q_{12}$ (icosahedron)@$T_{30}$. The paramagnetic cluster $[Cu_{43}Al_{12}](Cp^*)_{12}$ possesses a distinctive open-shell 67-electron superatomic configuration, characterized by a core Cu@$Cu_{12}$ and a shell $Al_{12}$ (icosahedron)@$Cu_{30}$ (icosidodecahedron), forming a double-shell Mackay cluster.[65] Additionally, the low-valent copper cluster $[(IDipp)_6Cu_{55}]$ ($Cu_{55}$, IDipp = 1,3-bis(2,6-diisopropylphenyl)imidazol-2-ylidene) exhibits approximately icosahedral symmetry similar to $Cu_{43}Al_{12}$, featuring a Cu@$Cu_{12}$@$Cu_{30}$@$Cu_{12}$ arrangement.[66] In contrast, among the numerous giant Ti-O clusters that have been constructed, only one exhibits polyhedral characteristics: the fullerene-like $H_6[Ti_{42}(\mu_3\text{-}O)_{60}(O^iPr)_{42}(OH)_{12})]$ ($Ti_{42}$), with a $Ti_{12}$@$Ti_{30}$ skeleton exhibiting $I_h$ symmetry similar to $C_{60}$.[67]

Co-dual clusters by Snub Dodecahedron: Taking the 12 pentagonal as dual faces, the co-dual cluster is represented as $Q_{12}$ (icosahedron)@$T_{60}$. The distinctive properties stem from the relativistic effects of the gold hollow cage, which shows attractiveness in various catalytic applications. The first member of the icosahedral series is the $Au_{12}$ icosahedron, followed by $Au_{32}$, which combines an icosahedron with a dodecahedron. Similarly, the icosahedral $Au_{42}$ combines an icosahedron with an icosidodecahedron. Following this construction principle, $A_{72}$ is anticipated to resemble an onion-skin-like spherical aromatic cage. This chiral $Au_{72}$ cage integrates an icosahedron (12 vertices) with the snub dodecahedron (60 vertices). Quantum chemical calculations indicate that this cage possesses high thermodynamic stability and a large HOMO-LUMO gap.[68] The $Q_{12}@T_{60}$ structure is prevalent in giant viruses. For instance, the capsids of Simian virus 40 ($SV_{40}$)[69] and murine polyomavirus (polyoma)[70] consist of pentamers formed by 72 major structural protein VP1, with each VP1 core containing a β-roll domain. Refinement of the structure through high-resolution cryo-electron microscopy (cryo-EM), at 3.1 Å resolution, reveals that an icosahedron is composed of 12 pentamers, while 60 pentamers occupy the vertices of the snub dodecahedron, interconnected via their C-terminal arms. Human papillomavirus (HPV)[71] and bovine papillomavirus type 1 (BPV-1)[72] share a similar structure, but in the $Q_{12}@T_{60}$ configuration, each vertex features a pentamer of the major capsid protein L1. Alternate HPV production methods using virus-like particles (VLPs) have been developed for virological and structural studies, effectively preserving the native capsid structure and have facilitating for vaccine development. Furthermore, cryo-EM studies of the cauliflower mosaic virus (CaMV) show an outer capsid formed by 72 capsomers, aligning with the $Q_{12}@T_{60}$ structure.[73]

Co-dual clusters by Truncated Icosahedron: The dual of the truncated icosahedron forms a dodecahedron, with each hexagon center serving a dual point. This configuration results in the co-dual cluster $Q_{20}$ (icosahedron)@$T_{60}$. Multi-vertex giant clusters with this structure are uncommon. $B_{80}$ is a cage with $I_h$ symmetry derived from the metastable $B_{60}$, achieved by placing an additional B atom at each hexagon center to form $B_{20}@B_{60}$. $B_{80}$ demonstrates a relatively large HOMO-LUMO gap (~1 eV) and higher energetic stability compared to the constituents of boron nanotubes (boron double rings).[74]

The practical applications of these clusters across diverse fields highlight the significance of their co-dual structures. Successful predictions and syntheses validate the utility of highly symmetric geometric structures in cluster design. In this paper, cages of concentric spheres designed based on co-dual structure are constructed. With empirical selections of co-dual clusters, it has achieved 68.5% stability rate of the candidate clusters. This high survival rate underscores the immense value of designing clusters using highly symmetrical co-dual structures. Inspired by this approach, future research could explore novel clusters by randomly arranging a broader range of elements or charged clusters based on co-dual structures.

Figure S2 clearly distinguishes primary categories of convex polyhedra: Platonic polyhedra, semi-regular polyhedra, and Johnson polyhedra and the others. The Platonic solids, known as the five regular polyhedra, are composed of identical regular polygons with all vertices being identical. Semi-regular polyhedra consists of two or more different types of regular polygons while maintain uniformity across all vertices. A notable subset includes the 13 Archimedean solids, which can all be derived from regular polyhedra through operations such as truncation, snubbing, and twisting.[75] The prism with lateral edges perpendicular to its polygonal base is termed a regular prism, showing an infinite variety. By slightly rotating the base of a regular prism so that each lateral face becomes an equilateral triangle, the antiprism is formed. The third category of Johnson solids[76] refers to convex polyhedra composed exclusively of regular polygons, excluding the regular and semi-regular polyhedra. Zalgaller demonstrated that there are precisely 92 such polyhedra. It is evident that among all the convex polyhedra discussed above, Platonic and Archimedean polyhedra exhibit the highest symmetry.

There is one more parameter should be considered, the pairs' relative sizes. Figure 2 shows the combination of octahedron and its dual (cube). The sizes of the octahedron and cube are represented by the radius of the sphere, which includes all the vertex. Seven combinations should be considered, as shown as Figure 2(a~g). The conjugated combination of the dual pairs would keep the original symmetry or even promote it. In this paper, we are more interested in those spherical clusters with shell structures.

In clusters research, empirical consideration such as bond lengths, should also be considered. Since Archimedean solids are characterized by multiple types of regular polygons, overlaying the original polyhedron with its dual polyhedron reveals significant differences in bond lengths represented by the dual-colored solid lines in Figure S4. The truncated cube, composed of 6 octagons and 8 triangles, provides a good example. In its full dual compound polyhedron f-$A_3$, the dual edges corresponding to octagons are significantly longer than those corresponding to triangles. Theoretical analysis suggests that such a pronounced contrast in bond lengths between vertices of the same atom in the full dual structure is highly unreasonable. Indeed, computational tests results indicate that when the dual vertices of f-$A_n$ (n=1~13) consist of the same atom, the relaxed structure frequently collapses into a nested double-shell configuration. These atoms struggle to stabilize under the conflicting demands of varying bond lengths, thereby forming separate shells internally. Clearly, these structures diverge from the anticipated sphere-like convex polyhedra.

To achieve more reasonable compound polyhedra for clusters, only the largest regular polygon from each Archimedean solids are selected as the dual face. The combined polyhedra of the partial dual structure with their original polyhedra are defined as co-dual structures. Their specific configurations are depicted in Figure 3, named $A_1$ to $A_{13}$ in increasing order of vertex count. An intriguing observation is that, for Archimedean solids, the partial dual structure represented by solid red lines are well-known distinctive structures, as summarized in Table 1. Partial dual structures and their original polyhedra share identical symmetries. By combining them, a set of 16 co-dual structures with enhanced spherical characteristics and high symmetry is achieved.

Utilizing the identified co-dual structure, occupying the vertices with all 118 elements from the periodic table provides initial objects for the required calculations. Among various multi-element hollow cage-like structures, the two fundamental composition modes are selected for search: shells composed of a single element and those composed of two elements, where one element occupies the vertices of the original polyhedron and the other occupies vertices of the partial dual polyhedron. Within the two kinds of shells, there is an additional way to embed another atom at the center point resulting in core-shell clusters with the same symmetry. Through these design approaches, four types of clusters are employed for calculations: $S_1$, $D_1$, $S_2$, $D_2$. If a simple permutation of the 118 elements applied to the four clusters is performed, the staggering data presented in Table 2 demonstrates the immense computational complexity. The largest number of $D_2$ clusters alone reaches 26,065,728, with a total of 26,511,296 clusters across all four types. It is imperative to apply rational rules in identifying appropriate computational targets.

Primarily, from an external geometric perspective, the partial dual modes of the co-dual structure effectively preselect suitable computational targets, constituting essential Rule 1. This process aims to ensure that chemical bonds for the same atomic species do not differ excessively as well as aligning bond lengths appropriately between distinct atoms. When a vertex of Platonic solid (or Archimedean solid) is occupied by one element, it is bonded to the other element on the vertex of the dual. Whether or not the chemical bonds of the two will match depends on their relative atomic sizes, since larger atoms typically forming longer bonds. Therefore, for dual-element shells, selecting pairs of elements with a radius difference within 0.2 Å is stipulated as Rule 2. Specially, for truncated cube, truncated cuboctahedron, and truncated dodecahedron, truncated icosidodecahedron containing octagons, and decagons, an upper limit on the radius difference between the two elements is not set due to the presence of excessively large dual faces.

Aside from these geometrically determined selection rules, further refinement is achievable through an examination of electronic properties. To link cluster stability with electronic structure and total electron count, the jellium model serves as Rule 3. Previously, Martins et al.[77] recognized similarities between electronic orbital from a molecular calculation with those from a jellium picture. In 1984, Knight et al.[78] found that the electronic structure of metal clusters reflects that of spherical potential wells, with clusters having a higher abundance of valence electrons matching closed-shell numbers. The complete filling of electronic shells in such spherical potential well is termed 'magic numbers'. Clusters with a total number of valence electrons corresponding to these magic numbers exhibit closed electronic shell configurations, thereby demonstrating heightened stability. The observation of magic numbers in experimental mass spectra occurs when electronic shells are completely filled by 2, 8, 18, 20, 34, 40, 58, 68, 70, 92, 106, 112, 138, and 156 electrons, respectively. We have discovered the stable cage structure of $Au_{32}$ with 32 valence electrons, thus incorporating 32 as a magic number.[35, 79]

By applying above three rules (dual surface selection, atomic radius differences, and valence electron rules), an increased likelihood of theoretically stable cages is derived. When embedding elements in these cages, those with radii greater than 1.9 Å, such as K, Rb, Cs, Fr, Sr, Ba, Ra, and Yb, are chosen for incorporation into these cages. It is important to note that certain non-metallic elements are excluded from the selectable range: H, He, C, N, P, some chalcogens (O, S, Se, Te), and halogens. Consequently, the total number of four types of co-dual clusters has been reduced to 6,786 as listed in Table 2.

In general, the co-dual clusters in this paper are composed through two approaches: 1) using a single element to form co-dual clusters; 2) positioning two elements respectively in the positions of Platonic solids (or Archimedean solids) and their dual polyhedra. With or without embedding large-size atom at the center, four types of clusters are formed: $S_1$ (co-dual clusters with single-element shell and no center-embed atom: $X_n$); $D_1$ (co-dual clusters with dual-element shell and no center-embed atom: $X_nY_m$); $S_2$ (co-dual clusters with single-element shell and center-embed atom: $M@X_n$); $D_2$ (co-dual clusters with dual-element shell and center-embed atom: $M@X_nY_m$). Here, X and Y represent different elements at the shell; M represents the embedded atom at the center; n and m respectively denote the number of X and Y. If all the possibility of such co-dual clusters are considered, one would get 26,511,296 clusters (18 co-dual structures, 118 elements: $18 \times P_{118}^{2} \times 118$). However, most of them are obvious unsuitable due to empirical conditions, such as valence electron counting rules, and radius matching. By those criteria, the total count of $S_1$, $D_1$, $S_2$ and $D_2$ was reduced from 26,511,296 to 6,786.

First-principle calculations were subsequently performed on these 6,786 co-dual clusters to identify stable structures, where 4,651 stable neutral clusters are achieved with a remarkable stability rate of 68.5%. The number of stable structures identified for each type is summarized in Table 3. Among these 16 co-dual structures, as vertex numbers increase, very few clusters adhere to the valence electron rule, and even fewer structures stabilized. The stability rates are as follows: $S_1$ (90%), $D_1$ (47.3%), $S_2$ (84.1%), and $D_2$ (73.4%). The stability rate of all clusters is 68.5%. Such high stability rates affirm the validity of selection rules. Six clusters from 70 with cage structures of only one element ($S_1$) are shown in Figure 4(a). For example, $Mg_{16}$ has the shell structure from $Li_{16}X$[25], and $Rb_{32}$ is found to be stable at the $Au_{32}$ fullerene structure ($Z_3$)[21]. It is surprising that we firstly found another $Au_{32}$ fullerene structure with $A_5$ co-dual structure, since a huge wave of gold fullerene hunting had been carried out after the $Au_{32}$ fullerene ($Z_3$) was found. Eight representatives from 1472 embedded clusters with one-element shell ($S_2$) are shown in Figure 4(b). There are a lot more stable clusters with two-element shell. The representatives of 416 $D_1$ and 4,828 $D_2$ stable clusters are shown in Figure 5. All stable clusters are provided in Table S1-16 (non-embedded clusters $S_1$, $D_1$) and Table S17-32 (embedded clusters $S_2$, $D_2$).

# Conclusion

A novel family of clusters (co-dual clusters) are constructed by the conjugated dual structures of regular and semi-regular polyhedra. Four types of co-dual clusters are formed using elements from the periodic table: $S_1$ ($X_n$), $D_1$ ($X_nY_m$), $S_2$ ($M@X_n$), and $D_2$ ($M@X_nY_m$). By applying three empirical rules (dual surface selection, atomic radius differences, and valence electron rules), the initial count of 26,511,296 configurations was reduced to 6,786. Computational analysis confirms the stability of up to 68.5% of these clusters. These results highlight the reliability of cluster design based on highly symmetric co-dual structures. The research value of co-dual structure extends far beyond our current study. Future investigations could explore alternative methods for predicting stable clusters, including the incorporation of a broader array of elements into co-dual structures. The co-dual structure introduces a novel perspective that promises to advance the field of cluster design research.

# Acknowledgments

This study was financially supported by the National Natural Science Foundation of China (No. 12274246). All the calculations were performed at the Supercomputing Center of Ningbo University, the Supercomputing Center of Fudan University and Ningbo Artificial Intelligence Supercomputing Center.

# Methods

Based on density functional theory (DFT), all calculations were carried out using the Vienna Ab initio Simulation Package (VASP),[80] employing the generalized gradient approximation in the Perdew-Burke-Ernzerhof (PBE) form.[81] The projector augmented wave (PAW)[82] were utilized to describe interactions between core and valence electrons. A plane-wave energy cutoff of 450 eV was employed, and the convergence criterion for electronic energy was set at $10^{-6}$ eV. In computations, individual clusters were placed within a cubic supercell of 35×35×35 $Å^3$. Using large supercells ensured that interactions between the cluster and its periodically adjacent clusters were negligible. Meanwhile, the gamma point was exclusively used for K-space sampling.

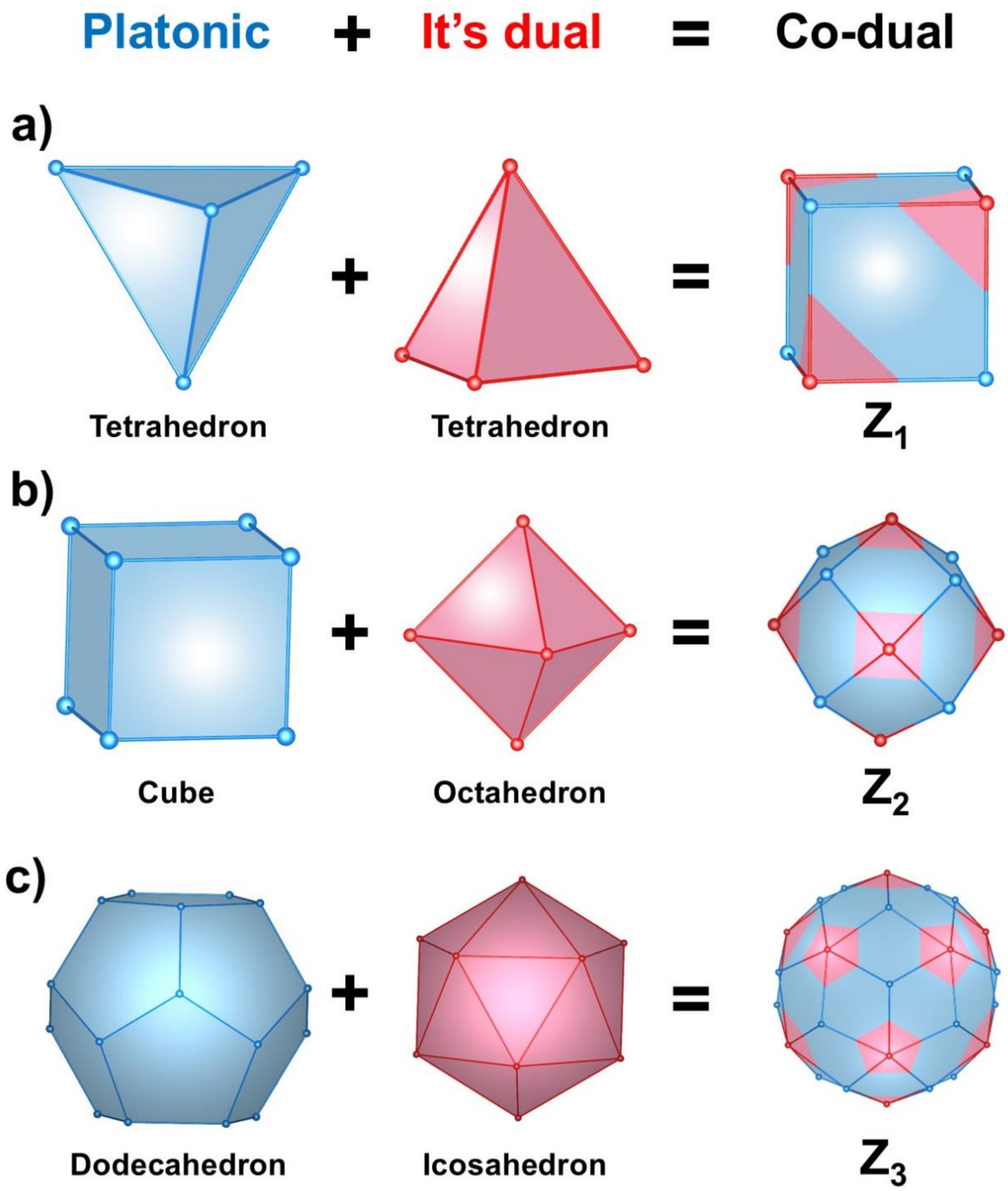

Figure 1. Dual polyhedra of Platonic solids and the corresponding co-dual structure. (a) The tetrahedron is self-dual, and it's co-dual is cube ($Z_1$). (b)The dual polyhedron of cube is octahedron, and $Z_2$ is the corresponding co-dual structure. (c) The dual of dodecahedron is icosahedron, and their co-dual structure ($Z_3$) is the structure of $Au_{32}$ gold Fullerene.

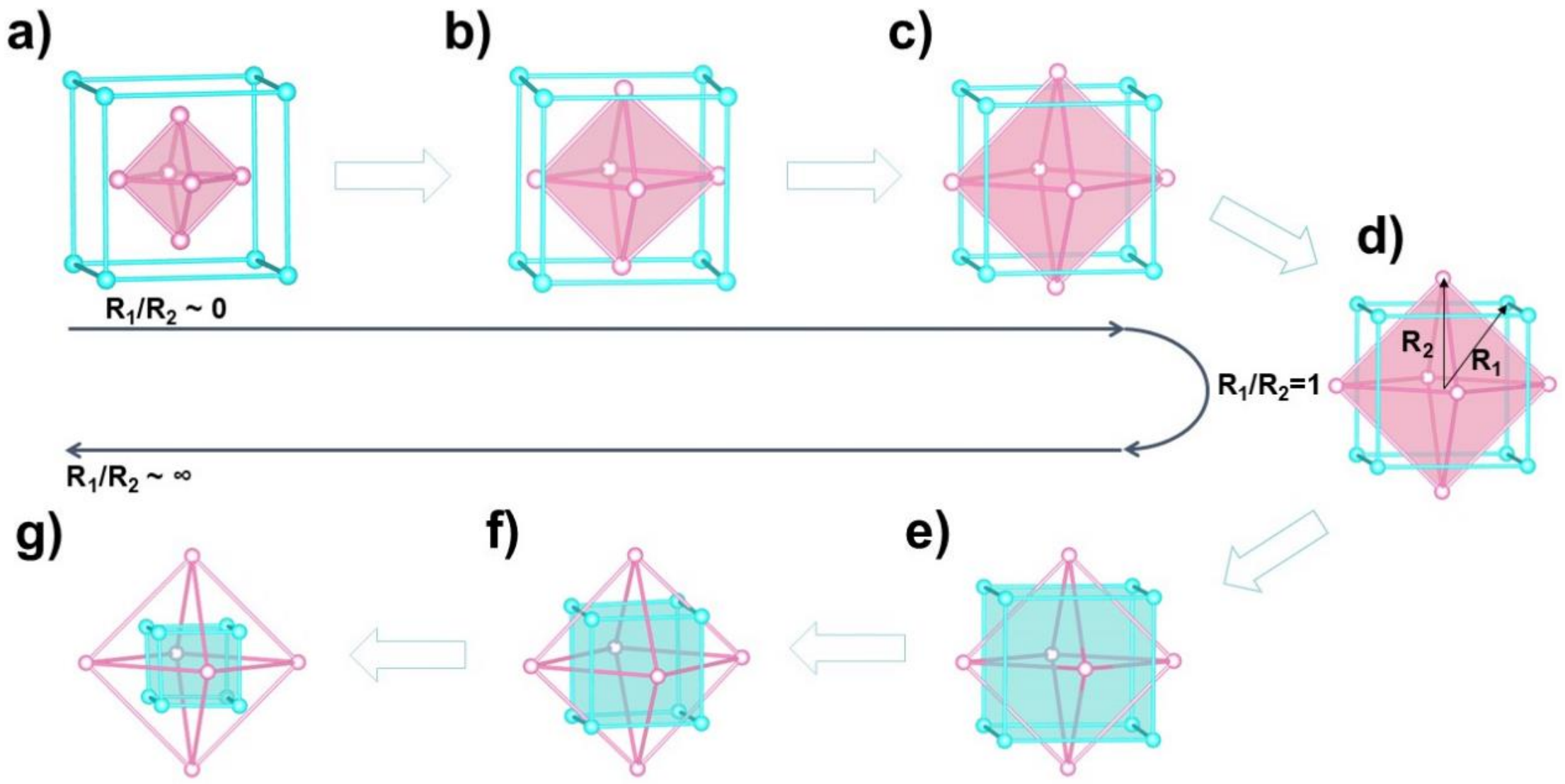

Figure 2. The conjugate combination of two dual pair with different sizes (cube and octahedron). $R_1$, $R_2$ are the radius of the spheres, which include all the vertex of the polyhedra, representing the relative size of the pair. a) and g) the two tetrahedra have the significant size difference. b) and f) The vertex of one tetrahedron are located on the faces of the other. c) and e) The sizes are comparable, but not the same. d) The dual pairs have the same size.

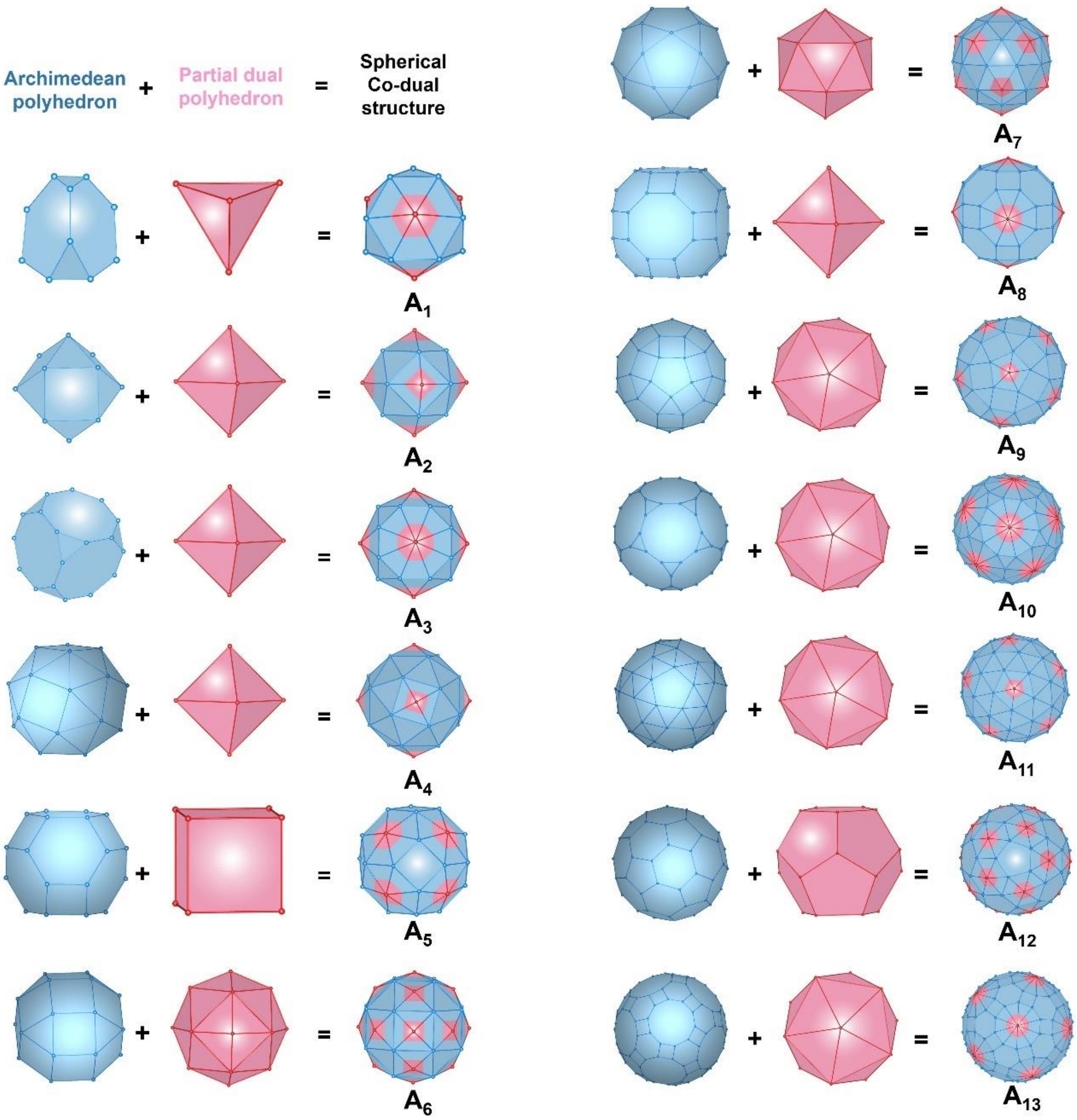

Figure 3. Spherical co-dual structures based on partial dual structures of all thirteen Archimedean (semi-regular) solids.

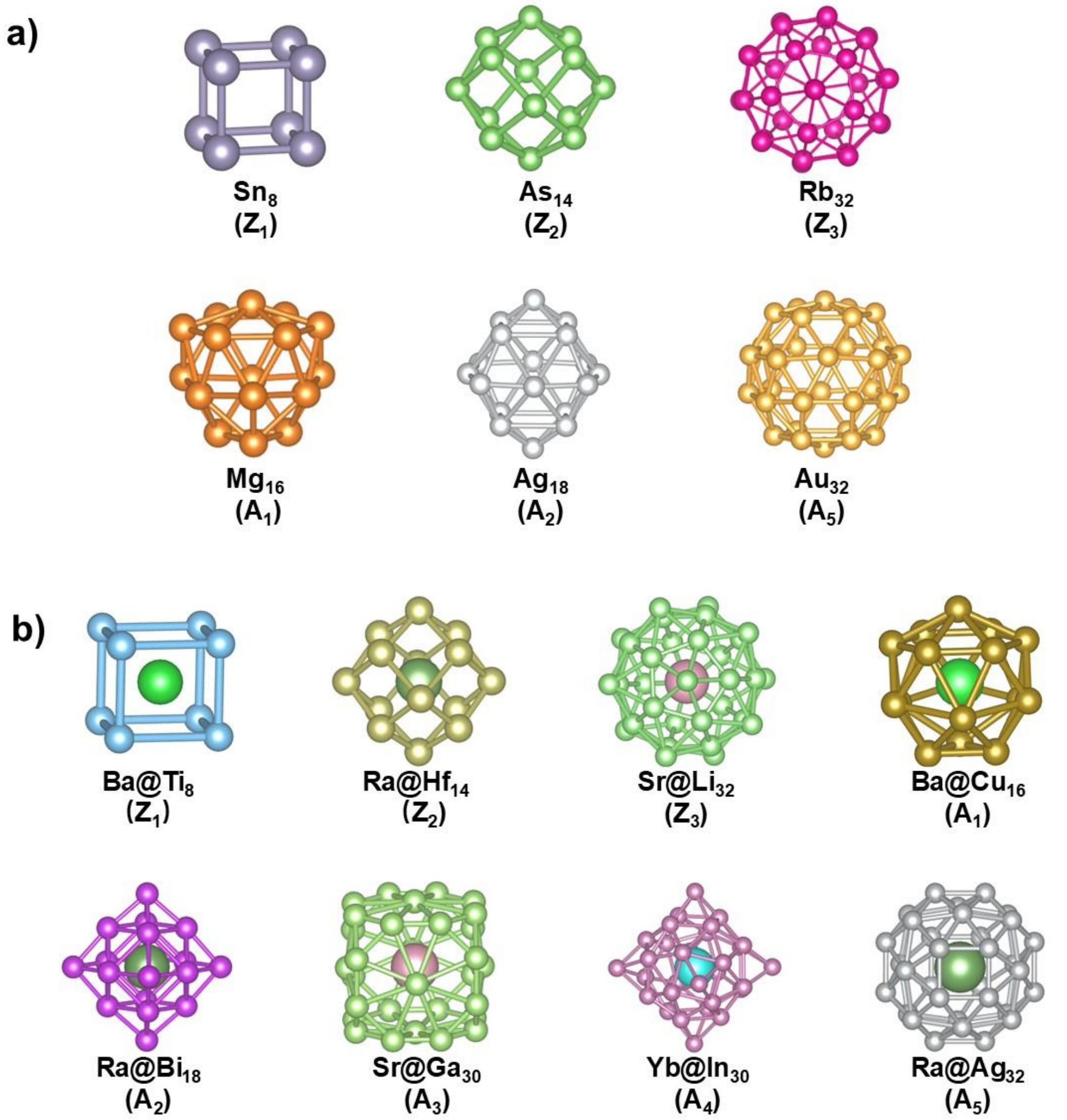

Figure 4. Representative clusters for co-dual structures with one-element shell. a) $S_1$ cage structures ($X_n$) and b) $S_2$ embedded cage structures ($M@X_n$).

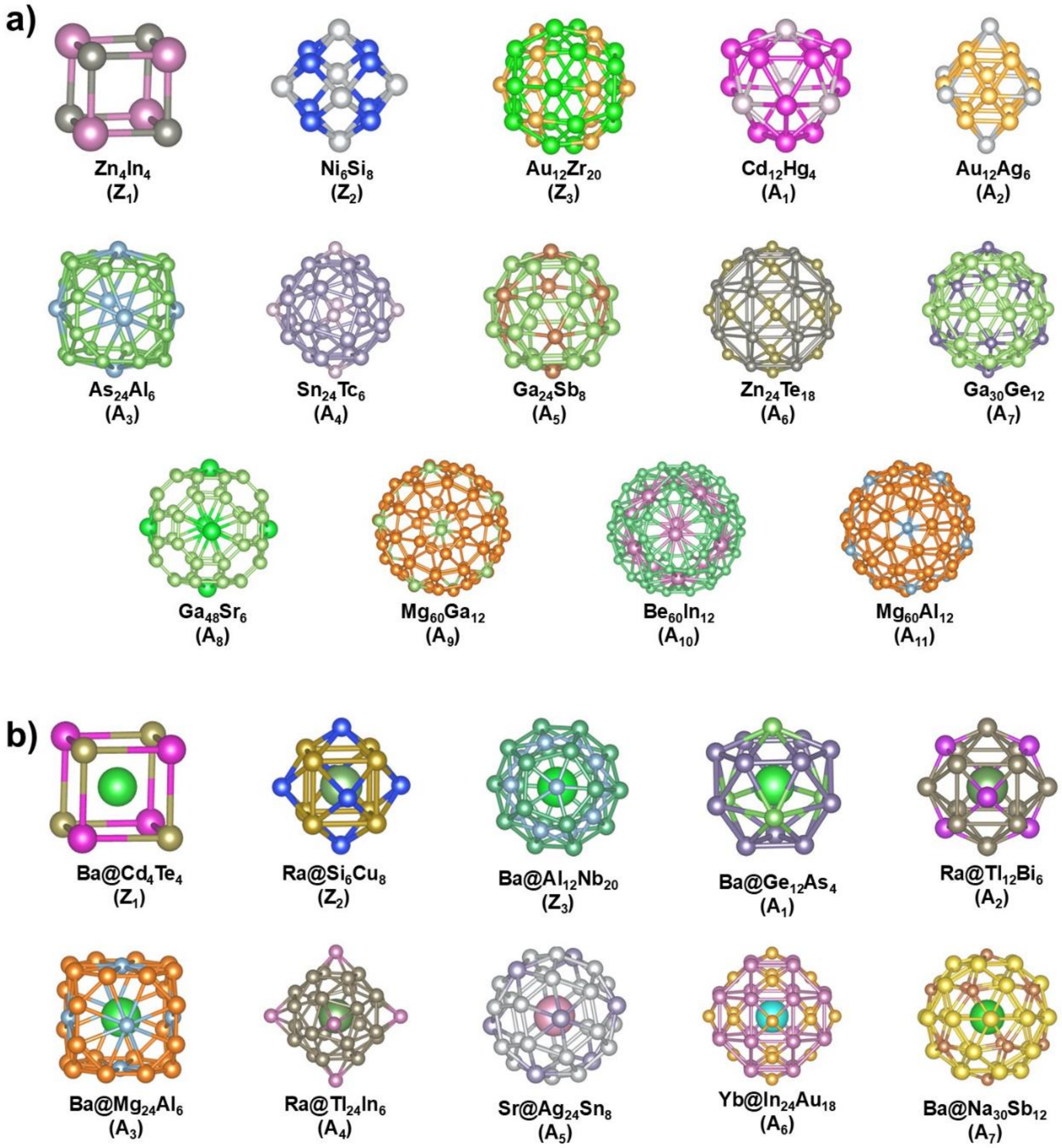

Figure 5. Representative clusters for co-dual structures with two-element shell. a) $D_1$ cage structures ($M@X_nY_n$) and b) $D_2$ embedded cage structures ($M@X_nY_n$).

Table 1. Platonic, Archimedean solids and their co-dual polyhedra by partial dual mode.

| Plantonic and Archimedean solids | Partial dual mode ($m_n$)[a] | Co-dual solids |
|---|---|---|
| Tetrahedron ($T_h$) | $4_3$ | Tetrahedron ($T_h$) |
| Cube ($O_h$) | $6_4$ | Octahedron ($O_h$) |
| Octahedron ($O_h$) | $8_3$ | Cube ($O_h$) |
| Dodecahedron ($I_h$) | $12_5$ | Icosahedron ($I_h$) |
| Icosahedron ($I_h$) | $20_3$ | Dodecahedron ($I_h$) |
| Truncated tetrahedron ($T_h$) | $4_6$ | Tetrahedron ($T_h$) |
| Cuboctahedron ($O_h$) | $6_4$ | Octahedron ($O_h$) |
| Truncated cube ($O_h$) | $6_8$ | Octahedron ($O_h$) |
| Truncated octahedron ($O_h$) | $8_6$ | Cube ($O_h$) |
| Rhombicuboctahedron ($O_h$) | $12_4$ | Cuboctahedron ($O_h$) |
| Truncated cuboctahedron ($O_h$) | $6_8$ | Octahedron ($O_h$) |
| Snub cube ($O$) | $6_4$ | Octahedron ($O_h$) |
| Icosidodecahedron ($I_h$) | $12_5$ | Icosahedron ($I_h$) |
| Truncated dodecahedron ($I_h$) | $12_{10}$ | Icosahedron ($I_h$) |
| Truncated icosahedron ($I_h$) | $20_6$ | Dodecahedron ($I_h$) |
| Rhombicosidodecahedron ($I_h$) | $12_5$ | Icosahedron ($I_h$) |
| Snub dodecahedron ($I$) | $12_5$ | Icosahedron ($I_h$) |
| Truncated icosidodecahedron ($I_h$) | $12_{10}$ | Icosahedron ($I_h$) |

[a] indicates a set of m n-gonal faces.[75]

Table 2. Number of clusters of different types with or without empirical conditions. N: number of co-dual structures.

| | Types | Number of clusters without empirical conditions | Number of clusters with three rules |
|---|---|---|---|
| **Cage clusters** | Single-element shell ($S_1$) | N×$P_{118}^{1}$=1,888 | 70 |
| | Dual-element shell ($D_1$) | N×$P_{118}^{2}$=220,896 | 1,472 |
| **Endohedral clusters** | Single-element shell ($S_2$) | N×$P_{118}^{1}$ × 118=222,784 | 416 |
| | Dual-element shell ($D_2$) | N×$P_{118}^{2}$ × 118=26,065,728 | 4,828 |
| **Total** | | 26,511,296 | 6,786 |

Table 3. Number of stable clusters in DFT calculations.

| Type of co-dual structures | $S_1$ ($X_n$) | $D_1$ ($X_nY_m$) | $S_2$ ($M@X_n$) | $D_2$ ($M@X_nY_m$) |
|---|---|---|---|---|
| $Z_1$ | 20 | 23 | 64 | 323 |
| $Z_2$ | 9 | 159 | 28 | 780 |
| $Z_3$ | 8 | 62 | 30 | 150 |
| $A_1$ | 11 | 161 | 94 | 772 |
| $A_2$ | 8 | 64 | 65 | 732 |
| $A_3$ | 0 | 10 | 22 | 79 |
| $A_4$ | 0 | 48 | 19 | 106 |
| $A_5$ | 7 | 65 | 28 | 470 |
| $A_6$ | 0 | 9 | 0 | 45 |
| $A_7$ | 0 | 46 | 0 | 85 |
| $A_8$ | 0 | 15 | 0 | 0 |
| $A_9$ | 0 | 13 | 0 | 0 |
| $A_{10}$ | 0 | 1 | 0 | 0 |
| $A_{11}$ | 0 | 20 | 0 | 0 |
| $A_{12}$ | 0 | 0 | 0 | 0 |
| $A_{13}$ | 0 | 0 | 0 | 0 |
| **Total** | 63 | 696 | 350 | 3,542 |
| **Stability rate** | 90% | 47.3% | 84.1% | 73.4% |

Graphic Abstract:

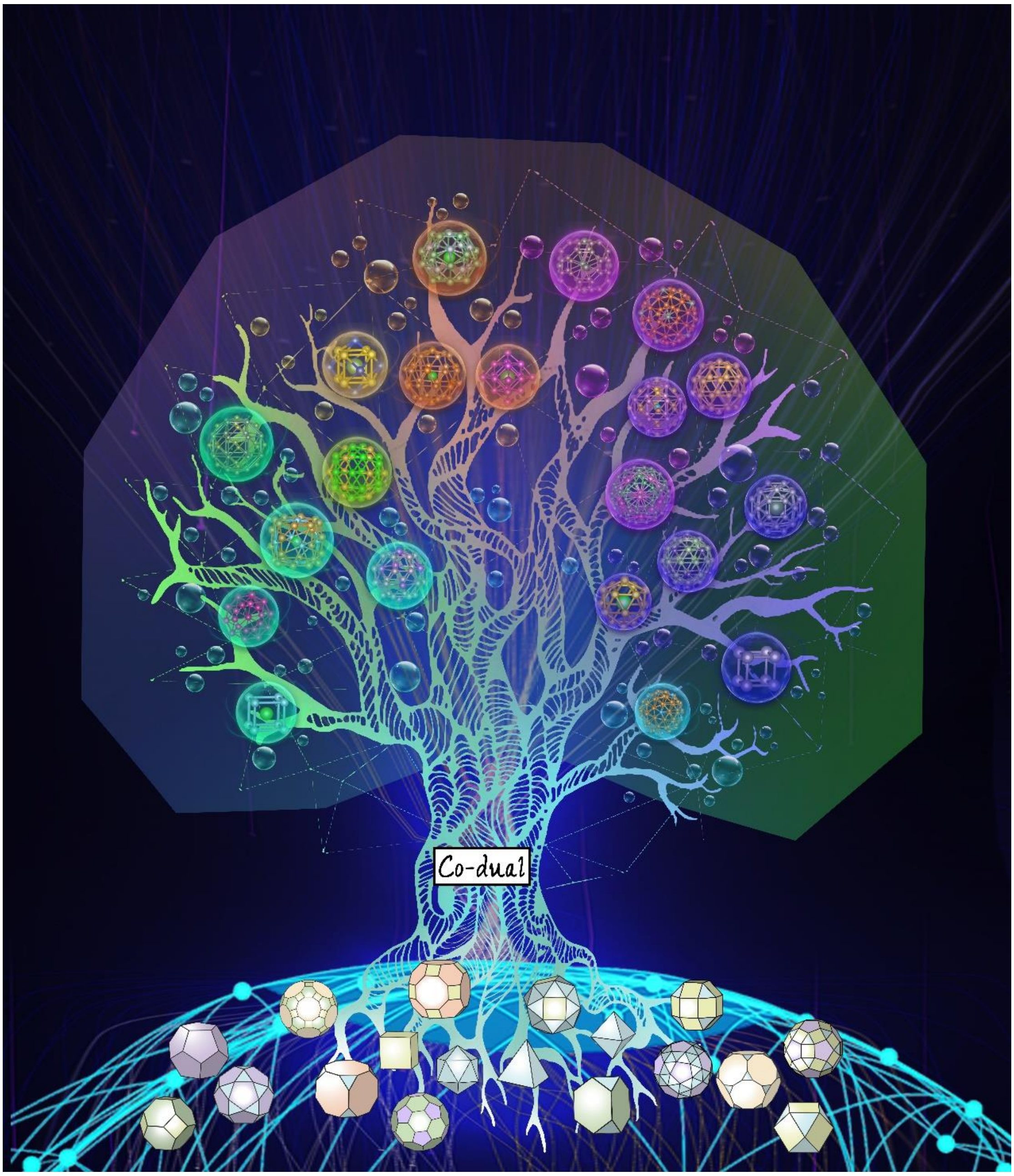

# Conjugate-dual clusters

Silei Wang[1], Jing Tian[1], Jiayu Li[1], Tian Gong[2], Xing Yan[3], Jijun Zhao[4] and Xin-Gao Gong[5], Xiao Gu[1*]

[1] School of Physical Science and Technology, Ningbo University, Ningbo 315211, China

[2] School of Mathematics and Statistics, Ningbo University, Ningbo 315211, China

[3] School of Civil & Environmental Engineering and Geography Science, Ningbo University, Ningbo 315211, China

[4] Guangdong Provincial Key Laboratory of Quantum Engineering and Quantum Materials, School of Physics, South China Normal University, Guangzhou 510006, China

[5] Department of Physics, Key Laboratory for Computational Science (MOE), State Key Laboratory of Surface Physics, Fudan University, Shanghai 200433, China

*Email: guxiao@nbu.edu.cn*

# Supporting Information

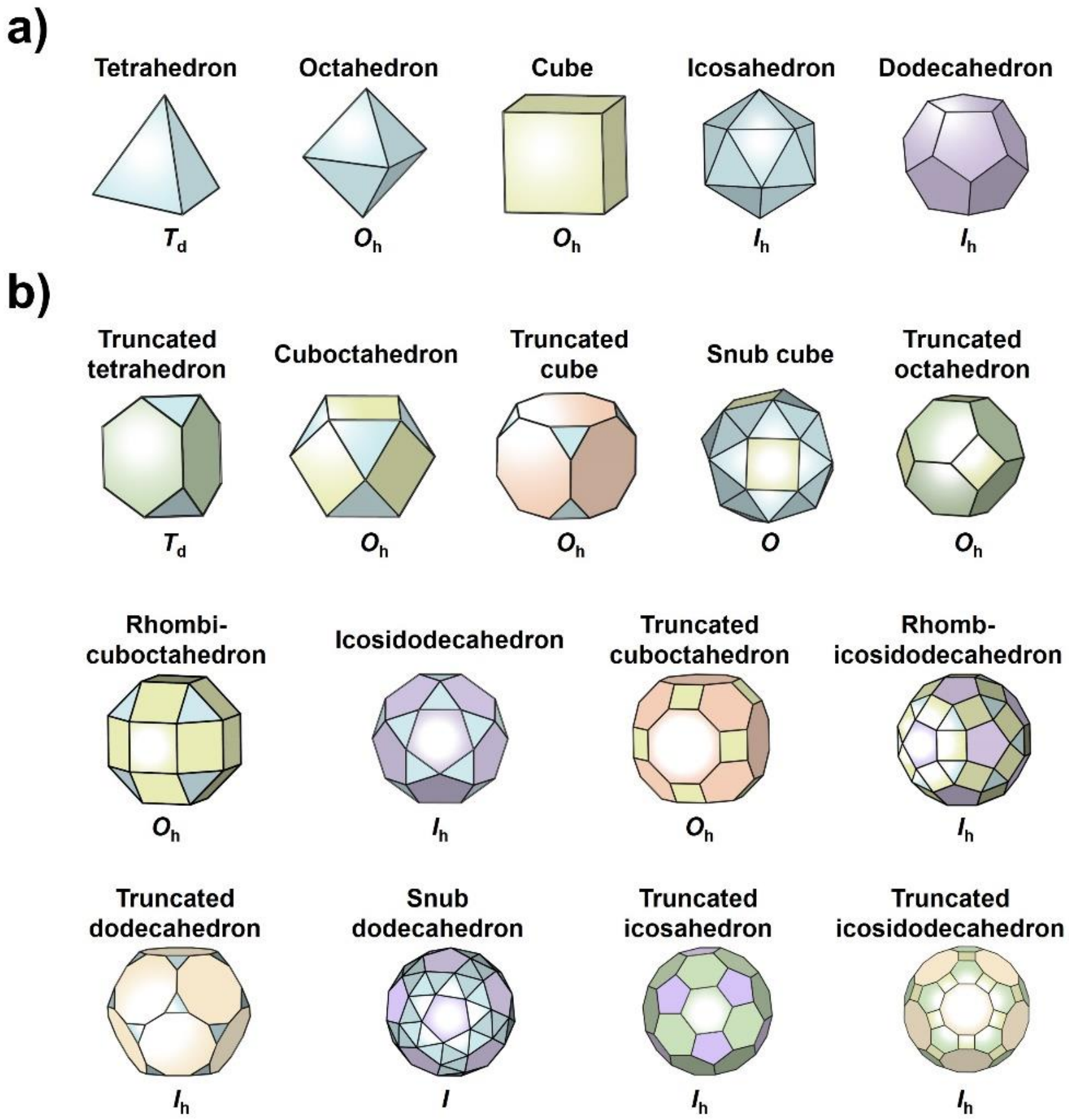

Figure S1. a) The five Platonic (regular) solids and their point group. b) All thirteen Archimedean (semi-regular) solids and their point group.

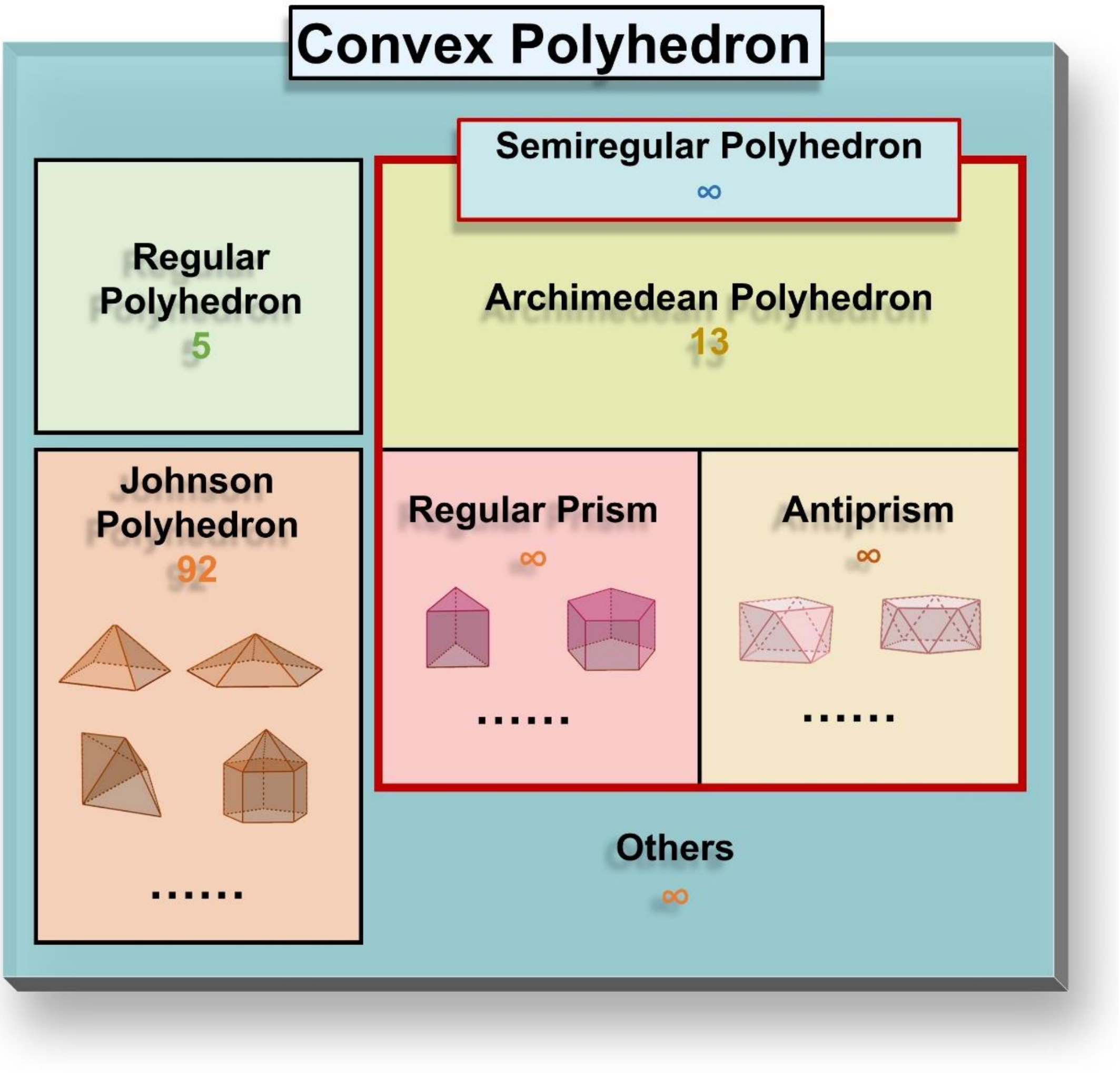

Figure S2. Category of convex polyhedra. Among them, 5 regular polyhedra and 13 Archimedean polyhedra have been shown in FigureS1.

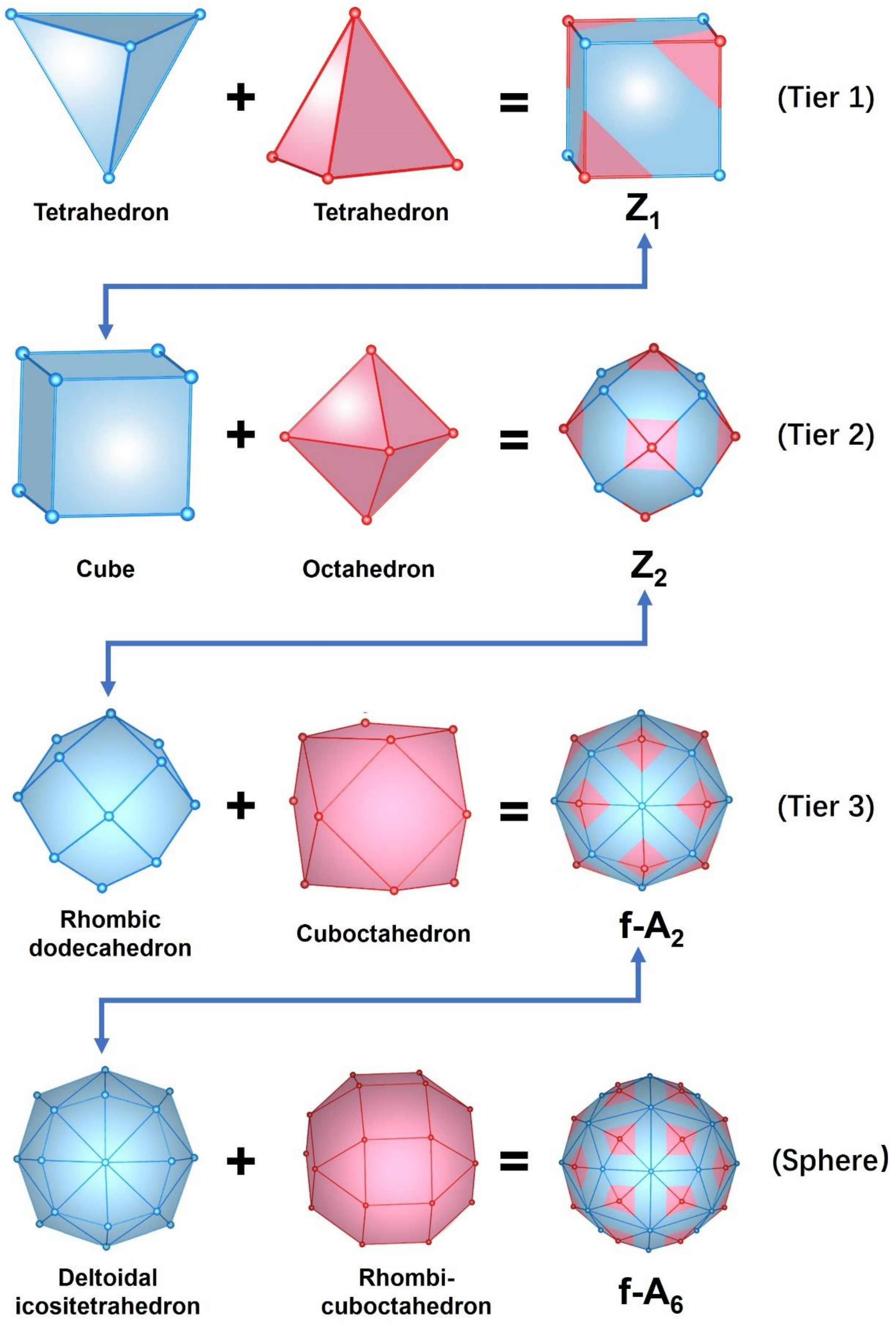

Figure S3 Hierarchy of conjugate combination of dual pair of polyhedra.

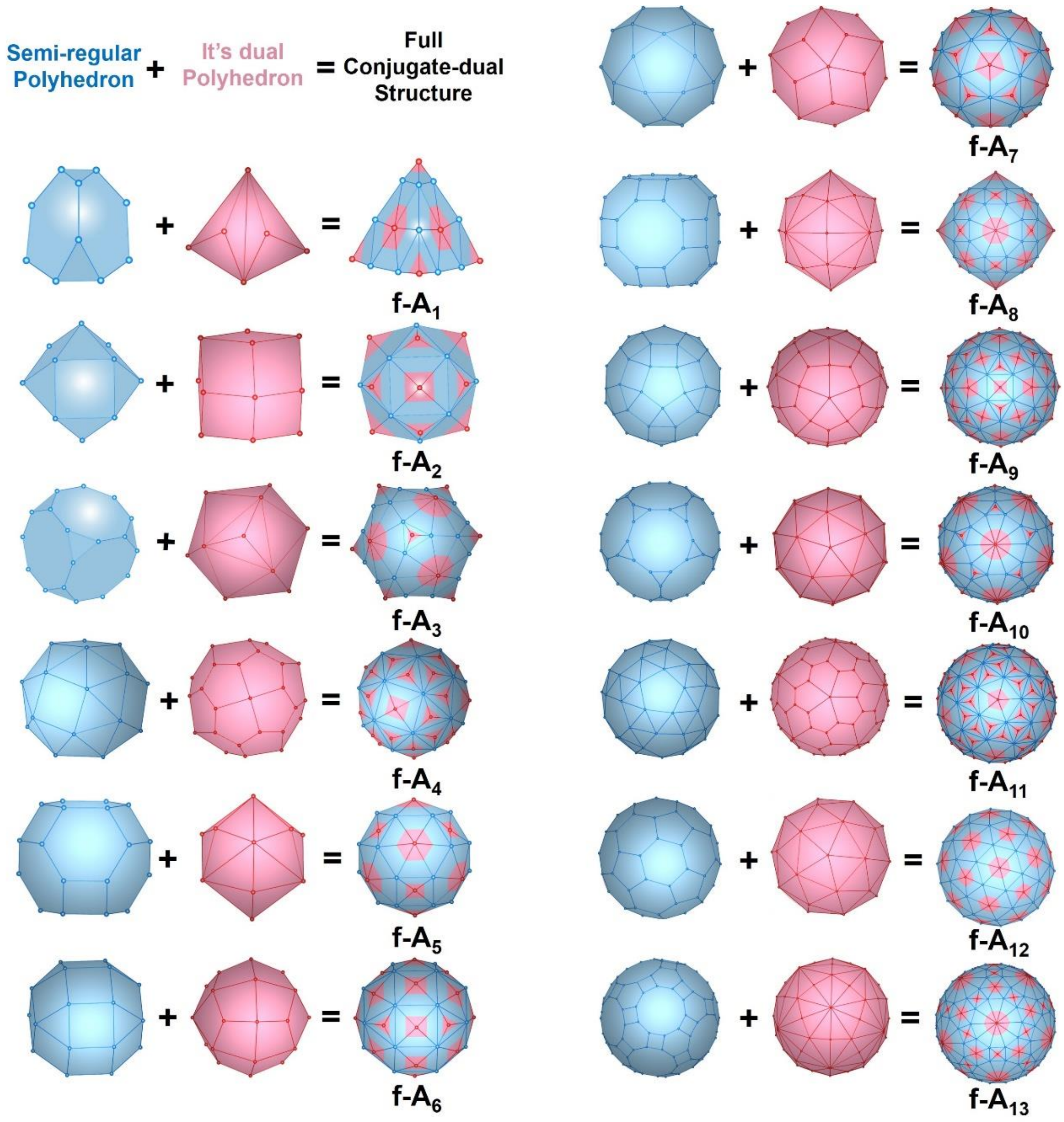

Figure S4 The dual polyhedron formed when each face of the 13 semiregular polyhedra takes a dyadic point form a conjugate-dual structure with the 13 semiregular polyhedra.

Table S1. The molecular formula of the stable structure of $Z_1$ when the co-dual structure is not embedded. In the displayed structural model, for a single-element structure, both colored balls represent X; for a two-element structure, the blue ball represents X and the pink ball represents Y. (This representation rule applies to TableS1-TableS16.)

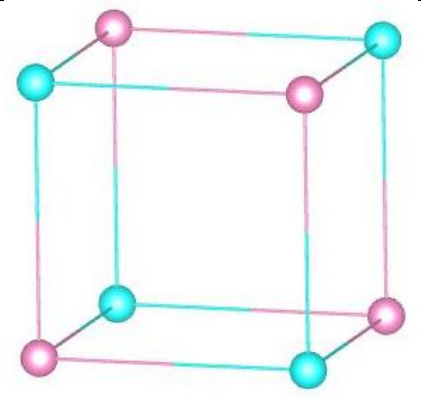

$Z_1$ Co-dual structure

| |
|---|
| **$X_8$ (20)** |
| $Ag_8$; $Au_8$; $Cs_8$; $Fr_8$; $K_8$; $Cu_8$; $Li_8$; $Na_8$; $Rb_8$; $Ge_8$; $Pb_8$; $Si_8$; $Sn_8$; $Bi_8$; $Sb_8$; $As_8$; $Hf_8$; $Zr_8$; $Nb_8$; $Ta_8$; |
| **$X_4Y_4$ (23)** |
| $Li_4Sn_4$; $Au_4Ge_4$; $Au_4Sn_4$; $Cu_4Ge_4$; $Zn_4In_4$; $Cu_4Rh_4$; $Cd_4Cr_4$; $Hg_4Mo_4$; $Ga_4V_4$; $Sc_4Tc_4$; $Pb_4Mo_4$; $Pb_4W_4$; $Mn_4Pt_4$; $Re_4Pt_4$; $Tc_4Ni_4$; $Tc_4Pt_4$; $Fe_4Co_4$; $Os_4Co_4$; $Os_4Ir_4$; $Os_4Rh_4$; $Ru_4Co_4$; $Rus_4Ir_4$; $Rus_4Rh_4$; |

Table S2. The molecular formula of the stable structure of $Z_2$ when the co-dual structure is not embedded.

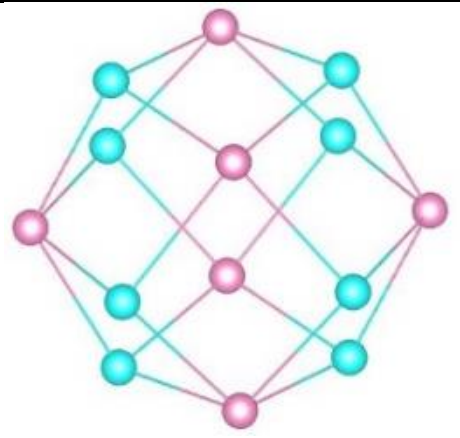

$Z_2$ Co-dual structure

| |
|---|
| **$X_{14}$ (9)** |
| $As_{14}$; $Bi_{14}$; $Sb_{14}$; $Nb_{14}$; $Ta_{14}$; $V_{14}$; $Fe_{14}$; $Os_{14}$; $Ru_{14}$; |
| **$X_6Y_8$ (159)** |
| $Mg_6Au_8$; $Hg_6Au_8$; $Mg_6Li_8$; $Cd_6Ag_8$; $Cd_6Au_8$; $Mg_6Cu_8$, $Hg_6Ag_8$; $Mg_6Ag_8$; $Zn_6Ag_8$; $Zn_6Au_8$; $Ni_6Ag_8$; $Ni_6Cu_8$; $Al_6Mg_8$; $Ga_6Mg_8$; $In_6Mg_8$; $In_6Zn_8$; $Tl_6Mg_8$; $Al_6Zn_8$; $Ga_6Hg_8$; $Ga_6Zn_8$; $In_6Cd_8$; $In_6Hg_8$; $Tl_6Cd_8$; $Tl_6Hg_8$; $Al_6As_8$; $Ga_6As_8$; $Ga_6Sb_8$; $In_6Bi_8$; $In_6Sb_8$; $Tl_6Bi_8$; $Tl_6Sb_8$; $Ge_6Ag_8$; $Ge_6Au_8$; $Pb_6Ag_8$; $Si_6Ag_8$; $Si_6Au_8$; $Sn_6Au_8$; $Ge_6Cu_8$; $Pb_6Au_8$; $Si_6Cu_8$; $Ge_6Mg_8$; $Ge_6Zn_8$; $Pb_6Cd_8$; $Pb_6Mg_8$; $Si_6Mg_8$; $Si_6Zn_8$; $Sn_6Cd_8$; $Sn_6Hg_8$; $Sn_6Mg_8$; $Po_6Pb_8$; $Po_6Sn_8$; $Te_6Ge_8$; $Te_6Pb_8$; $Te_6Sn_8$; $Ni_6Li_8$; $Pd_6Ag_8$; $Pd_6Au_8$; $Pd_6Cu_8$; $Pt_6Ag_8$; $Pt_6Au_8$; $Pt_6Cu_8$; $Pt_6Li_8$; $Ni_6Ge_8$; $Ni_6Si_8$; $Ni_6Ti_8$; $Pd_6Ge_8$; $Pd_6Hf_8$; $Pd_6Pb_8$; $Pd_6Si_8$; $Pd_6Sn_8$; $Pd_6Ti_8$; $Pd_6Zr_8$; $Pt_6Ge_8$; $Pt_6Hf_8$; $Pt_6Pb_8$; $Pt_6Si_8$; $Pt_6Sn_8$; $Pt_6Ti_8$; $Pt_6Zr_8$; $Cd_6Pt_8$; $Hg_6Pt_8$; $Mg_6Pd_8$; $Mg_6Pt_8$; $Zn_6Pd_8$; $Zn_6Pt_8$; $Zn_6Mn_8$; $Zn_6Tc_8$; $Sc_6Cd_8$; $Sc_6Hg_8$; $Sc_6Zn_8$; $Y_6Ca_8$; $Ga_6Nb_8$; $Ga_6Ta_8$; $Sc_6Bi_8$; $Sc_6Nb_8$; $Sc_6Sb_8$; $Sc_6Ta_8$; $Y_6Bi_8$; $Hf_6Ag_8$; $Hf_6Au_8$; $Hf_6Na_8$; $Ti_6Ag_8$; $Ti_6Au_8$; $Ti_6Cu_8$; $Zr_6Au_8$; $Hf_6Cd_8$; $Hf_6Hg_8$; $Hf_6Mg_8$; $Hf_6Zn_8$; $Ti_6Cd_8$; $Ti_6Hg_8$; $Ti_6Mg_8$; $Ti_6Zn_8$; $Zr_6Cd_8$; $Zr_6Hg_8$; $Zr_6Mg_8$; $Cr_6Ge_8$; $Cr_6Ti_8$; $Mo_6Ge_8$; $Mo_6Hf_8$; $Mo_6Pb_8$; $Mo_6Si_8$; $Mo_6Sn_8$; $Mo_6Ti_8$; $Mo_6Zr_8$; $Te_6Zr_8$; $W_6Ge_8$; $W_6Hf_8$; $W_6Pb_8$; $W_6Si_8$; $W_6Sn_8$; $W_6Ti_8$; $W_6Zr_8$; $Cr_6Tc_8$; $Mo_6Mn_8$; $Mo_6Re_8$; $Mo_6Tc_8$; $W_6Re_8$; $W_6Tc_8$; $Mn_6Mg_8$; $Re_6Mg_8$; $Re_6Zn_8$; $Tc_6Hg_8$; $Tc_6Mg_8$; $Tc_6Zn_8$; $Mn_6Os_8$; $Re_6Fe_8$; $Re_6Os_8$; $Re_6Ru_8$; $Tc_6Fe_8$; $Mn_6Os_8$; $Tc_6Ru_8$; $Co_6Zn_8$; $Ir_6Hg_8$; $Ir_6Mg_8$; $Ir_6Zn_8$; $Rh_6Hg_8$; $Rh_6Mg_8$; $Rh_6Zn_8$; |

Table S3. The molecular formula of the stable structure of $Z_3$ when the co-dual structure is not embedded.

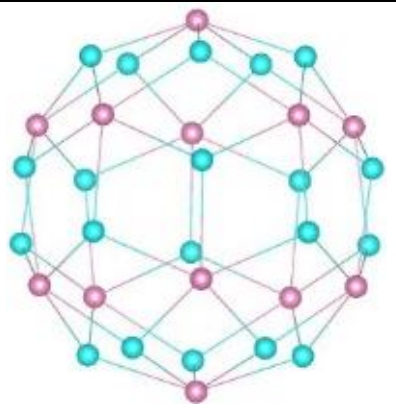

$Z_3$ Co-dual structure

| **$X_{32}$ (8)** |
|---|
| $Ag_{32}$; $Au_{32}$; $Cs_{32}$; $Cu_{32}$; $Fr_{32}$; $K_{32}$; $Na_{32}$; $Rb_{32}$; |
| **$X_{12}Y_{20}$ (62)** |
| $Ag_{12}Si_{20}$; $Au_{12}Ge_{20}$; $Au_{12}Pb_{20}$; $Au_{12}Si_{20}$; $Cu_{12}Ge_{20}$; $Li_{12}Ge_{20}$; $Li_{12}Sn_{20}$; $Au_{12}Sn_{20}$; $Ag_{12}As_{20}$; $Au_{12}As_{20}$; $Au_{12}Bi_{20}$; $Au_{12}Sb_{20}$; $Ga_{12}Te_{20}$; $Ge_{12}Ag_{20}$; $Pb_{12}Ag_{20}$; $Si_{12}Au_{20}$; $Si_{12}Cu_{20}$; $Sn_{12}Ag_{20}$; $Ge_{12}Au_{20}$; $Ge_{12}Cu_{20}$; $Si_{12}Ag_{20}$; $Sn_{12}Au_{20}$; $Po_{12}Na_{20}$; $Te_{12}Li_{20}$; $Te_{12}Na_{20}$; $Po_{12}Cd_{20}$; $Te_{12}Mg_{20}$; $Po_{12}Mg_{20}$; $Ag_{12}Hf_{20}$; $Ag_{12}Ti_{20}$; $Au_{12}Hf_{20}$; $Au_{12}Ti_{20}$; $Au_{12}Zr_{20}$; $Ag_{12}Ta_{20}$; $Ag_{12}V_{20}$; $Cu_{12}Ta_{20}$; $Cu_{12}V_{20}$; $Ga_{12}Cr_{20}$; $Y_{12}Po_{20}$; $Hf_{12}Ag_{20}$; $Hf_{12}Au_{20}$; $Cr_{12}Ag_{20}$; $Cr_{12}Mg_{20}$; $Cr_{12}Zn_{20}$; $Mo_{12}Cd_{20}$; $Mo_{12}Hg_{20}$; $Mo_{12}Mg_{20}$; $Mo_{12}Zn_{20}$; $W_{12}Cd_{20}$; $W_{12}Hg_{20}$; $W_{12}Mg_{20}$; $W_{12}Zn_{20}$; $Fe_{12}Al_{20}$; $Fe_{12}Ga_{20}$; $Os_{12}Al_{20}$; $Os_{12}Ga_{20}$; $Os_{12}In_{20}$; $Os_{12}Sc_{20}$; $Ru_{12}Al_{20}$; $Ru_{12}Ga_{20}$; $Ru_{12}In_{20}$; $Ru_{12}Sc_{20}$; |

Table S4. The molecular formula of the stable structure of $A_1$ when the co-dual structure is not embedded.

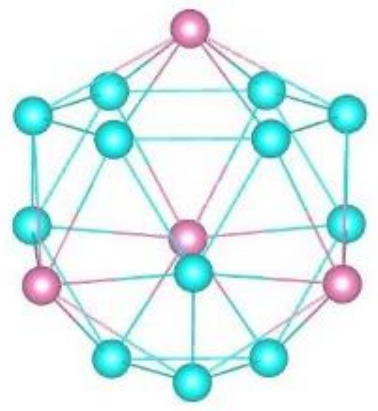

$A_1$ Co-dual structure

| **$X_{16}$ (11)** |
|---|
| $Be_{16}$; $Mg_{16}$; $Ba_{16}$; $Zn_{16}$; $Cd_{16}$; $Ca_{16}$; $Ra_{16}$; $Hg_{16}$; $Mn_{16}$; $Tc_{16}$; $Re_{16}$; |
| **$X_{12}Y_4$ (161)** |
| $Ag_{12}Hg_4$; $Ag_{12}Mg_4$; $Au_{12}Cd_4$; $Au_{12}Mg_4$; $Cs_{12}Ra_4$; $Cu_{12}Zn_4$; $K_{12}Ba_4$; $K_{12}Ra_4$; $Li_{12}Mg_4$; $Na_{12}Mg_4$; $Ag_{12}Cd_4$; $Au_{12}Zn_4$; $K_{12}Sr_4$; $Li_{12}Hg_4$; $Li_{12}Zn_4$; $Na_{12}Cd_4$; $Na_{12}Ca_4$; $Na_{12}Hg_4$; $Rb_{12}Ba_4$; $Rb_{12}Ra_4$; $Au_{12}Bi_4$; $Cu_{12}As_4$; $Li_{12}As_4$; $Li_{12}Sb_4$; $Na_{12}Bi_4$; $Na_{12}Sb_4$; $Cd_{12}Hg_4$; $Cd_{12}Mg_4$; $Cd_{12}Zn_4$; $Hg_{12}Mg_4$; $Mg_{12}Cd_4$; $Mg_{12}Hg_4$; $Mg_{12}Zn_4$; $Zn_{12}Mg_4$; $Ba_{12}Ra_4$; $Ba_{12}Sr_4$; $Ca_{12}Sr_4$; $Hg_{12}Cd_4$; $Hg_{12}Zn_4$; $Sr_{12}Ca_4$; $Sr_{12}Ba_4$; $Cd_{12}Ge_4$; $Cd_{12}Pb_4$; $Cd_{12}Sn_4$; $Mg_{12}Ge_4$; $Mg_{12}Pb_4$; $Mg_{12}Si_4$; $Mg_{12}Sn_4$; $Zn_{12}Ge_4$; $Zn_{12}Si_4$; $Zn_{12}Sn_4$; $Al_{12}Li_4$; $Ga_{12}Li_4$; $In_{12}Na_4$; $Al_{12}Ag_4$; $Al_{12}Au_4$; $Al_{12}Cu_4$; $Ga_{12}Ag_4$; $Ga_{12}Cu_4$; $Ge_{12}As_4$; $Ge_{12}Sb_4$; $Pb_{12}Bi_4$; $Pb_{12}Sb_4$; $Si_{12}AS_4$; $Sn_{12}Bi_4$; $Sn_{12}Sb_4$; $Li_{12}V_4$; $Li_{12}Ta_4$; $Cu_{12}Nb_4$; $Cu_{12}Ta_4$; $Ag_{12}V_4$; $Au_{12}Nb_4$; $Au_{12}Ta_4$; $Cu_{12}Mn_4$; $Cu_{12}Re_4$; $Ag_{12}Mn_4$; $Au_{12}Mn_4$; $Au_{12}Tc_4$; $Zn_{12}Ti_4$; $Zn_{12}Hf_4$; $Hg_{12}Ti_4$; $Sc_{12}Na_4$; $Sc_{12}Ag_4$; $Sc_{12}Ru_4$; $Ga_{12}Ru_4$; $Ga_{12}Os_4$; $Al_{12}Ru_4$; $Al_{12}Os_4$; $Ti_{12}Nb_4$; $Ti_{12}Ta_4$; $Ti_{12}As_4$; $Ti_{12}Sb_4$; $Ti_{12}Bi_4$; $Zr_{12}Nb_4$; $Zr_{12}Ta_4$; $Zr_{12}Sb_4$; $Zr_{12}Bi_4$; $Hf_{12}Ta_4$; $Hf_{12}Sb_4$; $Hf_{12}Bi_4$; $Si_{12}Ta_4$; $Sn_{12}Nb_4$; $V_{12}Zn_4$; $Nb_{12}Mg_4$; $Nb_{12}Zn_4$; $Nb_{12}Hg_4$; $Ta_{12}Mg_4$; $Ta_{12}Zn_4$; $Ta_{12}Cd_4$; $Ta_{12}Hg_4$; $Nb_{12}Fe_4$; $Ta_{12}Os_4$; $As_{12}Fe_4$; $As_{12}Ru_4$; $As_{12}Os_4$; $Cr_{12}Ta_4$; $Mo_{12}Ta_4$; $Mo_{12}As_4$; $Mo_{12}Sb_4$; $Mo_{12}Bi_4$; $W_{12}Nb_4$; $W_{12}Ta_4$; $W_{12}Sb_4$; $W_{12}Bi_4$; $Te_{12}Ta_4$; $Mo_{12}Ni_4$; $Mo_{12}Pt_4$; $W_{12}Pd_4$; $W_{12}Pt_4$; $Te_{12}Pd_4$; $Po_{12}Pd_4$; $Po_{12}Pt_4$; $Fe_{12}Ge_4$; $Ru_{12}Hf_4$; $Ru_{12}Si_4$; $Ru_{12}Ge_4$; $Ru_{12}Sn_4$; $Os_{12}Si_4$; $Tc_{12}Mg_4$; $Re_{12}Mg_4$; $Re_{12}Zn_4$; $Re_{12}Hg_4$; $Tc_{12}Mn_4$; $Tc_{12}Re_4$; $Re_{12}Tc_4$; $Co_{12}Li_4$; $Co_{12}Cu_4$; $Co_{12}Au_4$; $Rh_{12}Li_4$; $Rh_{12}Ag_4$; $Rh_{12}Au_4$; $Ir_{12}Li_4$; $Ir_{12}Ag_4$; $Ir_{12}Au_4$; $Ni_{12}Co_4$; $Ni_{12}Rh_4$; $Ni_{12}Ir_4$; $Pd_{12}Ir_4$; $Pt_{12}Co_4$; $Pt_{12}Rh_4$; $Pt_{12}Ir_4$; |

Table S5. The molecular formula of the stable structure of $A_2$ when the co-dual structure is not embedded.

| |
|---|
| 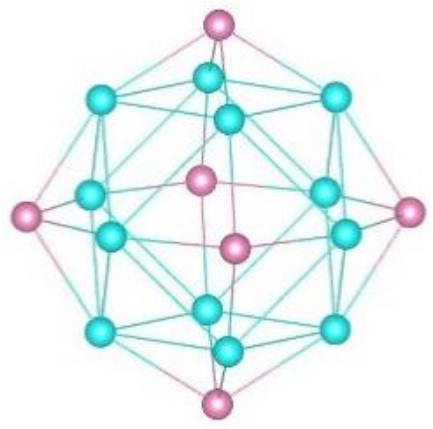 A$_2$ Co-dual structure |
| **$X_{18}$ (8)** |
| $Li_{18}$; $Na_{18}$; $K_{18}$; $Ag_{18}$; $Au_{18}$; $Rb_{18}$; $Cs_{18}$; $Fr_{18}$; |
| **$X_{12}Y_6$ (64)** |
| $Ag_{12}Au_6$; $Ag_{12}Na_6$; $Au_{12}Ag_6$; $Au_{12}Li_6$; $Cs_{12}K_6$; $Cs_{12}Rb_6$; $Cu_{12}Ag_6$; $Cu_{12}Au_6$; $Fr_{12}K_6$; $Fr_{12}Rb_6$; $K_{12}Rb_6$; $Li_{12}Ag_6$; $Rb_{12}Cs_6$; $Rb_{12}K_6$; $Cs_{12}Fr_6$; $Fr_{12}Cs_6$; $Li_{12}Cu_6$; $Na_{12}Ag_6$; $Mn_{12}Rh_6$; $Tc_{12}Rh_6$; $Tc_{12}Ir_6$; $Re_{12}Co_6$; $Re_{12}Rh_6$; $Re_{12}Ir_6$; $Fe_{12}Tc_6$; $Ru_{12}Mn_6$; $Os_{12}Mn_6$; $Os_{12}Tc_6$; $Os_{12}Re_6$; $Fe_{12}Ni_6$; $Fe_{12}Pd_6$; $Ru_{12}Pd_6$; $Ru_{12}Pt_6$; $Os_{12}Pd_6$; $Os_{12}Pt_6$; $Co_{12}Ta_6$; $Co_{12}As_6$; $Rh_{12}V_6$; $Rh_{12}Ta_6$; $Rh_{12}As_6$; $Rh_{12}Sb_6$; $Ir_{12}Nb_6$; $Ir_{12}As_6$; $Ir_{12}Sb_6$; $Ir_{12}Bi_6$; $Co_{12}Fe_6$; $Co_{12}Ru_6$; $Co_{12}Os_6$; $Rh_{12}Os_6$; $Ir_{12}Fe_6$; $Ni_{12}Al_6$; $Pd_{12}Ga_6$; $Pt_{12}Al_6$; $Pt_{12}Ga_6$; $Pt_{12}In_6$; $Pt_{12}Tl_6$; $Ni_{12}Cr_6$; $Pd_{12}Cr_6$; $Pd_{12}Mo_6$; $Pd_{12}W_6$; $Pd_{12}Te_6$; $Pd_{12}Po_6$; $Pt_{12}Te_6$; $Pt_{12}Po_6$; |

Table S6. The molecular formula of the stable structure of $A_3$ when the co-dual structure is not embedded.

| |
|---|
| 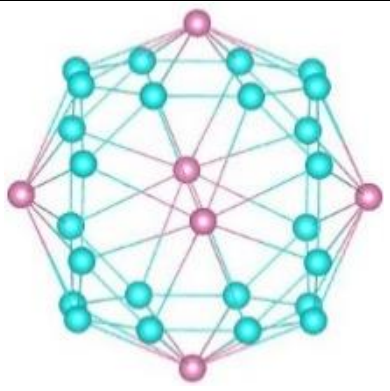 A$_3$ Co-dual structure |
| **$X_{30}$ (0)** |
| --- |
| **$X_{24}Y_6$ (10)** |
| $As_{24}Al_6$; $As_{24}Ga_6$; $Sb_{24}Ga_6$; $Sb_{24}In_6$; $Sb_{24}Tl_6$; $Hf_{24}Tc_6$; $Hf_{24}Pd_6$; $Hf_{24}Pt_6$; $Nb_{24}Ga_6$; $Nb_{24}Tl_6$; |

Table S7. The molecular formula of the stable structure of $A_4$ when the co-dual structure is not embedded.

| |
|---|
| 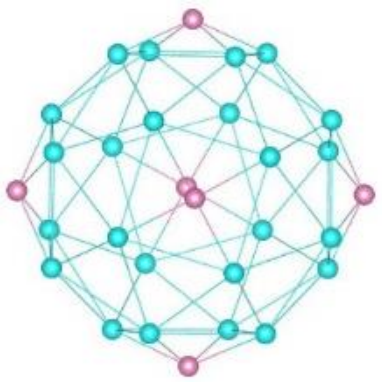 A$_4$ Co-dual structure |
| **$X_{30}$ (0)** |
| --- |
| **$X_{24}Y_6$ (48)** |
| $Sb_{24}Al_6$; $Sb_{24}Ga_6$; $Hf_{24}Tc_6$; $Si_{24}Tc_6$; $Ge_{24}Mn_6$; $Ge_{24}Tc_6$; $Ge_{24}Re_6$; $Sn_{24}Tc_6$; $Sn_{24}Re_6$; $Pb_{24}Re_6$; $Zr_{24}Pd_6$; $Zr_{24}Pt_6$; $Si_{24}Ni_6$; $Ge_{24}Ni_6$; $Ge_{24}Pd_6$; $Sn_{24}Pd_6$; $Sn_{24}Pt_6$; $Pb_{24}Pd_6$; $Pb_{24}Pt_6$; $V_{24}Ga_6$; $Nb_{24}Sc_6$; $Nb_{24}Al_6$; $Nb_{24}Ga_6$; $Nb_{24}In_6$; $Nb_{24}Tl_6$; $Ta_{24}Sc_6$; $Ta_{24}Al_6$; $Ta_{24}Ga_6$; $Ta_{24}In_6$; $Ta_{24}Tl_6$; $Sb_{24}Sc_6$; $Bi_{24}Sc_6$; $Bi_{24}Y_6$; $Nb_{24}Cr_6$; $Nb_{24}Te_6$; $Nb_{24}Po_6$; $As_{24}Mo_6$; $As_{24}W_6$; $As_{24}Te_6$; $Ta_{24}Te_6$; $Ta_{24}Po_6$; $Sb_{24}Mo_6$; $Sb_{24}W_6$; $Cr_{24}Mg_6$; $Mo_{24}Mg_6$; $Mo_{24}Cd_6$; $Te_{24}Cd_6$; $Te_{24}Hg_6$; |

Table S8. The molecular formula of the stable structure of $A_5$ when the co-dual structure is not embedded.

| |
|---|
| 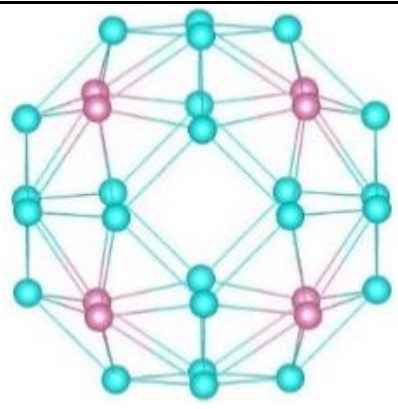 $A_5$ Co-dual structure |
| **$X_{32}$ (7)** |
| $Na_{32}$; $K_{32}$; $Cu_{32}$; $Ag_{32}$; $Au_{32}$; $Rb_{32}$; $Fr_{32}$; |
| **$X_{24}Y_8$ (65)** |
| $Ag_{24}Au_8$; $Ag_{24}Cu_8$; $Ag_{24}Li_8$; $Ag_{24}Na_8$; $Au_{24}Ag_8$; $Au_{24}Cu_8$; $Au_{24}Li_8$; $Au_{24}Na_8$; $Cs_{24}K_8$; $Cu_{24}Ag_8$; $Fr_{24}K_8$; $Fr_{24}Rb_8$; $Na_{24}Au_8$; $Cs_{24}Rb_8$; $Cu_{24}Au_8$; $K_{24}Rb_8$; $Li_{24}Ag_8$; $Ng_{24}Ag_8$; $Rb_{24}Cs_8$; $Rb_{24}K_8$; $Ag_{24}Cd_8$; $Ag_{24}Hg_8$; $Ag_{24}Mg_8$; $Ag_{24}Zn_8$; $Au_{24}Cd_8$; $Au_{24}Hg_8$; $Au_{24}Mg_8$; $Cu_{24}Zn_8$; $Li_{24}Hg_8$; $Na_{24}Hg_8$; $Na_{24}Mg_8$; $Au_{24}Zn_8$; $K_{24}Ra_8$; $Na_{24}Cd_8$; $Rb_{24}Ra_8$; $Al_{24}As_8$; $Ga_{24}As_8$; $Ga_{24}Sb_8$; $In_{24}Bi_8$; $In_{24}Sb_8$; $Tl_{24}Sb_8$; $Pb_{24}Cd_8$; $Si_{24}Mg_8$; $Pb_{24}Zn_8$; $Sn_{24}Zn_8$; $Pb_{24}Hg_8$; $Mg_{24}Fe_8$; $Mg_{24}Os_8$; $Zn_{24}Fe_8$; $Zn_{24}Ru_8$; $Zn_{24}Os_8$; $Hg_{24}Ru_8$; $Hg_{24}Os_8$; $Sc_{24}Sb_8$; $Sc_{24}Bi_8$; $Y_{24}Bi_8$; $Al_{24}Nb_8$; $Al_{24}Ta_8$; $Ga_{24}Nb_8$; $Ga_{24}Ta_8$; $In_{24}Nb_8$; $In_{24}Ta_8$; $Tl_{24}Nb_8$; $Ti_{24}Hg_8$; $Hf_{24}Hg_8$; |

Table S9. The molecular formula of the stable structure of $A_6$ when the co-dual structure is not embedded.

| |
|---|
| 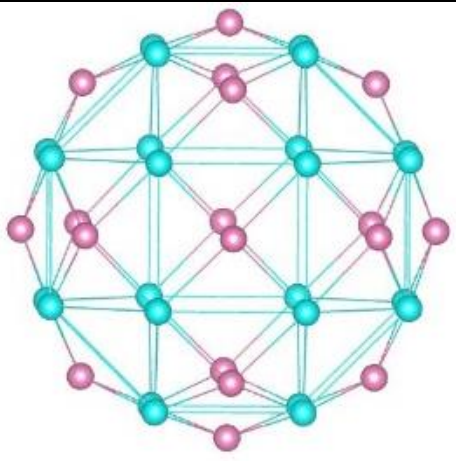 $A_6$ Co-dual structure |
| **$X_{42}$ (0)** |
| --- |
| **$X_{24}Y_{18}$ (9)** |
| $Cd_{24}As_{18}$; $Cd_{24}Sb_{18}$; $Cd_{24}Bi_{18}$; $Zn_{24}Te_{18}$; $Cd_{24}Cr_{18}$; $Cd_{24}Te_{18}$; $Cd_{24}Po_{18}$; $Hg_{24}Te_{18}$; $Hg_{24}Po_{18}$; |

Table S10. The molecular formula of the stable structure of $A_7$ when the co-dual structure is not embedded.

| |
|---|
| 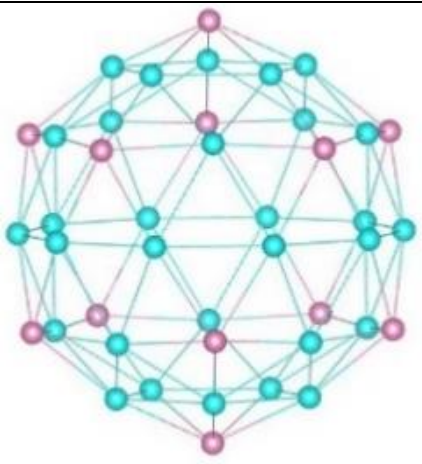 $A_7$ Co-dual structure |
| **$X_{42}$ (0)** |
| --- |
| **$X_{30}Y_{12}$ (46)** |
| $Al_{30}Ge_{12}$; $Ga_{30}Ge_{12}$; $Ga_{30}Si_{12}$; $Ga_{30}Sn_{12}$; $In_{30}Pb_{12}$; $In_{30}Sn_{12}$; $Tl_{30}Pb_{12}$; $Tl_{30}Sn_{12}$; $Cu_{30}Co_{12}$; $Cu_{30}Rh_{12}$; $Ag_{30}Co_{12}$; $Ag_{30}Rh_{12}$; $Ag_{30}Ir_{12}$; $Au_{30}Co_{12}$; $Au_{30}Rh_{12}$; $Au_{30}Ir_{12}$; $Mg_{30}Fe_{12}$; $Mg_{30}Ru_{12}$; $Hg_{30}Ru_{12}$; $Hg_{30}Os_{12}$; $Sc_{30}Ti_{12}$; $Sc_{30}Zr_{12}$; $Sc_{30}Hf_{12}$; $Sc_{30}Pb_{12}$; $Y_{30}Zr_{12}$; $Y_{30}Pb_{12}$; $Al_{30}Ti_{12}$; $Ga_{30}Zr_{12}$; $Ga_{30}Hf_{12}$; $In_{30}Ti_{12}$; $In_{30}Zr_{12}$; $In_{30}Hf_{12}$; $Tl_{30}Ti_{12}$; $Tl_{30}Zr_{12}$; $Tl_{30}Hf_{12}$; $Zr_{30}Sc_{12}$; $Zr_{30}Y_{12}$; $Zr_{30}Ga_{12}$; $Zr_{30}In_{12}$; $Hf_{30}Sc_{12}$; $Hf_{30}Y_{12}$; $Hf_{30}Ga_{12}$; $Hf_{30}Tl_{12}$; $Sn_{30}Sc_{12}$; $Pb_{30}Sc_{12}$; $Pb_{30}Y_{12}$; |

Table S11. The molecular formula of the stable structure of $A_8$ when the co-dual structure is not embedded.

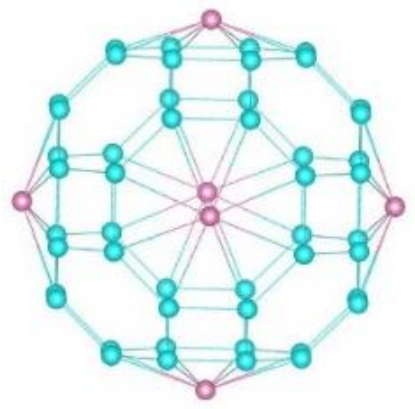

| $A_8$ Co-dual structure |
| --- |
| **$X_{54}$ (0)** |
| --- |
| **$X_{48}Y_6$ (15)** |
| $Al_{48}Ba_6$; $Al_{48}Ra_6$; $Al_{48}Sr_6$; $Tl_{48}Ba_6$; $Tl_{48}Ra_6$; $B_{48}Mg_6$; $Ga_{48}Ba_6$; $Ga_{48}Ra_6$; $Ga_{48}Sr_6$; $In_{48}Ba_6$; $In_{48}Ca_6$; $In_{48}Ra_6$; $In_{48}Sr_6$; $Hg_{48}Re_6$; $Zn_{48}Ni_6$; |

Table S12. The molecular formula of the stable structure of $A_9$ when the co-dual structure is not embedded.

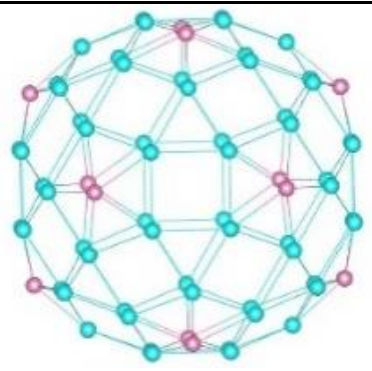

| $A_9$ Co-dual structure |
| --- |
| **$X_{72}$ (0)** |
| --- |
| **$X_{60}Y_{12}$ (13)** |
| $Be_{60}B_{12}$; $Hg_{60}Ga_{12}$; $Mg_{60}Al_{12}$; $Mg_{60}Ga_{12}$; $Na_{60}Ru_{12}$; $Au_{60}Ru_{12}$; $Au_{60}Os_{12}$; $Mg_{60}Sc_{12}$; $Ca_{60}Y_{12}$; $Cd_{60}Sc_{12}$; $Cd_{60}Y_{12}$; $Hg_{60}Sc_{12}$; $Hg_{60}Y_{12}$; |

Table S13. The molecular formula of the stable structure of $A_{10}$ when the co-dual structure is not embedded.

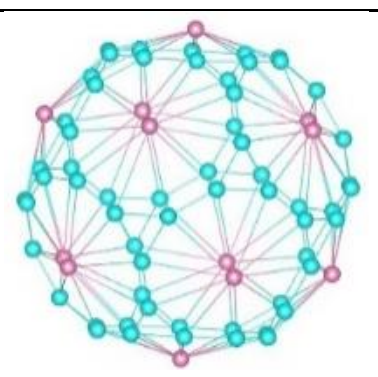

| $A_{10}$ Co-dual structure |
| --- |
| **$X_{72}$ (0)** |
| --- |
| **$X_{60}Y_{12}$ (1)** |
| $Be_{60}In_{12}$; |

Table S14. The molecular formula of the stable structure of $A_{11}$ when the co-dual structure is not embedded.

| |
|---|
| 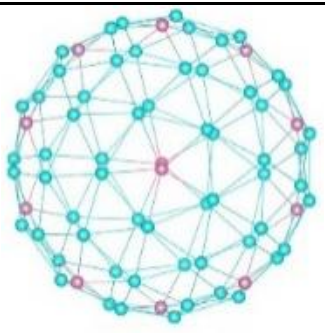 $A_{11}$ Co-dual structure |
| **$X_{72}$ (0)** |
| --- |
| **$X_{60}Y_{12}$ (20)** |
| $Be_{60}B_{12}$; $Hg_{60}Ga_{12}$; $Mg_{60}Al_{12}$; $Mg_{60}Ga_{12}$; $Na_{60}Os_{12}$; $Cu_{60}Fe_{12}$; $Cu_{60}Ru_{12}$; $Cu_{60}Os_{12}$; $Ag_{60}Fe_{12}$; $Ag_{60}Ru_{12}$; $Ag_{60}Os_{12}$; $Au_{60}Fe_{12}$; $Au_{60}Ru_{12}$; $Au_{60}Os_{12}$; $Mg_{60}Sc_{12}$; $Ca_{60}Y_{12}$; $Cd_{60}Sc_{12}$; $Cd_{60}Y_{12}$; $Hg_{60}Sc_{12}$; $Hg_{60}Y_{12}$; |

Table S15. The molecular formula of the stable structure of $A_{12}$ when the co-dual structure is not embedded.

| |
|---|
| 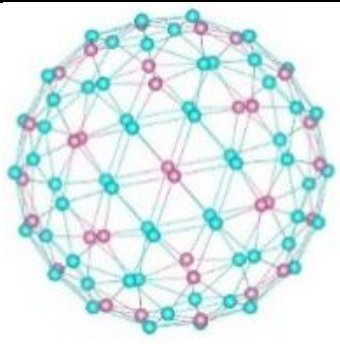 $A_{12}$ Co-dual structure |
| **$X_{80}$ (0)** |
| --- |
| **$X_{60}Y_{20}$ (0)** |
| --- |

Table S16. The molecular formula of the stable structure of $A_{13}$ when the co-dual structure is not embedded.

| |
|---|
| 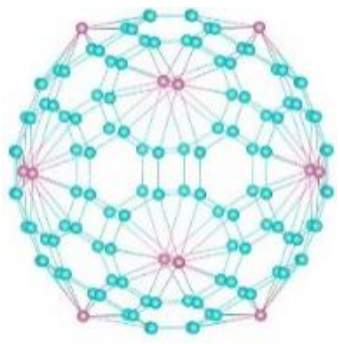 $A_{13}$ Co-dual structure |
| **$X_{132}$ (0)** |
| --- |
| **$X_{120}Y_{12}$ (0)** |
| --- |

Table S17. The molecular formula of the stable structure of $Z_1$ when the co-dual structure is embedded. In the displayed structural model, for a single-element structure, both colored balls represent X; for a two-element structure, the blue ball represents X and the pink ball represents Y. The green ball in the center represents the embedded element (This representation rule applies to TableS17-TableS32.)

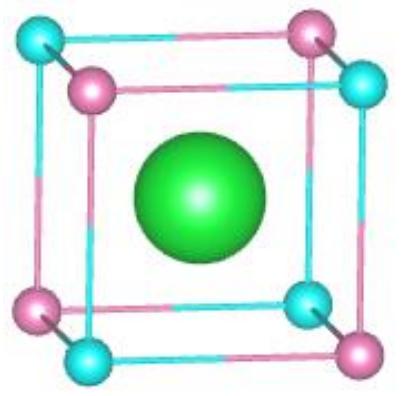

$Z_1$ Co-dual structure

| **M@$X_8$ (64)** |
| --- |
| Ba@$Ba_8$; Ra@$Ba_8$; Sr@$Ba_8$; Yb@$Ba_8$; Ba@$Be_8$; Ra@$Be_8$; Sr@$Be_8$; Yb@$Be_8$; Ba@$Ca_8$; Ra@$Ca_8$; Sr@$Ca_8$; Yb@$Ca_8$; Ba@$Cd_8$; Ra@$Cd_8$; Sr@$Cd_8$; Yb@$Cd_8$; Ba@$Hg_8$; Ra@$Hg_8$; Sr@$Hg_8$; Yb@$Hg_8$; Ba@$Mg_8$; Ra@$Mg_8$; Sr@$Mg_8$; Yb@$Mg_8$; Ba@$Ra_8$; Ra@$Ra_8$; Sr@$Ra_8$; Yb@$Ra_8$; Ba@$Sr_8$; Ra@$Sr_8$; Sr@$Sr_8$; Yb@$Sr_8$; Ba@$Zn_8$; Ra@$Zn_8$; Sr@$Zn_8$; Yb@$Zn_8$; Ba@$Ge_8$; Ra@$Ge_8$; Sr@$Ge_8$; Ba@$Pb_8$; Ra@$Pb_8$; Sr@$Pb_8$; Yb@$Pb_8$; Ba@$Si_8$; Ra@$Si_8$; Sr@$Si_8$; Yb@$Si_8$; Ba@$Sn_8$; Ra@$Sn_8$; Sr@$Sn_8$; Yb@$Sn_8$; Ba@$Hf_8$; Ra@$Hf_8$; Sr@$Hf_8$; Yb@$Hf_8$; Ba@$Ti_8$; Ra@$Ti_8$; Sr@$Ti_8$; Yb@$Ti_8$; Ba@$Zr_8$; Ra@$Zr_8$; Sr@$Zr_8$; Yb@$Zr_8$; Sr@$Mn_8$; |
| **M@$X_4Y_4$ (323)** |
| Ba@$Ag_4Al_4$; Ra@$Ag_4Al_4$; Sr@$Ag_4Al_4$; Yb@$Ag_4Al_4$; Ba@$Ag_4Ga_4$; Ra@$Ag_4Ga_4$; Sr@$Ag_4Ga_4$; Yb@$Ag_4Ga_4$; Ba@$Ag_4In_4$; Ra@$Ag_4In_4$; Sr@$Ag_4In_4$; Yb@$Ag_4In_4$; Ba@$Ag_4Tl_4$; Ra@$Ag_4Tl_4$; Sr@$Ag_4Tl_4$; Yb@$Ag_4Tl_4$; Ba@$Au_4Al_4$; Ra@$Au_4Al_4$; Sr@$Au_4Al_4$; Yb@$Au_4Al_4$; Ba@$Au_4Ga_4$; Ra@$Au_4Ga_4$; Sr@$Au_4Ga_4$; Yb@$Au_4Ga_4$; Ba@$Au_4In_4$; Ra@$Au_4In_4$; Yb@$Au_4In_4$; Ba@$Au_4Tl_4$; Ra@$Au_4Tl_4$; Yb@$Au_4Tl_4$; Ba@$Cu_4Al_4$; Sr@$Cu_4Al_4$; Yb@$Cu_4Al_4$; Ba@$Cu_4Ga_4$; Ra@$Cu_4Ga_4$; Sr@$Cu_4Ga_4$; Yb@$Cu_4Ga_4$; Ba@$Li_4Al_4$; Ra@$Li_4Al_4$; Sr@$Li_4Al_4$; Yb@$Li_4Al_4$; Ba@$Li_4Ga_4$; Sr@$Li_4Ga_4$; Yb@$Li_4Ga_4$; Ba@$Na_4In_4$; Ra@$Na_4In_4$; Sr@$Na_4In_4$; Yb@$Na_4In_4$; Ba@$Na_4Tl_4$; Ra@$Na_4Tl_4$; Sr@$Na_4Tl_4$; Yb@$Na_4Tl_4$; Ba@$Cd_4Po_4$; Sr@$Cd_4Po_4$; Yb@$Cd_4Po_4$; Ba@$Cd_4Te_4$; Ra@$Cd_4Te_4$; Sr@$Cd_4Te_4$; Yb@$Cd_4Te_4$; Ba@$Hg_4Po_4$; Ra@$Hg_4Po_4$; Sr@$Hg_4Po_4$; Yb@$Hg_4Po_4$; Ba@$Hg_4Te_4$; Ra@$Hg_4Te_4$; Sr@$Hg_4Te_4$; Yb@$Hg_4Te_4$; Ba@$Mg_4Po_4$; Ra@$Mg_4Po_4$; Sr@$Mg_4Po_4$; Yb@$Mg_4Po_4$; Ba@$Mg_4Te_4$; Ra@$Mg_4Te_4$; Sr@$Mg_4Te_4$; Yb@$Mg_4Te_4$; Yb@$Zn_4Te_4$; Ba@$Al_4As_4$; Ra@$Al_4As_4$; Sr@$Al_4As_4$; Yb@$Al_4As_4$; Ba@$Ga_4As_4$; Ra@$Ga_4As_4$; Sr@$Ga_4As_4$; Yb@$Ga_4As_4$; Ba@$Ga_4Sb_4$; Ra@$Ga_4Sb_4$; Sr@$Ga_4Sb_4$; Yb@$Ga_4Sb_4$; Ba@$In_4Bi_4$; Ra@$In_4Bi_4$; Sr@$In_4Bi_4$; Yb@$In_4Bi_4$; Ba@$In_4Sb_4$; Ra@$In_4Sb_4$; Sr@$In_4Sb_4$; Yb@$In_4Sb_4$; Ba@$Tl_4Bi_4$; Ra@$Tl_4Bi_4$; Sr@$Tl_4Bi_4$; Yb@$Tl_4Bi_4$; Ba@$Tl_4Sb_4$; Ra@$Tl_4Sb_4$; Sr@$Tl_4Sb_4$; Yb@$Tl_4Sb_4$; Ba@$Ag_4Sc_4$; Sr@$Ag_4Sc_4$; Yb@$Ag_4Sc_4$; Ba@$Au_4Sc_4$; Ra@$Au_4Sc_4$; Sr@$Au_4Sc_4$; Yb@$Au_4Sc_4$; Ba@$Na_4Sc_4$; Ra@$Na_4Sc_4$; Ba@$Na_4Y_4$; Ra@$Na_4Y_4$; Sr@$Na_4Y_4$; Yb@$Na_4Y_4$; Ba@$Cd_4Cr_4$; Sr@$Cd_4Cr_4$; Yb@$Cd_4Cr_4$; Sr@$Cd_4Mo_4$; Yb@$Cd_4Mo_4$; Ba@$Cd_4W_4$; Ba@$Hg_4Mo_4$; Ra@$Hg_4Mo_4$; Sr@$Hg_4Mo_4$; Ba@$Hg_4W_4$; Ra@$Hg_4W_4$; Sr@$Hg_4W_4$; Yb@$Hg_4W_4$; Ba@$Mg_4Cr_4$; Sr@$Mg_4Cr_4$; Yb@$Mg_4Cr_4$; Ba@$Mg_4Mo_4$; Sr@$Mg_4Mo_4$; Yb@$Mg_4Mo_4$; Ba@$Mg_4W_4$; Yb@$Zn_4W_4$; Yb@$Mg_4W_4$; Sr@$Al_4Nb_4$; Yb@$Al_4Nb_4$; Ra@$Al_4Ta_4$; Sr@$Ga_4Nb_4$; Yb@$Ga_4Nb_4$; Ra@$Ga_4Ta_4$; Ra@$In_4Nb_4$; Sr@$In_4Nb_4$; Ra@$In_4Ta_4$; Ba@$Sc_4Bi_4$; Ra@$Sc_4Bi_4$; Sr@$Sc_4Bi_4$; Yb@$Sc_4Bi_4$; Ba@$Sc_4Nb_4$; Ra@$Sc_4Nb_4$; Sr@$Sc_4Nb_4$; Yb@$Sc_4Nb_4$; Ba@$Sc_4Sb_4$; Ra@$Sc_4Sb_4$; Sr@$Sc_4Sb_4$; Yb@$Sc_4Sb_4$; Ba@$Sc_4Ta_4$; Ra@$Sc_4Ta_4$; Sr@$Sc_4Ta_4$; Yb@$Sc_4Ta_4$;Sr@$Tl_4Nb_4$; Yb@$Tl_4Nb_4$; Yb@$Tl_4Ta_4$; Ba@$Y_4Bi_4$; Ra@$Y_4Bi_4$; Sr@$Y_4Bi_4$; Yb@$Y_4Bi_4$; Ba@$Ge_4Ni_4$; Ra@$Ge_4Ni_4$; Sr@$Ge_4Ni_4$; Ba@$Ge_4Pd_4$; Ra@$Ge_4Pd_4$; Sr@$Ge_4Pd_4$; Yb@$Ge_4Pd_4$; Ba@$Ge_4Pt_4$; Ra@$Ge_4Pt_4$; Sr@$Ge_4Pt_4$; Yb@$Ge_4Pt_4$; Ba@$Hf_4Pd_4$; Ra@$Hf_4Pd_4$; Sr@$Hf_4Pd_4$; Yb@$Hf_4Pd_4$; Ba@$Hf_4Pt_4$; Ra@$Hf_4Pt_4$; Sr@$Hf_4Pt_4$; Yb@$Hf_4Pt_4$; Ba@$Pb_4Pd_4$; Ra@$Pb_4Pd_4$; Sr@$Pb_4Pd_4$; Yb@$Pb_4Pd_4$; Ba@$Pb_4Pt_4$; Ra@$Pb_4Pt_4$; Sr@$Pb_4Pt_4$; Yb@$Pb_4Pt_4$; Ba@$Si_4Ni_4$; Ra@$Si_4Ni_4$; Sr@$Si_4Ni_4$; Ba@$Si_4Pd_4$; Ra@$Si_4Pd_4$; Sr@$Si_4Pd_4$; Yb@$Si_4Pd_4$; Ba@$Si_4Pt_4$; Ra@$Si_4Pt_4$; Sr@$Si_4Pt_4$; Yb@$Si_4Pt_4$; Ba@$Sn_4Pd_4$; Ra@$Sn_4Pd_4$; Sr@$Sn_4Pd_4$; Yb@$Sn_4Pd_4$; Ba@$Sn_4Pt_4$; Ra@$Sn_4Pt_4$; Sr@$Sn_4Pt_4$; Yb@$Sn_4Pt_4$; Ba@$Ti_4Ni_4$; Ra@$Ti_4Ni_4$; Sr@$Ti_4Ni_4$; Ra@$Ti_4Pd_4$; Sr@$Ti_4Pd_4$; Yb@$Ti_4Pd_4$; Ba@$Ti_4Pt_4$; Sr@$Ti_4Pt_4$; Ba@$Zr_4Pd_4$; Ra@$Zr_4Pd_4$; Sr@$Zr_4Pd_4$; Yb@$Zr_4Pd_4$; Ba@$Zr_4Pt_4$; Ra@$Zr_4Pt_4$; Sr@$Zr_4Pt_4$; Yb@$Zr_4Pt_4$; Ra@$As_4Co_4$; Yb@$As_4Co_4$; Ba@$As_4Ir_4$; Ra@$As_4Ir_4$; Sr@$As_4Ir_4$; Yb@$As_4Ir_4$; Ba@$As_4Rh_4$; Ra@$As_4Rh_4$; Sr@$As_4Rh_4$; Yb@$As_4Rh_4$; Ba@$Bi_4Ir_4$; Ra@$Bi_4Ir_4$; Sr@$Bi_4Ir_4$; Yb@$Bi_4Ir_4$; Ba@$Nb_4Ir_4$; Ra@$Nb_4Ir_4$; Sr@$Nb_4Ir_4$; Yb@$Nb_4Ir_4$; Ba@$Nb_4Rh_4$; Ra@$Nb_4Rh_4$; Sr@$Nb_4Rh_4$; Yb@$Nb_4Rh_4$; Ba@$Sb_4Ir_4$; Ra@$Sb_4Ir_4$; Sr@$Sb_4Ir_4$; Yb@$Sb_4Ir_4$; Ba@$Sb_4Rh_4$; Ra@$Sb_4Rh_4$; Sr@$Sb_4Rh_4$; Yb@$Sb_4Rh_4$; Ba@$Ta_4Co_4$; Sr@$Ta_4Co_4$; Ba@$Ta_4Ir_4$; Ra@$Ta_4Ir_4$; Sr@$Ta_4Ir_4$; Yb@$Ta_4Ir_4$; Ba@$Ta_4Rh_4$; Ra@$Ta_4Rh_4$; Sr@$Ta_4Rh_4$; Yb@$Ta_4Rh_4$; Sr@$V_4Rh_4$; Ba@$Te_4Ru_4$; Ra@$Te_4Ru_4$; Yb@$Te_4Ru_4$; Sr@$W_4Os_4$; Yb@$W_4Os_4$; Ba@$Mn_4Ni_4$; Ra@$Mn_4Ni_4$; Ba@$Mn_4Pd_4$; Ra@$Mn_4Pd_4$; Sr@$Mn_4Pd_4$;Yb@$Mn_4Pd_4$; Ra@$Mn_4Pt_4$; Sr@$Mn_4Pt_4$; Yb@$Mn_4Pt_4$; Ba@$Tc_4Ni_4$; Ra@$Tc_4Ni_4$; Sr@$Tc_4Ni_4$; Yb@$Tc_4Ni_4$; Ba@$Tc_4Pd_4$; Ra@$Tc_4Pd_4$; Sr@$Tc_4Pd_4$; Yb@$Tc_4Pd_4$; Ba@$Tc_4Pt_4$; Sr@$Tc_4Pt_4$; Yb@$Tc_4Pt_4$; Ba@$Fe_4Ir_4$; Sr@$Fe_4Rh_4$; Yb@$Fe_4Rh_4$; Ba@$Os_4Co_4$; Ra@$Os_4Co_4$; Ba@$Os_4Ir_4$; Ra@$Os_4Ir_4$; Sr@$Os_4Ir_4$; Yb@$Os_4Ir_4$; Ba@$Os_4Rh_4$; Ra@$Os_4Rh_4$; Sr@$Os_4Rh_4$; Yb@$Os_4Rh_4$; Ba@$Ru_4Co_4$; Ra@$Ru_4Co_4$; Sr@$Ru_4Co_4$; Ba@$Ru_4Ir_4$; Ra@$Ru_4Ir_4$; Sr@$Ru_4Ir_4$; Yb@$Ru_4Ir_4$; Ba@$Ru_4Rh_4$; Ra@$Ru_4Rh_4$; Sr@$Ru_4Rh_4$; Yb@$Ru_4Rh_4$; |

Table S18. The molecular formula of the stable structure of $Z_2$ when the co-dual structure is embedded.

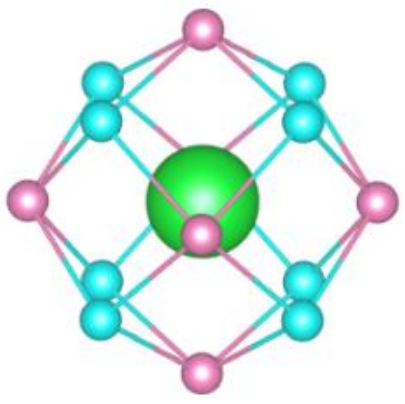

$Z_2$ Co-dual structure

| **$M@X_{14}$ (28)** |
|---|
| $Ba@Ge_{14}$; $Ra@Ge_{14}$; $Sr@Ge_{14}$; $Yb@Ge_{14}$; $Ba@Pb_{14}$; $Ra@Pb_{14}$; $Sr@Pb_{14}$; $Yb@Pb_{14}$; $Ba@Si_{14}$; $Ra@Si_{14}$; $Sr@Si_{14}$; $Yb@Si_{14}$; $Ba@Sn_{14}$; $Ra@Sn_{14}$; $Sr@Sn_{14}$; $Yb@Sn_{14}$; $Ba@Hf_{14}$; $Ra@Hf_{14}$; $Sr@Hf_{14}$; $Yb@Hf_{14}$; $Ba@Ti_{14}$; $Ra@Ti_{14}$; $Sr@Ti_{14}$; $Yb@Ti_{14}$; $Ba@Zr_{14}$; $Ra@Zr_{14}$; $Sr@Zr_{14}$; $Yb@Zr_{14}$; |
| **$M@X_6Y_8$ (780)** |
| $Ba@Ag_6Ge_8$; $Ra@Ag_6Ge_8$; $Sr@Ag_6Ge_8$; $Yb@Ag_6Ge_8$; $Ba@Ag_6Pb_8$; $Ra@Ag_6Pb_8$; $Sr@Ag_6Pb_8$; $Yb@Ag_6Pb_8$; $Ba@Ag_6Si_8$; $Sr@Ag_6Si_8$; $Yb@Ag_6Si_8$; $Ba@Ag_6Sn_8$; $Ra@Ag_6Sn_8$; $Sr@Ag_6Sn_8$; $Yb@Ag_6Sn_8$; $Ba@Au_6Ge_8$; $Ra@Au_6Ge_8$; $Sr@Au_6Ge_8$; $Yb@Au_6Ge_8$; $Ba@Au_6Pb_8$; $Ra@Au_6Pb_8$; $Sr@Au_6Pb_8$; $Yb@Au_6Pb_8$; $Ba@Au_6Si_8$; $Ra@Au_6Si_8$; $Sr@Au_6Si_8$; $Yb@Au_6Si_8$; $Ba@Au_6Sn_8$; $Ra@Au_6Sn_8$; $Sr@Au_6Sn_8$; $Yb@Au_6Sn_8$; $Ba@Cu_6Ge_8$; $Ra@Cu_6Ge_8$; $Sr@Cu_6Ge_8$; $Yb@Cu_6Ge_8$; $Ba@Cu_6Si_8$; $Ra@Cu_6Si_8$; $Sr@Cu_6Si_8$; $Yb@Cu_6Si_8$; $Ba@Li_6Ge_8$; $Ra@Li_6Ge_8$; $Sr@Li_6Ge_8$; $Yb@Li_6Ge_8$; $Ba@Li_6Si_8$; $Ra@Li_6Si_8$; $Sr@Li_6Si_8$; $Yb@Li_6Si_8$; $Ba@Li_6Sn_8$; $Ra@Li_6Sn_8$; $Sr@Li_6Sn_8$; $Yb@Li_6Sn_8$; $Ba@Na_6Pb_8$; $Ra@Na_6Pb_8$; $Sr@Na_6Pb_8$; $Yb@Na_6Pb_8$; $Ba@Na_6Sn_8$; $Ra@Na_6Sn_8$; $Sr@Na_6Sn_8$; $Yb@Na_6Sn_8$; $Ba@Ge_6Ag_8$; $Ra@Ge_6Ag_8$; $Sr@Ge_6Ag_8$; $Yb@Ge_6Ag_8$; $Ba@Ge_6Au_8$; $Ra@Ge_6Au_8$; $Sr@Ge_6Au_8$; $Yb@Ge_6Au_8$; $Ba@Ge_6Cu_8$; $Ra@Ge_6Cu_8$; $Sr@Ge_6Cu_8$; $Yb@Ge_6Cu_8$; $Ba@Ge_6Li_8$; $Ra@Ge_6Li_8$; $Sr@Ge_6Li_8$; $Yb@Ge_6Li_8$; $Ba@Pb_6Ag_8$; $Ra@Pb_6Ag_8$; $Sr@Pb_6Ag_8$; $Yb@Pb_6Ag_8$;$Ba@Pb_6Au_8$; $Ra@Pb_6Au_8$; $Sr@Pb_6Au_8$; $Yb@Pb_6Au_8$; $Ba@Pb_6Na_8$; $Ra@Pb_6Na_8$; $Sr@Pb_6Na_8$; $Yb@Pb_6Na_8$; $Ba@Si_6Ag_8$; $Ra@Si_6Ag_8$; $Sr@Si_6Ag_8$; $Yb@Si_6Ag_8$; $Ba@Si_6Au_8$; $Ra@Si_6Au_8$; $Sr@Si_6Au_8$; $Yb@Si_6Au_8$; $Ba@Si_6Cu_8$; $Ra@Si_6Cu_8$; $Sr@Si_6Cu_8$; $Yb@Si_6Cu_8$; $Ba@Si_6Li_8$; $Ra@Si_6Li_8$; $Sr@Si_6Li_8$; $Yb@Si_6Li_8$; $Ba@Sn_6Ag_8$; $Ra@Sn_6Ag_8$; $Sr@Sn_6Ag_8$; $Yb@Sn_6Ag_8$; $Ba@Sn_6Au_8$; $Ra@Sn_6Au_8$; $Sr@Sn_6Au_8$; $Yb@Sn_6Au_8$; $Ba@Sn_6Li_8$; $Ra@Sn_6Li_8$; $Sr@Sn_6Li_8$; $Yb@Sn_6Li_8$; $Ba@Sn_6Na_8$; $Ra@Sn_6Na_8$; $Sr@Sn_6Na_8$; $Yb@Sn_6Na_8$; $Ba@As_6Ag_8$; $Ra@As_6Ag_8$; $Sr@As_6Ag_8$; $Yb@As_6Ag_8$; $Ba@As_6Au_8$; $Ra@As_6Au_8$; $Sr@As_6Au_8$; $Yb@As_6Au_8$; $Ba@As_6Cu_8$; $Ra@As_6Cu_8$; $Sr@As_6Cu_8$; $Yb@As_6Cu_8$; $Ba@As_6Li_8$; $Ra@As_6Li_8$; $Sr@As_6Li_8$; $Yb@As_6Li_8$; $Ba@Bi_6Ag_8$; $Ra@Bi_6Ag_8$; $Sr@Bi_6Ag_8$; $Yb@Bi_6Ag_8$; $Ba@Bi_6Au_8$; $Ra@Bi_6Au_8$; $Sr@Bi_6Au_8$; $Yb@Bi_6Au_8$; $Ba@Bi_6Na_8$; $Ra@Bi_6Na_8$; $Sr@Bi_6Na_8$; $Yb@Bi_6Na_8$; $Ba@Sb_6Ag_8$; $Ra@Sb_6Ag_8$; $Sr@Sb_6Ag_8$; $Yb@Sb_6Ag_8$; $Ba@Sb_6Au_8$; $Ra@Sb_6Au_8$; $Sr@Sb_6Au_8$; $Yb@Sb_6Au_8$; $Ba@Sb_6Li_8$; $Ra@Sb_6Li_8$; $Sr@Sb_6Li_8$; $Yb@Sb_6Li_8$; $Ba@Sb_6Na_8$; $Ra@Sb_6Na_8$; $Sr@Sb_6Na_8$; $Yb@Sb_6Na_8$; $Ba@Po_6Pb_8$; $Ra@Po_6Pb_8$; $Sr@Po_6Pb_8$; $Yb@Po_6Pb_8$; $Ba@Po_6Sn_8$; $Ra@Po_6Sn_8$; $Sr@Po_6Sn_8$; $Yb@Po_6Sn_8$; $Ba@Te_6Ga_8$; $Ra@Te_6Ga_8$; $Sr@Te_6Ga_8$; $Yb@Te_6Ge_8$; $Ba@Te_6Pb_8$; $Ra@Te_6Pb_8$; $Sr@Te_6Pb_8$; $Yb@Te_6Pb_8$; $Ba@Te_6Sn_8$; $Ra@Te_6Sn_8$; $Sr@Te_6Sn_8$;$Yb@Te_6Sn_8$; $Ba@Ag_6Hf_8$; $Ra@Ag_6Hf_8$; $Sr@Ag_6Hf_8$; $Yb@Ag_6Hf_8$; $Ba@Ag_6Ti_8$; $Ra@Ag_6Ti_8$; $Sr@Ag_6Ti_8$; $Yb@Ag_6Ti_8$; $Ba@Ag_6Zr_8$; $Ra@Ag_6Zr_8$; $Sr@Ag_6Zr_8$; $Yb@Ag_6Zr_8$; $Ba@Ag_6Hf_8$; $Ra@Ag_6Hf_8$; $Sr@Ag_6Hf_8$; $Yb@Ag_6Hf_8$; $Ba@Ag_6Ti_8$; $Ra@Ag_6Ti_8$; $Sr@Ag_6Ti_8$; $Yb@Ag_6Ti_8$; $Ba@Au_6Zr_8$; $Ra@Au_6Zr_8$; $Sr@Au_6Zr_8$; $Yb@Au_6Zr_8$; $Ba@Cu_6Ti_8$; $Ra@Cu_6Ti_8$; $Sr@Cu_6Ti_8$; $Yb@Cu_6Ti_8$; $Ba@Ni_6Ag_8$; $Ra@Ni_6Ag_8$; $Sr@Ni_6Ag_8$; $Yb@Ni_6Ag_8$; $Ba@Ni_6Au_8$; $Ra@Ni_6Au_8$; $Sr@Ni_6Au_8$; $Yb@Ni_6Au_8$; $Ba@Ni_6Cu_8$; $Ra@Ni_6Cu_8$; $Sr@Ni_6Cu_8$; $Yb@Ni_6Cu_8$; $Ba@Ni_6Li_8$; $Ra@Ni_6Li_8$; $Sr@Ni_6Li_8$; $Yb@Ni_6Li_8$; $Ba@Pd_6Ag_8$; $Ra@Pd_6Ag_8$; $Sr@Pd_6Ag_8$; $Yb@Pd_6Ag_8$; $Ba@Pd_6Au_8$; $Ra@Pd_6Au_8$; $Sr@Pd_6Au_8$; $Yb@Pd_6Au_8$; $Ba@Pd_6Cu_8$; $Ra@Pd_6Cu_8$; $Sr@Pd_6Cu_8$; $Yb@Pd_6Cu_8$; $Ba@Pd_6Li_8$; $Ra@Pd_6Li_8$; $Sr@Pd_6Li_8$; $Yb@Pd_6Li_8$; $Ba@Pt_6Ag_8$; $Ra@Pt_6Ag_8$; $Sr@Pt_6Ag_8$; $Yb@Pt_6Ag_8$; $Ba@Pt_6Au_8$; $Ra@Pt_6Au_8$; $Sr@Pt_6Au_8$; $Yb@Pt_6Au_8$; $Ba@Pt_6Cu_8$; $Ra@Pt_6Cu_8$; $Sr@Pt_6Cu_8$; $Yb@Pt_6Cu_8$; $Ba@Pt_6Li_8$; $Ra@Pt_6Li_8$; $Sr@Pt_6Li_8$; $Yb@Pt_6Li_8$; $Ba@Hg_6Tc_8$; $Ra@Hg_6Tc_8$; $Sr@Hg_6Tc_8$; $Ba@Mg_6Re_8$; $Ba@Mg_6Tc_8$; $Ra@Mg_6Tc_8$; $Ba@Zn_6Mn_8$; $Ra@Zn_6Mn_8$; $Sr@Zn_6Mn_8$; $Yb@Zn_6Mn_8$; $Sr@Zn_6Re_8$; $Yb@Zn_6Re_8$; $Ra@Zn_6Tc_8$; $Yb@Mg_6Mn_8$; $Ba@Al_6Co_8$; $Ra@Al_6Co_8$; $Sr@Al_6Co_8$; $Yb@Al_6Co_8$; $Ba@Al_6Ir_8$; $Ra@Al_6Ir_8$; $Sr@Al_6Ir_8$; $Yb@Al_6Ir_8$; $Ba@Al_6Rh_8$; $Ra@Al_6Rh_8$; $Sr@Al_6Rh_8$; $Yb@Al_6Rh_8$; $Ba@Ga_6Co_8$; $Ra@Ga_6Co_8$; $Sr@Ga_6Co_8$; $Yb@Ga_6Co_8$; $Ba@Ga_6Ir_8$; $Ra@Ga_6Ir_8$; $Sr@Ga_6Ir_8$; $Yb@Ga_6Ir_8$; $Ba@Ga_6Rh_8$; $Ra@Ga_6Rh_8$; $Sr@Ga_6Rh_8$; $Yb@Ga_6Rh_8$; $Ba@In_6Ir_8$; $Ra@In_6Ir_8$; $Sr@In_6Ir_8$; $Yb@In_6Ir_8$; $Ba@In_6Rh_8$; $Ra@In_6Rh_8$; $Sr@In_6Rh_8$; $Yb@In_6Rh_8$; $Ba@Sc_6Ir_8$; $Ra@Sc_6Ir_8$; $Sr@Sc_6Ir_8$; $Yb@Sc_6Ir_8$; $Ba@Sc_6Rh_8$; $Ra@Sc_6Rh_8$; $Sr@Sc_6Rh_8$; $Yb@Sc_6Rh_8$; $Ba@Hf_6Ag_8$; $Ra@Hf_6Ag_8$; $Sr@Hf_6Ag_8$;$Yb@Hf_6Ag_8$; $Ba@Hf_6Au_8$; $Ra@Hf_6Au_8$; $Sr@Hf_6Au_8$; $Yb@Hf_6Au_8$; $Ba@Hf_6Na_8$; $Ra@Hf_6Na_8$; $Sr@Hf_6Na_8$; $Yb@Hf_6Na_8$; $Ba@Ti_6Ag_8$; $Ra@Ti_6Ag_8$; $Sr@Ti_6Ag_8$; $Yb@Ti_6Ag_8$; $Ba@Ti_6Au_8$; $Ra@Ti_6Au_8$; $Sr@Ti_6Au_8$; $Yb@Ti_6Au_8$; $Ba@Ti_6Cu_8$; $Ra@Ti_6Cu_8$; $Sr@Ti_6Cu_8$; $Yb@Ti_6Cu_8$; $Ba@Ti_6Li_8$; $Ra@Ti_6Li_8$; $Sr@Ti_6Li_8$; $Yb@Ti_6Li_8$; $Ba@Zr_6Ag_8$; $Ra@Zr_6Ag_8$; $Sr@Zr_6Ag_8$; $Yb@Zr_6Ag_8$; $Ba@Zr_6Au_8$; $Ra@Zr_6Au_8$; $Sr@Zr_6Au_8$; $Yb@Zr_6Au_8$; $Ba@Zr_6Na_8$; $Ra@Zr_6Na_8$; $Sr@Zr_6Na_8$; $Yb@Zr_6Na_8$; $Ba@Ge_6Ni_8$; $Ra@Ge_6Ni_8$; $Sr@Ge_6Ni_8$; $Yb@Ge_6Ni_8$; $Ba@Ge_6Pd_8$; $Ra@Ge_6Pd_8$; $Sr@Ge_6Pd_8$; $Yb@Ge_6Pd_8$; $Ba@Ge_6Pt_8$; $Ra@Ge_6Pt_8$; $Sr@Ge_6Pt_8$; $Yb@Ge_6Pt_8$; $Ba@Hf_6Pd_8$; $Ra@Hf_6Pd_8$; $Sr@Hf_6Pd_8$; $Yb@Hf_6Pd_8$; $Ba@Hf_6Pt_8$; $Ra@Hf_6Pt_8$; $Sr@Hf_6Pt_8$; $Yb@Hf_6Pt_8$; $Ba@Pb_6Pd_8$; $Ra@Pb_6Pd_8$; $Sr@Pb_6Pd_8$; $Yb@Pb_6Pd_8$; $Ba@Pb_6Pt_8$; $Ra@Pb_6Pt_8$; $Sr@Pb_6Pt_8$; $Yb@Pb_6Pt_8$; $Ba@Si_6Ni_8$; $Ra@Si_6Ni_8$; $Sr@Si_6Ni_8$; $Yb@Si_6Ni_8$; $Ba@Si_6Pd_8$; $Ra@Si_6Pd_8$; $Sr@Si_6Pd_8$; $Yb@Si_6Pd_8$; $Ba@Si_6Pt_8$; $Ra@Si_6Pt_8$; $Sr@Si_6Pt_8$; $Yb@Si_6Pt_8$; $Ba@Sn_6Pd_8$; $Ra@Sn_6Pd_8$; $Sr@Sn_6Pd_8$; $Yb@Sn_6Pd_8$; $Ba@Sn_6Pt_8$; $Ra@Sn_6Pt_8$; $Sr@Sn_6Pt_8$; $Yb@Sn_6Pt_8$; $Ba@Ti_6Ni_8$; $Ra@Ti_6Ni_8$; |

$Sr@Ti_6Ni_8$; $Yb@Ti_6Ni_8$; $Ba@Ti_6Pd_8$; $Ra@Ti_6Pd_8$; $Sr@Ti_6Pd_8$; $Yb@Ti_6Pd_8$; $Ba@Ti_6Pt_8$; $Ra@Ti_6Pt_8$; $Sr@Ti_6Pt_8$;
$Yb@Ti_6Pt_8$; $Ba@Zr_6Pd_8$; $Ra@Zr_6Pd_8$; $Sr@Zr_6Pd_8$; $Yb@Zr_6Pd_8$; $Ba@Zr_6Pt_8$; $Ra@Zr_6Pt_8$; $Sr@Zr_6Pt_8$;
$Yb@Zr_6Pt_8$; $Ba@Nb_6Ag_8$; $Ra@Nb_6Ag_8$; $Sr@Nb_6Ag_8$; $Yb@Nb_6Ag_8$; $Ba@Nb_6Au_8$; $Ra@Nb_6Au_8$; $Sr@Nb_6Au_8$;
$Yb@Nb_6Au_8$; $Ba@Nb_6Cu_8$; $Ra@Nb_6Cu_8$; $Sr@Nb_6Cu_8$; $Yb@Nb_6Cu_8$; $Ba@Nb_6Li_8$; $Ra@Nb_6Li_8$; $Sr@Nb_6Li_8$;
$Yb@Nb_6Li_8$; $Ba@Ta_6Ag_8$; $Ra@Ta_6Ag_8$; $Sr@Ta_6Ag_8$; $Yb@Ta_6Ag_8$; $Ba@Ta_6Au_8$; $Ra@Ta_6Au_8$; $Sr@Ta_6Au_8$;
$Yb@Ta_6Au_8$; $Ba@Ta_6Cu_8$; $Ra@Ta_6Cu_8$; $Sr@Ta_6Cu_8$; $Yb@Ta_6Cu_8$; $Ba@Ta_6Li_8$; $Ra@Ta_6Li_8$; $Sr@Ta_6Li_8$;
$Yb@Ta_6Li_8$; $Ba@V_6Ag_8$; $Ra@V_6Ag_8$; $Sr@V_6Ag_8$; $Yb@V_6Ag_8$; $Ba@V_6Au_8$; $Ra@V_6Au_8$; $Sr@V_6Au_8$;
$Yb@V_6Au_8$; $Ba@V_6Cu_8$; $Ra@V_6Cu_8$; $Sr@V_6Cu_8$; $Yb@V_6Cu_8$; $Ba@V_6Li_8$; $Ra@V_6Li_8$; $Sr@V_6Li_8$; $Yb@V_6Li_8$;
$Ba@As_6Ni_8$; $Ra@As_6Ni_8$; $Sr@As_6Ni_8$; $Yb@As_6Ni_8$; $Ba@As_6Pd_8$; $Ra@As_6Pd_8$; $Sr@As_6Pd_8$; $Yb@As_6Pd_8$;
$Ba@As_6Pt_8$; $Ra@As_6Pt_8$; $Sr@As_6Pt_8$; $Yb@As_6Pt_8$; $Ba@Bi_6Pd_8$; $Ra@Bi_6Pd_8$; $Sr@Bi_6Pd_8$; $Yb@Bi_6Pd_8$;
$Ba@Bi_6Pt_8$; $Ra@Bi_6Pt_8$; $Sr@Bi_6Pt_8$; $Yb@Bi_6Pt_8$; $Ba@Nb_6Ni_8$; $Ra@Nb_6Ni_8$; $Sr@Nb_6Ni_8$; $Yb@Nb_6Ni_8$;
$Ba@Nb_6Pd_8$; $Ra@Nb_6Pd_8$; $Sr@Nb_6Pd_8$; $Yb@Nb_6Pd_8$; $Ba@Nb_6Pt_8$; $Ra@Nb_6Pt_8$; $Sr@Nb_6Pt_8$; $Yb@Nb_6Pt_8$;
$Ba@Sb_6Pd_8$; $Ra@Sb_6Pd_8$; $Sr@Sb_6Pd_8$; $Yb@Sb_6Pd_8$; $Ba@Sb_6Pt_8$; $Ra@Sb_6Pt_8$; $Sr@Sb_6Pt_8$; $Yb@Sb_6Pt_8$;
$Ba@Ta_6Ni_8$; $Ra@Ta_6Ni_8$; $Sr@Ta_6Ni_8$; $Yb@Ta_6Ni_8$; $Ba@Ta_6Pd_8$; $Ra@Ta_6Pd_8$; $Sr@Ta_6Pd_8$; $Yb@Ta_6Pd_8$;
$Ba@Ta_6Pt_8$; $Ra@Ta_6Pt_8$; $Sr@Ta_6Pt_8$;$Yb@Ta_6Pt_8$; $Ba@V_6Ni_8$; $Ra@V_6Ni_8$; $Sr@V_6Ni_8$;$Yb@V_6Ni_8$; $Ba@V_6Pd_8$;
$Ra@V_6Pd_8$; $Sr@V_6Pd_8$; $Yb@V_6Pd_8$; $Ba@V_6Pt_8$; $Ra@V_6Pt_8$; $Sr@V_6Pt_8$; $Yb@V_6Pt_8$; $Ba@Cr_6Ge_8$; $Ra@Cr_6Ge_8$;
$Sr@Cr_6Ge_8$; $Yb@Cr_6Ge_8$; $Ba@Cr_6Si_8$; $Ra@Cr_6Si_8$; $Sr@Cr_6Si_8$; $Yb@Cr_6Si_8$; $Ba@Cr_6Ti_8$; $Ra@Cr_6Ti_8$;
$Sr@Cr_6Ti_8$; $Yb@Cr_6Ti_8$; $Ba@Mo_6Ge_8$; $Ra@Mo_6Ge_8$; $Sr@Mo_6Ge_8$; $Yb@Mo_6Ge_8$; $Ba@Mo_6Hf_8$; $Ra@Mo_6Hf_8$;
$Sr@Mo_6Hf_8$; $Yb@Mo_6Hf_8$; $Ba@Mo_6Pb_8$; $Ra@Mo_6Pb_8$; $Sr@Mo_6Pb_8$; $Yb@Mo_6Pb_8$; $Ba@Mo_6Si_8$; $Ra@Mo_6Si_8$;
$Sr@Mo_6Si_8$; $Yb@Mo_6Si_8$; $Ba@Mo_6Sn_8$; $Ra@Mo_6Sn_8$; $Sr@Mo_6Sn_8$; $Yb@Mo_6Sn_8$; $Ba@Mo_6Ti_8$; $Ra@Mo_6Ti_8$;
$Sr@Mo_6Ti_8$; $Yb@Mo_6Ti_8$; $Ba@Mo_6Zr_8$; $Ra@Mo_6Zr_8$; $Sr@Mo_6Zr_8$; $Yb@Mo_6Zr_8$; $Ba@Po_6Hf_8$; $Ra@Po_6Hf_8$;
$Sr@Po_6Hf_8$; $Yb@Po_6Hf_8$; $Ba@Po_6Ti_8$; $Ra@Po_6Ti_8$; $Sr@Po_6Ti_8$; $Yb@Po_6Ti_8$; $Ba@Po_6Zr_8$; $Ra@Po_6Zr_8$;
$Sr@Po_6Zr_8$; $Yb@Po_6Zr_8$; $Ba@Te_6Hf_8$; $Ra@Te_6Hf_8$; $Sr@Te_6Hf_8$; $Yb@Te_6Hf_8$; $Ba@Te_6Ti_8$; $Ra@Te_6Ti_8$;
$Sr@Te_6Ti_8$; $Yb@Te_6Ti_8$; $Ba@Te_6Zr_8$; $Ra@Te_6Zr_8$; $Sr@Te_6Zr_8$; $Yb@Te_6Zr_8$; $Ba@W_6Ge_8$; $Ra@W_6Ge_8$;
$Sr@W_6Ge_8$; $Yb@W_6Ge_8$; $Ba@W_6Hf_8$; $Ra@W_6Hf_8$; $Sr@W_6Hf_8$; $Yb@W_6Hf_8$; $Ba@W_6Pb_8$; $Ra@W_6Pb_8$;
$Sr@W_6Pb_8$; $Yb@W_6Pb_8$; $Ba@W_6Si_8$; $Ra@W_6Si_8$; $Sr@W_6Si_8$; $Yb@W_6Si_8$; $Ba@W_6Sn_8$; $Ra@W_6Sn_8$;
$Sr@W_6Sn_8$; $Yb@W_6Sn_8$; $Ba@W_6Ti_8$; $Ra@W_6Ti_8$; $Sr@W_6Ti_8$; $Yb@W_6Ti_8$; $Ba@W_6Zr_8$; $Ra@W_6Zr_8$; $Sr@W_6Zr_8$;
$Yb@W_6Zr_8$; $Ba@Mn_6Al_8$; $Ra@Mn_6Al_8$; $Sr@Mn_6Al_8$; $Yb@Mn_6Al_8$; $Ba@Mn_6Ga_8$; $Ra@Mn_6Ga_8$; $Sr@Mn_6Ga_8$;
$Yb@Mn_6Ga_8$; $Ba@Re_6Al_8$; $Ra@Re_6Al_8$; $Sr@Re_6Al_8$; $Yb@Re_6Al_8$; $Ba@Re_6Ga_8$; $Ra@Re_6Ga_8$; $Sr@Re_6Ga_8$;
$Yb@Re_6Ga_8$; $Ba@Re_6In_8$; $Ra@Re_6In_8$; $Sr@Re_6In_8$; $Yb@Re_6In_8$; $Ba@Re_6Sc_8$; $Ra@Re_6Sc_8$; $Sr@Re_6Sc_8$;
$Yb@Re_6Sc_8$; $Ba@Tc_6Al_8$; $Ra@Tc_6Al_8$; $Sr@Tc_6Al_8$; $Yb@Tc_6Al_8$; $Ba@Tc_6Ga_8$; $Ra@Tc_6Ga_8$; $Sr@Tc_6Ga_8$;
$Yb@Tc_6Ga_8$; $Ba@Tc_6In_8$; $Ra@Tc_6In_8$; $Sr@Tc_6In_8$; $Yb@Tc_6In_8$; $Ba@Tc_6Sc_8$; $Ra@Tc_6Sc_8$; $Sr@Tc_6Sc_8$;
$Yb@Tc_6Sc_8$; $Ba@Fe_6Li_8$; $Ra@Fe_6Li_8$; $Sr@Fe_6Li_8$; $Yb@Fe_6Li_8$; $Ba@Fe_6Ag_8$; $Ra@Fe_6Ag_8$; $Sr@Fe_6Ag_8$;
$Yb@Fe_6Ag_8$; $Ba@Fe_6Au_8$; $Ra@Fe_6Au_8$; $Sr@Fe_6Au_8$; $Yb@Fe_6Au_8$; $Ba@Fe_6Cu_8$; $Ra@Fe_6Cu_8$; $Sr@Fe_6Cu_8$;
$Yb@Fe_6Cu_8$; $Ba@Os_6Ag_8$; $Ra@Os_6Ag_8$; $Sr@Os_6Ag_8$; $Yb@Os_6Ag_8$; $Ba@Os_6Au_8$; $Ra@Os_6Au_8$; $Sr@Os_6Au_8$;
$Yb@Os_6Au_8$; $Ba@Os_6Cu_8$; $Ra@Os_6Cu_8$; $Sr@Os_6Cu_8$; $Yb@Os_6Cu_8$; $Ba@Os_6Li_8$; $Ra@Os_6Li_8$; $Sr@Os_6Li_8$;
$Yb@Os_6Li_8$; $Ba@Ru_6Ag_8$; $Ra@Ru_6Ag_8$; $Sr@Ru_6Ag_8$; $Yb@Ru_6Ag_8$; $Ba@Ru_6Au_8$; $Ra@Ru_6Au_8$; $Sr@Ru_6Au_8$;
$Yb@Ru_6Au_8$; $Ba@Ru_6Cu_8$; $Ra@Ru_6Cu_8$; $Sr@Ru_6Cu_8$; $Yb@Ru_6Cu_8$; $Ba@Ru_6Li_8$; $Ra@Ru_6Li_8$; $Sr@Ru_6Li_8$;
$Yb@Ru_6Li_8$; $Ba@Fe_6Mn_8$; $Ra@Fe_6Mn_8$; $Sr@Fe_6Mn_8$; $Yb@Fe_6Mn_8$; $Ra@Fe_6Re_8$; $Sr@Fe_6Re_8$; $Yb@Fe_6Re_8$;
$Ba@Fe_6Tc_8$; $Ra@Fe_6Tc_8$; $Sr@Fe_6Tc_8$; $Yb@Fe_6Tc_8$; $Ba@Os_6Mn_8$; $Ra@Os_6Mn_8$; $Sr@Os_6Mn_8$; $Yb@Os_6Mn_8$;
$Ba@Os_6Re_8$; $Ra@Os_6Re_8$; $Sr@Os_6Re_8$; $Yb@Os_6Re_8$; $Ba@Os_6Tc_8$; $Ra@Os_6Tc_8$; $Sr@Os_6Tc_8$; $Yb@Os_6Tc_8$;
$Ba@Ru_6Mn_8$; $Ra@Ru_6Mn_8$; $Sr@Ru_6Mn_8$; $Yb@Ru_6Mn_8$; $Ba@Ru_6Re_8$; $Ra@Ru_6Re_8$; $Sr@Ru_6Re_8$; $Yb@Ru_6Re_8$;
$Ba@Ru_6Tc_8$; $Ra@Ru_6Tc_8$; $Sr@Ru_6Tc_8$; $Yb@Ru_6Tc_8$; $Ba@Co_6Mn_8$; $Ra@Co_6Mn_8$; $Sr@Co_6Mn_8$; $Yb@Co_6Mn_8$;
$Ba@Co_6Re_8$; $Sr@Co_6Re_8$; $Ba@Co_6Tc_8$; $Ra@Co_6Tc_8$; $Sr@Co_6Tc_8$; $Yb@Co_6Tc_8$; $Ba@Ir_6Mn_8$; $Ra@Ir_6Mn_8$;
$Sr@Ir_6Mn_8$; $Yb@Ir_6Mn_8$; $Ba@Ir_6Re_8$; $Sr@Ir_6Re_8$; $Yb@Ir_6Re_8$; $Ba@Ir_6Tc_8$; $Ra@Ir_6Tc_8$; $Sr@Ir_6Tc_8$; $Yb@Ir_6Tc_8$;
$Ba@Rh_6Mn_8$; $Ra@Rh_6Mn_8$; $Sr@Rh_6Mn_8$; $Yb@Rh_6Mn_8$; $Ba@Rh_6Re_8$; $Ra@Rh_6Re_8$; $Sr@Rh_6Re_8$; $Ba@Rh_6Tc_8$;
$Ra@Rh_6Tc_8$; $Sr@Rh_6Tc_8$; $Yb@Rh_6Tc_8$;

Table S19. The molecular formula of the stable structure of Z$_3$ when the co-dual structure is embedded.

| |
|---|
| 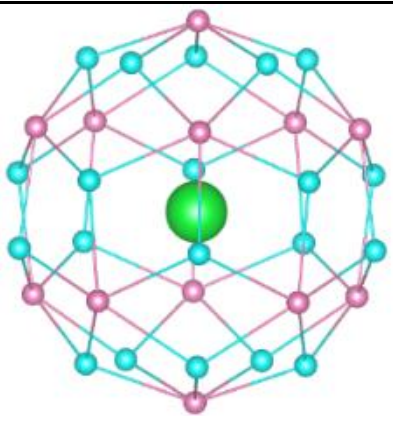<br>Z$_3$ Co-dual structure |
| **M@X$_{34}$ (30)** |
| Ba@Ag$_{34}$; Ra@Ag$_{34}$; Sr@Ag$_{34}$; Yb@Ag$_{34}$; Ba@Au$_{34}$; Ra@Au$_{34}$; Sr@Au$_{34}$; Yb@Au$_{34}$; Ba@Cs$_{34}$; Yb@Cs$_{34}$; Ba@Cu$_{34}$; Ra@Cu$_{34}$; Sr@Cu$_{34}$; Yb@Cu$_{34}$; Ba@Fr$_{34}$; Ra@Fr$_{34}$; Sr@Fr$_{34}$; Yb@Fr$_{34}$; Ba@K$_{34}$; Ra@K$_{34}$; Sr@K$_{34}$; Yb@K$_{34}$; Ba@Na$_{34}$; Ra@Na$_{34}$; Sr@Na$_{34}$;Yb@Na$_{34}$; Ba@Rb$_{34}$; Ra@Rb$_{34}$; Sr@Rb$_{34}$; Yb@Rb$_{34}$; |
| **M@X$_{12}$Y$_{20}$ (150)** |
| Ba@Al$_{12}$Ag$_{20}$; Ra@Al$_{12}$Ag$_{20}$; Sr@Al$_{12}$Ag$_{20}$; Yb@Al$_{12}$Ag$_{20}$; Ba@Ga$_{12}$Ag$_{20}$; Sr@Ga$_{12}$Ag$_{20}$; Yb@Ga$_{12}$Ag$_{20}$; Ba@Al$_{12}$Au$_{20}$; Ra@Al$_{12}$Au$_{20}$; Sr@Al$_{12}$Au$_{20}$; Yb@Al$_{12}$Au$_{20}$; Ba@Al$_{12}$Cu$_{20}$; Ra@Al$_{12}$Cu$_{20}$; Sr@Al$_{12}$Cu$_{20}$; Yb@Al$_{12}$Cu$_{20}$; Ba@Ga$_{12}$Au$_{20}$; Ra@Ga$_{12}$Au$_{20}$; Yb@Ga$_{12}$Au$_{20}$; Ba@Al$_{12}$As$_{20}$; Ra@Al$_{12}$As$_{20}$; Ra@Ga$_{12}$As$_{20}$; Ba@In$_{12}$Bi$_{20}$; Ra@In$_{12}$Bi$_{20}$; Sr@In$_{12}$Bi$_{20}$; Yb@In$_{12}$Bi$_{20}$; Ba@Tl$_{12}$Bi$_{20}$; Ra@Tl$_{12}$Bi$_{20}$; Sr@Tl$_{12}$Bi$_{20}$; Yb@Tl$_{12}$Bi$_{20}$; Ba@Ga$_{12}$Sb$_{20}$; Ra@Ga$_{12}$Sb$_{20}$; Sr@Ga$_{12}$Sb$_{20}$; Yb@Ga$_{12}$Sb$_{20}$; Ba@Ge$_{12}$Ag$_{20}$; Ra@Ge$_{12}$Ag$_{20}$; Sr@Ge$_{12}$Ag$_{20}$; Yb@Ge$_{12}$Ag$_{20}$; Ba@Ge$_{12}$Au$_{20}$; Ra@Ge$_{12}$Au$_{20}$; Sr@Ge$_{12}$Au$_{20}$; Yb@Ge$_{12}$Au$_{20}$; Ba@Ge$_{12}$Cu$_{20}$; Ra@Ge$_{12}$Cu$_{20}$; Sr@Ge$_{12}$Cu$_{20}$; Yb@Ge$_{12}$Cu$_{20}$; Ba@Si$_{12}$Ag$_{20}$; Ra@Si$_{12}$Ag$_{20}$; Sr@Si$_{12}$Ag$_{20}$; Yb@Si$_{12}$Ag$_{20}$; Ba@Si$_{12}$Au$_{20}$; Ra@Si$_{12}$Au$_{20}$; Sr@Si$_{12}$Au$_{20}$; Yb@Si$_{12}$Au$_{20}$; Ba@Si$_{12}$Cu$_{20}$; Ra@Si$_{12}$Cu$_{20}$; Sr@Si$_{12}$Cu$_{20}$; Yb@Si$_{12}$Cu$_{20}$; Ba@Sn$_{12}$Ag$_{20}$; Ra@Sn$_{12}$Ag$_{20}$; Sr@Sn$_{12}$Ag$_{20}$; Yb@Sn$_{12}$Ag$_{20}$; Ba@Sc$_{12}$Ag$_{20}$; Ra@Sc$_{12}$Ag$_{20}$; Sr@Sc$_{12}$Ag$_{20}$; Ba@Sc$_{12}$Au$_{20}$; Ra@Sc$_{12}$Au$_{20}$; Sr@Sc$_{12}$Au$_{20}$; Ba@Al$_{12}$Nb$_{20}$; Ra@Al$_{12}$Nb$_{20}$; Sr@Al$_{12}$Nb$_{20}$; Yb@Al$_{12}$Nb$_{20}$; Ba@Al$_{12}$Ta$_{20}$; Ra@Al$_{12}$Ta$_{20}$; Yb@Al$_{12}$Ta$_{20}$; Ba@Ga$_{12}$Nb$_{20}$; Ra@Ga$_{12}$Nb$_{20}$; Sr@Ga$_{12}$Nb$_{20}$; Yb@Ga$_{12}$Nb$_{20}$; Ba@Ga$_{12}$Ta$_{20}$; Ra@Ga$_{12}$Ta$_{20}$; Sr@Ga$_{12}$Ta$_{20}$; Yb@Ga$_{12}$Ta$_{20}$; Ba@Ga$_{12}$V$_{20}$; Ra@Ga$_{12}$V$_{20}$; Yb@Ga$_{12}$V$_{20}$; Ba@In$_{12}$Nb$_{20}$; Ra@In$_{12}$Nb$_{20}$; Sr@In$_{12}$Nb$_{20}$; Yb@In$_{12}$Nb$_{20}$; Ba@In$_{12}$Ta$_{20}$; Ra@In$_{12}$Ta$_{20}$; Sr@In$_{12}$Ta$_{20}$; Yb@In$_{12}$Ta$_{20}$; Ba@Hf$_{12}$Ag$_{20}$; Ra@Hf$_{12}$Ag$_{20}$; Sr@Hf$_{12}$Ag$_{20}$; Yb@Hf$_{12}$Ag$_{20}$; Ba@Hf$_{12}$Au$_{20}$; Ra@Hf$_{12}$Au$_{20}$; Sr@Hf$_{12}$Au$_{20}$; Ba@Ti$_{12}$Au$_{20}$; Ra@Ti$_{12}$Au$_{20}$; Sr@Ti$_{12}$Au$_{20}$; Yb@Ti$_{12}$Au$_{20}$; Sr@Ti$_{12}$Li$_{20}$; Ba@Zr$_{12}$Ag$_{20}$; Ba@Zr$_{12}$Au$_{20}$; Ra@Zr$_{12}$Au$_{20}$; Sr@Zr$_{12}$Au$_{20}$; Yb@Zr$_{12}$Au$_{20}$; Ba@Mn$_{12}$Ag$_{20}$; Ra@Mn$_{12}$Ag$_{20}$; Sr@Mn$_{12}$Ag$_{20}$; Yb@Mn$_{12}$Ag$_{20}$; Ba@Mn$_{12}$Au$_{20}$; Ra@Mn$_{12}$Au$_{20}$; Sr@Mn$_{12}$Au$_{20}$; Yb@Mn$_{12}$Au$_{20}$; Ba@Mn$_{12}$Cu$_{20}$; Ra@Mn$_{12}$Cu$_{20}$; Sr@Mn$_{12}$Cu$_{20}$; Yb@Mn$_{12}$Cu$_{20}$; Ba@Re$_{12}$Ag$_{20}$; Ra@Re$_{12}$Ag$_{20}$; Sr@Re$_{12}$Ag$_{20}$; Yb@Re$_{12}$Ag$_{20}$; Ba@Re$_{12}$Au$_{20}$; Ra@Re$_{12}$Au$_{20}$; Yb@Re$_{12}$Au$_{20}$; Ba@Tc$_{12}$Ag$_{20}$; Ra@Tc$_{12}$Ag$_{20}$; Sr@Tc$_{12}$Ag$_{20}$; Yb@Tc$_{12}$Ag$_{20}$; Ba@Tc$_{12}$Au$_{20}$; Ra@Tc$_{12}$Au$_{20}$; Sr@Tc$_{12}$Au$_{20}$; Yb@Tc$_{12}$Au$_{20}$; Ba@Tc$_{12}$Cu$_{20}$; Ra@Tc$_{12}$Cu$_{20}$; Sr@Tc$_{12}$Cu$_{20}$; Yb@Tc$_{12}$Cu$_{20}$; Ba@Fe$_{12}$Zn$_{20}$; Ba@Os$_{12}$Hg$_{20}$; Ra@Os$_{12}$Hg$_{20}$; Sr@Os$_{12}$Hg$_{20}$; Ba@Os$_{12}$Zn$_{20}$; Ra@Os$_{12}$Zn$_{20}$; Ba@Ru$_{12}$Hg$_{20}$; Ra@Ru$_{12}$Hg$_{20}$; Sr@Ru$_{12}$Hg$_{20}$; |

Table S20. The molecular formula of the stable structure of A$_1$ when the co-dual structure is embedded.

| |
|---|
| 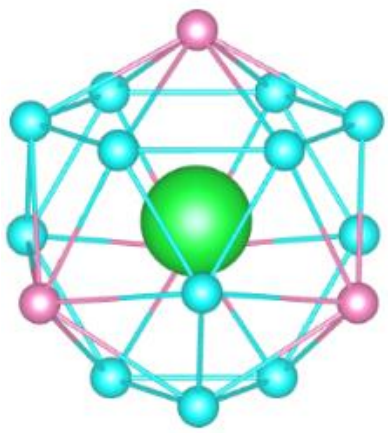 A$_1$ Co-dual structure |
| **M@X$_{16}$ (94)** |
| Ba@Au$_{16}$; Ra@Au$_{16}$; Sr@Au$_{16}$; Yb@Au$_{16}$; Ba@Ag$_{16}$; Ra@Ag$_{16}$; Sr@Ag$_{16}$; Yb@Ag$_{16}$; Ba@Cs$_{16}$; Ra@Cs$_{16}$; Ba@Cu$_{16}$; Ra@Cu$_{16}$; Sr@Cu$_{16}$; Yb@Cu$_{16}$; Ba@Li$_{16}$; Ra@Li$_{16}$; Sr@Li$_{16}$; Yb@Li$_{16}$; Ba@Na$_{16}$; Ra@Na$_{16}$; Sr@Na$_{16}$; Yb@Na$_{16}$; Ba@Rb$_{16}$; Ra@Rb$_{16}$; Yb@Rb$_{16}$; Sr@Cs$_{16}$; Yb@Cs$_{16}$; Sr@Rb$_{16}$; Ba@Fr$_{16}$; Ra@Fr$_{16}$; Sr@Fr$_{16}$; Yb@Fr$_{16}$; Ba@K$_{16}$; Ra@K$_{16}$; Sr@K$_{16}$; Yb@K$_{16}$; Yb@Ba$_{16}$; Ba@Be$_{16}$; Ra@Be$_{16}$; Sr@Be$_{16}$; Yb@Be$_{16}$; Ba@Cd$_{16}$; Yb@Cd$_{16}$; Ra@Cd$_{16}$; Sr@Cd$_{16}$; Yb@Hg$_{16}$; Ba@Mg$_{16}$; Ra@Mg$_{16}$; Sr@Mg$_{16}$; Yb@Mg$_{16}$; Ra@Ra$_{16}$; Yb@Ra$_{16}$; Yb@Sr$_{16}$; Ra@Zn$_{16}$; Sr@Zn$_{16}$; Yb@Zn$_{16}$; Sr@Hg$_{16}$; Ba@Hg$_{16}$; Ra@Hg$_{16}$; Ba@Ba$_{16}$; Ra@Ba$_{16}$; Sr@Ba$_{16}$; Ba@Ra$_{16}$; Sr@Ra$_{16}$; Ba@Sr$_{16}$; Ra@Sr$_{16}$; Sr@Sr$_{16}$; Ba@Ca$_{16}$; Ra@Ca$_{16}$; Sr@Ca$_{16}$; Yb@Ca$_{16}$; Sr@Cu$_{16}$; Ba@Cu$_{16}$; Ra@Cu$_{16}$; Yb@Cu$_{16}$; Sr@Ag$_{16}$; Ba@Ag$_{16}$; Ra@Ag$_{16}$; Yb@Ag$_{16}$; Sr@Au$_{16}$; Ba@Au$_{16}$; Ra@Au$_{16}$; Yb@Au$_{16}$; Sr@Sr$_{16}$; Ba@Sr$_{16}$; Ra@Sr$_{16}$; Sr@Cd$_{16}$; Ba@Cd$_{16}$; Ra@Cd$_{16}$; Yb@Cd$_{16}$; Sr@Hg$_{16}$; Ba@Hg$_{16}$; Ra@Hg$_{16}$; Yb@Hg$_{16}$; |
| **M@X$_{12}$Y$_4$ (772)** |
| Ba@Ag$_{12}$Au$_4$; Ra@Ag$_{12}$Au$_4$; Sr@Ag$_{12}$Au$_4$; Yb@Ag$_{12}$Au$_4$; Ba@Ag$_{12}$Cu$_4$; Ra@Ag$_{12}$Cu$_4$; Sr@Ag$_{12}$Cu$_4$; Yb@Ag$_{12}$Cu$_4$; Ba@Ag$_{12}$Li$_4$; Ra@Ag$_{12}$Li$_4$; Sr@Ag$_{12}$Li$_4$; Yb@Ag$_{12}$Li$_4$; Ba@Ag$_{12}$Na$_4$; Ra@Ag$_{12}$Na$_4$; Sr@Ag$_{12}$Na$_4$; Yb@Ag$_{12}$Na$_4$; Ba@Au$_{12}$Ag$_4$; Ra@Au$_{12}$Ag$_4$; Sr@Au$_{12}$Ag$_4$; Yb@Au$_{12}$Ag$_4$; Ba@Au$_{12}$Cu$_4$; Ra@Au$_{12}$Cu$_4$; Sr@Au$_{12}$Cu$_4$; Yb@Au$_{12}$Cu$_4$; Ba@Au$_{12}$Li$_4$; Ra@Au$_{12}$Li$_4$; Sr@Au$_{12}$Li$_4$; Yb@Au$_{12}$Li$_4$; Ba@Au$_{12}$Na$_4$; Ra@Au$_{12}$Na$_4$; Sr@Au$_{12}$Na$_4$; Sr@Au$_{12}$Na$_4$; Ba@Cs$_{12}$Fr$_4$; Ra@Cs$_{12}$Fr$_4$; Sr@Cs$_{12}$Fr$_4$; Yb@Cs$_{12}$Fr$_4$; Ba@Cs$_{12}$Rb$_4$; Ra@Cs$_{12}$Rb$_4$; Sr@Cs$_{12}$Rb$_4$; Yb@Cs$_{12}$Rb$_4$; Ba@Cu$_{12}$Ag$_4$; Ra@Cu$_{12}$Ag$_4$; Sr@Cu$_{12}$Ag$_4$; Yb@Cu$_{12}$Ag$_4$; Ba@Cu$_{12}$Au$_4$; Ra@Cu$_{12}$Au$_4$; Sr@Cu$_{12}$Au$_4$; Yb@Cu$_{12}$Au$_4$; Ba@Cu$_{12}$Li$_4$; Ra@Cu$_{12}$Li$_4$; Sr@Cu$_{12}$Li$_4$; Yb@Cu$_{12}$Li$_4$; Ba@Li$_{12}$Ag$_4$; Ra@Li$_{12}$Ag$_4$; Sr@Li$_{12}$Ag$_4$; Yb@Li$_{12}$Ag$_4$; Ba@Li$_{12}$Au$_4$; Ra@Li$_{12}$Au$_4$; Sr@Li$_{12}$Au$_4$; Yb@Li$_{12}$Au$_4$; Ba@Li$_{12}$Cu$_4$; Ra@Li$_{12}$Cu$_4$; Sr@Li$_{12}$Cu$_4$; Yb@Li$_{12}$Cu$_4$; Ba@Na$_{12}$Ag$_4$; Ra@Na$_{12}$Ag$_4$; Sr@Na$_{12}$Ag$_4$; Yb@Na$_{12}$Ag$_4$; Ba@Na$_{12}$Au$_4$; Ra@Na$_{12}$Au$_4$; Sr@Na$_{12}$Au$_4$; Yb@Na$_{12}$Au$_4$; Ba@Rb$_{12}$K$_4$; Ra@Rb$_{12}$K$_4$; Sr@Rb$_{12}$K$_4$; Yb@Rb$_{12}$K$_4$; Ba@Cs$_{12}$K$_4$; Ra@Cs$_{12}$K$_4$; Sr@Cs$_{12}$K$_4$; Yb@Cs$_{12}$K$_4$; Ba@Fr$_{12}$Cs$_4$; Ra@Fr$_{12}$Cs$_4$; Sr@Fr$_{12}$Cs$_4$; Yb@Fr$_{12}$Cs$_4$; Ba@Fr$_{12}$K$_4$; Ra@Fr$_{12}$K$_4$; Sr@Fr$_{12}$K$_4$; Yb@Fr$_{12}$K$_4$; Ba@Fr$_{12}$Rb$_4$; Ra@Fr$_{12}$Rb$_4$; Sr@Fr$_{12}$Rb$_4$; Yb@Fr$_{12}$Rb$_4$; Ba@K$_{12}$Rb$_4$; Ra@K$_{12}$Rb$_4$; Sr@K$_{12}$Rb$_4$; Yb@K$_{12}$Rb$_4$; Ba@Rb$_{12}$Cs$_4$; Ra@Rb$_{12}$Cs$_4$; Sr@Rb$_{12}$Cs$_4$; Yb@Rb$_{12}$Cs$_4$; Ba@Ag$_{12}$As$_4$; Ra@Ag$_{12}$As$_4$; Sr@Ag$_{12}$As$_4$; Yb@Ag$_{12}$As$_4$; Ba@Ag$_{12}$Bi$_4$; Ra@Ag$_{12}$Bi$_4$; Sr@Ag$_{12}$Bi$_4$; Yb@Ag$_{12}$Bi$_4$; Ba@Ag$_{12}$Sb$_4$; Ra@Ag$_{12}$Sb$_4$; Sr@Ag$_{12}$Sb$_4$; Yb@Ag$_{12}$Sb$_4$; Ba@Au$_{12}$As$_4$; Ba@Au$_{12}$Bi$_4$; Ra@Au$_{12}$Bi$_4$; Sr@Au$_{12}$Bi$_4$; Yb@Au$_{12}$Bi$_4$; Ba@Au$_{12}$Sb$_4$; Ra@Au$_{12}$Sb$_4$; Yb@Au$_{12}$Sb$_4$; Ba@Cu$_{12}$As$_4$; Sr@Cu$_{12}$As$_4$; Yb@Cu$_{12}$As$_4$; Ba@Li$_{12}$As$_4$; Ra@Li$_{12}$As$_4$; Sr@Li$_{12}$As$_4$; Yb@Li$_{12}$As$_4$; Ba@Li$_{12}$Sb$_4$; Ra@Li$_{12}$Sb$_4$; Sr@Li$_{12}$Sb$_4$; Yb@Li$_{12}$Sb$_4$; Ba@Na$_{12}$Bi$_4$; Ra@Na$_{12}$Bi$_4$; Sr@Na$_{12}$Bi$_4$; Yb@Na$_{12}$Bi$_4$; Ba@Na$_{12}$Sb$_4$; Ra@Na$_{12}$Sb$_4$; Sr@Na$_{12}$Sb$_4$; Yb@Na$_{12}$Sb$_4$; Ba@Al$_{12}$As$_4$; Ra@Al$_{12}$As$_4$; Sr@Al$_{12}$As$_4$; Yb@Al$_{12}$As$_4$; Ba@Ga$_{12}$As$_4$; Ra@Ga$_{12}$As$_4$; Sr@Ga$_{12}$As$_4$; Yb@Ga$_{12}$As$_4$; Ba@Ga$_{12}$Sb$_4$; Ra@Ga$_{12}$Sb$_4$; Sr@Ga$_{12}$Sb$_4$; Yb@Ga$_{12}$Sb$_4$; Ba@In$_{12}$Bi$_4$; Ra@In$_{12}$Bi$_4$; Sr@In$_{12}$Bi$_4$; Yb@In$_{12}$Bi$_4$; Ba@In$_{12}$Sb$_4$; Ra@In$_{12}$Sb$_4$; Sr@In$_{12}$Sb$_4$; Yb@In$_{12}$Sb$_4$; Ba@Tl$_{12}$Bi$_4$; Ra@Tl$_{12}$Bi$_4$; Sr@Tl$_{12}$Bi$_4$; Yb@Tl$_{12}$Bi$_4$; Ba@Tl$_{12}$Sb$_4$; Ra@Tl$_{12}$Sb$_4$; Sr@Tl$_{12}$Sb$_4$; Yb@Tl$_{12}$Sb$_4$; Ba@Ge$_{12}$Mg$_4$; Ra@Ge$_{12}$Mg$_4$; Sr@Ge$_{12}$Mg$_4$; Yb@Ge$_{12}$Mg$_4$; Ba@Ge$_{12}$Zn$_4$; Ra@Ge$_{12}$Zn$_4$; Sr@Ge$_{12}$Zn$_4$; Yb@Ge$_{12}$Zn$_4$; Ba@Pb$_{12}$Cd$_4$; Ra@Pb$_{12}$Cd$_4$; Sr@Pb$_{12}$Cd$_4$; Yb@Pb$_{12}$Cd$_4$; Ba@Pb$_{12}$Hg$_4$; Ra@Pb$_{12}$Hg$_4$; Sr@Pb$_{12}$Hg$_4$; Yb@Pb$_{12}$Hg$_4$; Ba@Pb$_{12}$Mg$_4$; Ra@Pb$_{12}$Mg$_4$; Sr@Pb$_{12}$Mg$_4$; Yb@Pb$_{12}$Mg$_4$; Ba@Pb$_{12}$Zn$_4$; Ra@Pb$_{12}$Zn$_4$; Sr@Pb$_{12}$Zn$_4$; Yb@Pb$_{12}$Zn$_4$; Ba@Si$_{12}$Mg$_4$; Ra@Si$_{12}$Mg$_4$; Sr@Si$_{12}$Mg$_4$; Yb@Si$_{12}$Mg$_4$; Ba@Si$_{12}$Zn$_4$; Sr@Si$_{12}$Zn$_4$; Ba@Sn$_{12}$Cd$_4$; Ra@Sn$_{12}$Cd$_4$; Sr@Sn$_{12}$Cd$_4$; Yb@Sn$_{12}$Cd$_4$; Ba@Sn$_{12}$Hg$_4$; Ra@Sn$_{12}$Hg$_4$; Sr@Sn$_{12}$Hg$_4$; Yb@Sn$_{12}$Hg$_4$; Ba@Sn$_{12}$Mg$_4$; Ra@Sn$_{12}$Mg$_4$; Sr@Sn$_{12}$Mg$_4$; Yb@Sn$_{12}$Mg$_4$; Ba@Sn$_{12}$Zn$_4$; Ra@Sn$_{12}$Zn$_4$; Sr@Sn$_{12}$Zn$_4$; Yb@Sn$_{12}$Zn$_4$; Ba@Ge$_{12}$As$_4$; Ra@Ge$_{12}$As$_4$; Sr@Ge$_{12}$As$_4$; Yb@Ge$_{12}$As$_4$; Ba@Ge$_{12}$Sb$_4$; Ra@Ge$_{12}$Sb$_4$; Sr@Ge$_{12}$Sb$_4$; Yb@Ge$_{12}$Sb$_4$; Ba@Pb$_{12}$Bi$_4$; Ra@Pb$_{12}$Bi$_4$; Sr@Pb$_{12}$Bi$_4$; Yb@Pb$_{12}$Bi$_4$; Ba@Pb$_{12}$Sb$_4$; Ra@Pb$_{12}$Sb$_4$; Sr@Pb$_{12}$Sb$_4$; Yb@Pb$_{12}$Sb$_4$; Ba@Si$_{12}$As$_4$; Ra@Si$_{12}$As$_4$; Sr@Si$_{12}$As$_4$; Yb@Si$_{12}$As$_4$; Ba@Sn$_{12}$Bi$_4$; Ra@Sn$_{12}$Bi$_4$; Sr@Sn$_{12}$Bi$_4$; Yb@Sn$_{12}$Bi$_4$; Ba@Sn$_{12}$Sb$_4$; Ra@Sn$_{12}$Sb$_4$; Sr@Sn$_{12}$Sb$_4$; Yb@Sn$_{12}$Sb$_4$; Ba@As$_{12}$Mg$_4$; Ra@As$_{12}$Mg$_4$; Ra@Bi$_{12}$Mg$_4$; Ba@Sb$_{12}$Mg$_4$; Ra@Sb$_{12}$Mg$_4$; Sr@Sb$_{12}$Mg$_4$; Yb@Sb$_{12}$Mg$_4$; Ba@Li$_{12}$V$_4$; Ra@Li$_{12}$V$_4$; Sr@Li$_{12}$V$_4$; Yb@Li$_{12}$V$_4$; Ba@Li$_{12}$Nb$_4$; Ra@Li$_{12}$Nb$_4$; Sr@Li$_{12}$Nb$_4$; Yb@Li$_{12}$Nb$_4$; Ba@Li$_{12}$Ta$_4$; Ra@Li$_{12}$Ta$_4$; Sr@Li$_{12}$Ta$_4$; Yb@Li$_{12}$Ta$_4$; Ba@Cu$_{12}$V$_4$; Ra@Cu$_{12}$V$_4$; Sr@Cu$_{12}$V$_4$; Yb@Cu$_{12}$V$_4$; Ba@Cu$_{12}$Nb$_4$; Ra@Cu$_{12}$Nb$_4$; Sr@Cu$_{12}$Nb$_4$; Yb@Cu$_{12}$Nb$_4$; Ra@Cu$_{12}$Ta$_4$; Sr@Cu$_{12}$Ta$_4$; Yb@Cu$_{12}$Ta$_4$; Ba@Ag$_{12}$V$_4$; Ra@Ag$_{12}$V$_4$; Sr@Ag$_{12}$V$_4$; Yb@Ag$_{12}$V$_4$; Ba@Ag$_{12}$Nb$_4$; Ra@Ag$_{12}$Nb$_4$; Sr@Ag$_{12}$Nb$_4$; Yb@Ag$_{12}$Nb$_4$; Ba@Ag$_{12}$Ta$_4$; Ra@Ag$_{12}$Ta$_4$; Sr@Ag$_{12}$Ta$_4$; Yb@Ag$_{12}$Ta$_4$; Ba@Au$_{12}$V$_4$; Ra@Au$_{12}$V$_4$; Sr@Au$_{12}$V$_4$; Yb@Au$_{12}$V$_4$; Ba@Au$_{12}$Nb$_4$; Ra@Au$_{12}$Nb$_4$; Sr@Au$_{12}$Nb$_4$; Yb@Au$_{12}$Nb$_4$; Ba@Au$_{12}$Ta$_4$; Ra@Au$_{12}$Ta$_4$; Yb@Au$_{12}$Ta$_4$; Ra@Mg$_{12}$Zn$_4$; Sr@Mg$_{12}$Zn$_4$; Ba@Mg$_{12}$Cd$_4$; Ra@Mg$_{12}$Cd$_4$; Sr@Mg$_{12}$Cd$_4$; Yb@Mg$_{12}$Cd$_4$; Ba@Mg$_{12}$Hg$_4$; Ra@Mg$_{12}$Hg$_4$; Sr@Mg$_{12}$Hg$_4$; Yb@Mg$_{12}$Hg$_4$; Ra@Zn$_{12}$Mg$_4$; Sr@Zn$_{12}$Mg$_4$; Ba@Cd$_{12}$Mg$_4$; Ra@Cd$_{12}$Mg$_4$; Sr@Cd$_{12}$Mg$_4$; Yb@Cd$_{12}$Mg$_4$; Ba@Cd$_{12}$Zn$_4$; |

Ra@$Cd_{12}Zn_4$; Sr@$Cd_{12}Zn_4$; Yb@$Cd_{12}Zn_4$; Ba@$Cd_{12}Hg_4$; Ra@$Cd_{12}Hg_4$; Sr@$Cd_{12}Hg_4$; Yb@$Cd_{12}Hg_4$;
Ba@$Hg_{12}Mg_4$; Ra@$Hg_{12}Mg_4$; Sr@$Hg_{12}Mg_4$; Yb@$Hg_{12}Mg_4$; Ba@$Hg_{12}Zn_4$; Ra@$Hg_{12}Zn_4$; Sr@$Hg_{12}Zn_4$;
Yb@$Hg_{12}Zn_4$; Ba@$Hg_{12}Cd_4$; Ra@$Hg_{12}Cd_4$; Sr@$Hg_{12}Cd_4$; Yb@$Hg_{12}Cd_4$; Ba@$Mg_{12}Fe_4$; Ra@$Mg_{12}Fe_4$;
Sr@$Mg_{12}Fe_4$; Yb@$Mg_{12}Fe_4$; Ba@$Mg_{12}Ru_4$; Ra@$Mg_{12}Ru_4$; Sr@$Mg_{12}Ru_4$; Yb@$Mg_{12}Ru_4$; Ba@$Mg_{12}Os_4$;
Ra@$Mg_{12}Os_4$; Sr@$Mg_{12}Os_4$; Yb@$Mg_{12}Os_4$; Ba@$Zn_{12}Ru_4$; Ra@$Zn_{12}Ru_4$; Sr@$Zn_{12}Ru_4$; Yb@$Zn_{12}Ru_4$;
Ba@$Zn_{12}Os_4$; Ra@$Zn_{12}Os_4$; Sr@$Zn_{12}Os_4$; Yb@$Zn_{12}Os_4$; Ba@$Hg_{12}Ru_4$; Ra@$Hg_{12}Ru_4$; Sr@$Hg_{12}Ru_4$;
Yb@$Hg_{12}Ru_4$; Sr@$Hg_{12}Os_4$; Ba@$Sc_{12}Nb_4$; Ra@$Sc_{12}Nb_4$; Sr@$Sc_{12}Nb_4$; Yb@$Sc_{12}Nb_4$; Ba@$Sc_{12}Ta_4$;
Ra@$Sc_{12}Ta_4$; Sr@$Sc_{12}Ta_4$; Yb@$Sc_{12}Ta_4$; Ba@$Sc_{12}Sb_4$; Ra@$Sc_{12}Sb_4$; Sr@$Sc_{12}Sb_4$; Yb@$Sc_{12}Sb_4$; Ba@$Sc_{12}Bi_4$;
Ra@$Sc_{12}Bi_4$; Sr@$Sc_{12}Bi_4$; Yb@$Sc_{12}Bi_4$; Ba@$Y_{12}Bi_4$; Ra@$Y_{12}Bi_4$; Sr@$Y_{12}Bi_4$; Yb@$Y_{12}Bi_4$; Ba@$Al_{12}V_4$;
Ra@$Al_{12}V_4$; Sr@$Al_{12}V_4$; Yb@$Al_{12}V_4$; Ba@$Al_{12}Nb_4$; Ra@$Al_{12}Nb_4$; Sr@$Al_{12}Nb_4$; Yb@$Al_{12}Nb_4$; Ba@$Al_{12}Ta_4$;
Ra@$Al_{12}Ta_4$; Sr@$Al_{12}Ta_4$; Yb@$Al_{12}Ta_4$; Ba@$Ga_{12}V_4$; Ra@$Ga_{12}V_4$; Yb@$Ga_{12}V_4$; Ba@$Ga_{12}Nb_4$; Ra@$Ga_{12}Nb_4$;
Sr@$Ga_{12}Nb_4$; Yb@$Ga_{12}Nb_4$; Ba@$Ga_{12}Ta_4$; Ra@$Ga_{12}Ta_4$; Sr@$Ga_{12}Ta_4$; Yb@$Ga_{12}Ta_4$; Ba@$In_{12}Nb_4$;
Ra@$In_{12}Nb_4$; Sr@$In_{12}Nb_4$; Yb@$In_{12}Nb_4$; Ba@$In_{12}Ta_4$; Ra@$In_{12}Ta_4$; Sr@$In_{12}Ta_4$; Yb@$In_{12}Ta_4$; Ba@$Tl_{12}Nb_4$;
Ra@$Tl_{12}Nb_4$; Sr@$Tl_{12}Nb_4$; Yb@$Tl_{12}Nb_4$; Ba@$Tl_{12}Ta_4$; Ra@$Tl_{12}Ta_4$; Sr@$Tl_{12}Ta_4$; Yb@$Tl_{12}Ta_4$; Ba@$Ti_{12}Mg_4$;
Ra@$Ti_{12}Mg_4$; Sr@$Ti_{12}Mg_4$; Yb@$Ti_{12}Mg_4$; Ba@$Ti_{12}Zn_4$; Ra@$Ti_{12}Zn_4$; Sr@$Ti_{12}Zn_4$; Yb@$Ti_{12}Zn_4$; Ba@$Ti_{12}Cd_4$;
Ra@$Ti_{12}Cd_4$; Sr@$Ti_{12}Cd_4$; Yb@$Ti_{12}Cd_4$; Ba@$Ti_{12}Hg_4$; Ra@$Ti_{12}Hg_4$; Sr@$Ti_{12}Hg_4$; Yb@$Ti_{12}Hg_4$; Ba@$Zr_{12}Mg_4$;
Ra@$Zr_{12}Mg_4$; Sr@$Zr_{12}Mg_4$; Yb@$Zr_{12}Mg_4$; Ba@$Zr_{12}Cd_4$; Ra@$Zr_{12}Cd_4$; Sr@$Zr_{12}Cd_4$; Yb@$Zr_{12}Cd_4$;
Ba@$Zr_{12}Hg_4$; Ra@$Zr_{12}Hg_4$; Sr@$Zr_{12}Hg_4$; Yb@$Zr_{12}Hg_4$; Ba@$Hf_{12}Mg_4$; Ra@$Hf_{12}Mg_4$; Sr@$Hf_{12}Mg_4$;
Yb@$Hf_{12}Mg_4$; Ba@$Hf_{12}Zn_4$; Ra@$Hf_{12}Zn_4$; Sr@$Hf_{12}Zn_4$; Yb@$Hf_{12}Zn_4$; Ba@$Hf_{12}Cd_4$; Ra@$Hf_{12}Cd_4$;
Sr@$Hf_{12}Cd_4$; Yb@$Hf_{12}Cd_4$; Ba@$Hf_{12}Hg_4$; Ra@$Hf_{12}Hg_4$; Sr@$Hf_{12}Hg_4$; Yb@$Hf_{12}Hg_4$; Ba@$Ti_{12}V_4$; Ra@$Ti_{12}V_4$;
Sr@$Ti_{12}V_4$; Yb@$Ti_{12}V_4$; Ba@$Ti_{12}Nb_4$; Ra@$Ti_{12}Nb_4$; Sr@$Ti_{12}Nb_4$; Yb@$Ti_{12}Nb_4$; Ba@$Ti_{12}Ta_4$; Ra@$Ti_{12}Ta_4$;
Sr@$Ti_{12}Ta_4$; Yb@$Ti_{12}Ta_4$; Sr@$Ti_{12}As_4$; Yb@$Ti_{12}As_4$; Ba@$Ti_{12}Sb_4$; Ra@$Ti_{12}Sb_4$; Sr@$Ti_{12}Sb_4$; Yb@$Ti_{12}Sb_4$;
Ba@$Ti_{12}Bi_4$; Ra@$Ti_{12}Bi_4$; Sr@$Ti_{12}Bi_4$; Yb@$Ti_{12}Bi_4$; Ba@$Zr_{12}Nb_4$; Ra@$Zr_{12}Nb_4$; Sr@$Zr_{12}Nb_4$; Yb@$Zr_{12}Nb_4$;
Ba@$Zr_{12}Ta_4$; Ra@$Zr_{12}Ta_4$; Sr@$Zr_{12}Ta_4$; Yb@$Zr_{12}Ta_4$; Ba@$Zr_{12}Sb_4$; Ra@$Zr_{12}Sb_4$; Sr@$Zr_{12}Sb_4$; Yb@$Zr_{12}Sb_4$;
Ba@$Zr_{12}Bi_4$; Ra@$Zr_{12}Bi_4$; Sr@$Zr_{12}Bi_4$; Yb@$Zr_{12}Bi_4$; Ba@$Hf_{12}Nb_4$; Ra@$Hf_{12}Nb_4$; Sr@$Hf_{12}Nb_4$; Yb@$Hf_{12}Nb_4$;
Ba@$Hf_{12}Ta_4$; Ra@$Hf_{12}Ta_4$; Sr@$Hf_{12}Ta_4$; Yb@$Hf_{12}Ta_4$; Ba@$Hf_{12}Sb_4$; Ra@$Hf_{12}Sb_4$; Sr@$Hf_{12}Sb_4$; Yb@$Hf_{12}Sb_4$;
Ba@$Hf_{12}Bi_4$; Ra@$Hf_{12}Bi_4$; Sr@$Hf_{12}Bi_4$; Yb@$Hf_{12}Bi_4$; Ba@$Si_{12}V_4$; Ra@$Si_{12}V_4$; Sr@$Si_{12}V_4$; Yb@$Si_{12}V_4$;
Ba@$Si_{12}Nb_4$; Ra@$Si_{12}Nb_4$; Sr@$Si_{12}Nb_4$; Yb@$Si_{12}Nb_4$; Ba@$Si_{12}Ta_4$; Ra@$Si_{12}Ta_4$; Sr@$Si_{12}Ta_4$; Yb@$Si_{12}Ta_4$;
Ba@$Ge_{12}V_4$; Ra@$Ge_{12}V_4$; Sr@$Ge_{12}V_4$; Yb@$Ge_{12}V_4$; Ba@$Ge_{12}Nb_4$; Ra@$Ge_{12}Nb_4$; Sr@$Ge_{12}Nb_4$; Yb@$Ge_{12}Nb_4$;
Ba@$Ge_{12}Ta_4$; Ra@$Ge_{12}Ta_4$; Sr@$Ge_{12}Ta_4$; Yb@$Ge_{12}Ta_4$; Ba@$Sn_{12}V_4$; Ra@$Sn_{12}V_4$; Sr@$Sn_{12}V_4$; Yb@$Sn_{12}V_4$;
Ba@$Sn_{12}Nb_4$; Ra@$Sn_{12}Nb_4$; Sr@$Sn_{12}Nb_4$; Yb@$Sn_{12}Nb_4$; Ba@$Sn_{12}Ta_4$; Ra@$Sn_{12}Ta_4$; Sr@$Sn_{12}Ta_4$; Yb@$Sn_{12}Ta_4$;
Ba@$Pb_{12}Nb_4$; Ra@$Pb_{12}Nb_4$; Sr@$Pb_{12}Nb_4$; Yb@$Pb_{12}Nb_4$; Ba@$Pb_{12}Ta_4$; Ra@$Pb_{12}Ta_4$; Sr@$Pb_{12}Ta_4$; Yb@$Pb_{12}Ta_4$;
Ra@$V_{12}Mg_4$; Ra@$V_{12}Zn_4$; Ba@$Nb_{12}Mg_4$; Ra@$Nb_{12}Mg_4$; Sr@$Nb_{12}Mg_4$; Yb@$Nb_{12}Mg_4$; Ba@$Nb_{12}Zn_4$;
Ra@$Nb_{12}Zn_4$; Sr@$Nb_{12}Zn_4$; Yb@$Nb_{12}Zn_4$; Ba@$Nb_{12}Cd_4$; Ra@$Nb_{12}Cd_4$; Sr@$Nb_{12}Cd_4$; Yb@$Nb_{12}Cd_4$;
Ba@$Nb_{12}Hg_4$; Ra@$Nb_{12}Hg_4$; Sr@$Nb_{12}Hg_4$; Yb@$Nb_{12}Hg_4$; Ba@$Ta_{12}Mg_4$; Ra@$Ta_{12}Mg_4$; Sr@$Ta_{12}Mg_4$;
Yb@$Ta_{12}Mg_4$; Ba@$Ta_{12}Zn_4$; Ra@$Ta_{12}Zn_4$; Sr@$Ta_{12}Zn_4$; Yb@$Ta_{12}Zn_4$; Ba@$Ta_{12}Cd_4$; Ra@$Ta_{12}Cd_4$; Sr@$Ta_{12}Cd_4$;
Yb@$Ta_{12}Cd_4$; Ba@$Ta_{12}Hg_4$; Ra@$Ta_{12}Hg_4$; Sr@$Ta_{12}Hg_4$; Yb@$Ta_{12}Hg_4$; Ba@$Cr_{12}Fe_4$; Sr@$Cr_{12}Fe_4$; Yb@$Cr_{12}Fe_4$;
Sr@$Mo_{12}Fe_4$; Yb@$Mo_{12}Fe_4$; Yb@$Mo_{12}Ru_4$; Yb@$Mo_{12}Os_4$; Sr@$W_{12}Fe_4$; Yb@$W_{12}Fe_4$; Yb@$W_{12}Os_4$;
Sr@$Te_{12}Os_4$; Yb@$Te_{12}Os_4$; Yb@$Mn_{12}V_4$; Sr@$Mn_{12}Nb_4$; Yb@$Mn_{12}Nb_4$; Ra@$Mn_{12}Ta_4$; Sr@$Mn_{12}Ta_4$;
Yb@$Mn_{12}Ta_4$; Ra@$Mn_{12}As_4$; Yb@$Tc_{12}Ta_4$; Sr@$Tc_{12}As_4$; Yb@$Tc_{12}As_4$; Ra@$Tc_{12}Sb_4$; Sr@$Tc_{12}Sb_4$; Yb@$Tc_{12}Sb_4$;
Ra@$Tc_{12}Bi_4$; Sr@$Tc_{12}Bi_4$; Yb@$Tc_{12}Bi_4$; Ra@$Re_{12}Nb_4$; Yb@$Re_{12}Nb_4$; Yb@$Re_{12}Ta_4$; Sr@$Re_{12}As_4$; Ba@$Re_{12}Sb_4$;
Ra@$Re_{12}Sb_4$; Sr@$Re_{12}Sb_4$; Yb@$Re_{12}Sb_4$; Ba@$Re_{12}Bi_4$; Sr@$Re_{12}Bi_4$; Yb@$Re_{12}Bi_4$; Ba@$Fe_{12}Mg_4$;
Ra@$Fe_{12}Mg_4$; Sr@$Fe_{12}Mg_4$; Yb@$Fe_{12}Mg_4$; Ba@$Fe_{12}Zn_4$; Ra@$Fe_{12}Zn_4$; Sr@$Fe_{12}Zn_4$; Yb@$Fe_{12}Zn_4$;
Ba@$Ru_{12}Mg_4$; Ra@$Ru_{12}Mg_4$; Sr@$Ru_{12}Mg_4$; Yb@$Ru_{12}Mg_4$; Ba@$Ru_{12}Zn_4$; Ra@$Ru_{12}Zn_4$; Sr@$Ru_{12}Zn_4$;
Yb@$Ru_{12}Zn_4$; Ba@$Ru_{12}Hg_4$; Ra@$Ru_{12}Hg_4$; Sr@$Ru_{12}Hg_4$; Yb@$Ru_{12}Hg_4$; Ba@$Os_{12}Mg_4$; Ra@$Os_{12}Mg_4$;
Sr@$Os_{12}Mg_4$; Yb@$Os_{12}Mg_4$; Ba@$Os_{12}Zn_4$; Ra@$Os_{12}Zn_4$; Sr@$Os_{12}Zn_4$; Yb@$Os_{12}Zn_4$; Ba@$Os_{12}Hg_4$;
Sr@$Os_{12}Hg_4$; Yb@$Os_{12}Hg_4$; Ra@$Fe_{12}Ni_4$; Sr@$Fe_{12}Ni_4$; Yb@$Fe_{12}Ni_4$; Ba@$Fe_{12}Pd_4$; Ra@$Fe_{12}Pd_4$; Sr@$Fe_{12}Pd_4$;
Yb@$Fe_{12}Pd_4$; Ba@$Fe_{12}Pt_4$; Ra@$Fe_{12}Pt_4$; Sr@$Fe_{12}Pt_4$; Yb@$Fe_{12}Pt_4$; Ba@$Ru_{12}Ni_4$; Sr@$Ru_{12}Ni_4$; Yb@$Ru_{12}Ni_4$;
Ba@$Ru_{12}Pd_4$; Ra@$Ru_{12}Pd_4$; Sr@$Ru_{12}Pd_4$; Yb@$Ru_{12}Pd_4$; Ba@$Ru_{12}Pt_4$; Sr@$Ru_{12}Pt_4$; Yb@$Ru_{12}Pt_4$; Ba@$Os_{12}Ni_4$;
Sr@$Os_{12}Ni_4$; Yb@$Os_{12}Ni_4$; Ba@$Os_{12}Pd_4$; Sr@$Os_{12}Pd_4$; Yb@$Os_{12}Pd_4$; Ba@$Os_{12}Pt_4$; Ra@$Os_{12}Pt_4$; Sr@$Os_{12}Pt_4$;
Yb@$Os_{12}Pt_4$; Ba@$Co_{12}Mn_4$; Ra@$Co_{12}Mn_4$; Sr@$Co_{12}Mn_4$; Yb@$Co_{12}Mn_4$; Ba@$Co_{12}Tc_4$; Ra@$Co_{12}Tc_4$;
Yb@$Co_{12}Tc_4$; Sr@$Co_{12}Re_4$; Yb@$Co_{12}Re_4$; Ra@$Rh_{12}Mn_4$; Sr@$Rh_{12}Mn_4$; Ra@$Ir_{12}Mn_4$; Sr@$Ir_{12}Mn_4$;
Yb@$Ir_{12}Mn_4$; Ba@$Ir_{12}Tc_4$; Ra@$Ir_{12}Tc_4$; Sr@$Ir_{12}Tc_4$; Yb@$Ir_{12}Tc_4$; Sr@$Ir_{12}Re_4$; Yb@$Ir_{12}Re_4$; Ba@$Ni_{12}Si_4$;
Ra@$Ni_{12}Si_4$; Sr@$Ni_{12}Si_4$; Yb@$Ni_{12}Si_4$; Ba@$Ni_{12}Ge_4$; Ra@$Ni_{12}Ge_4$; Sr@$Ni_{12}Ge_4$; Yb@$Ni_{12}Ge_4$; Sr@$Pd_{12}Ti_4$;
Ba@$Pd_{12}Zr_4$; Ra@$Pd_{12}Zr_4$; Sr@$Pd_{12}Zr_4$; Yb@$Pd_{12}Zr_4$; Ba@$Pd_{12}Hf_4$; Ra@$Pd_{12}Hf_4$; Sr@$Pd_{12}Hf_4$; Yb@$Pd_{12}Hf_4$;
Ba@$Pd_{12}Si_4$; Ra@$Pd_{12}Si_4$; Sr@$Pd_{12}Si_4$; Yb@$Pd_{12}Si_4$; Ba@$Pd_{12}Ge_4$; Ra@$Pd_{12}Ge_4$; Sr@$Pd_{12}Ge_4$; Yb@$Pd_{12}Ge_4$;
Ba@$Pd_{12}Sn_4$; Ra@$Pd_{12}Sn_4$; Sr@$Pd_{12}Sn_4$; Yb@$Pd_{12}Sn_4$; Ba@$Pd_{12}Pb_4$; Ra@$Pd_{12}Pb_4$; Sr@$Pd_{12}Pb_4$; Yb@$Pd_{12}Pb_4$;
Ba@$Pt_{12}Ti_4$; Ra@$Pt_{12}Ti_4$; Sr@$Pt_{12}Ti_4$; Yb@$Pt_{12}Ti_4$; Ba@$Pt_{12}Zr_4$; Ra@$Pt_{12}Zr_4$; Sr@$Pt_{12}Zr_4$; Yb@$Pt_{12}Zr_4$;
Ba@$Pt_{12}Hf_4$; Ra@$Pt_{12}Hf_4$; Sr@$Pt_{12}Hf_4$; Yb@$Pt_{12}Hf_4$; Ba@$Pt_{12}Si_4$; Ra@$Pt_{12}Si_4$; Sr@$Pt_{12}Si_4$; Yb@$Pt_{12}Si_4$;
Ba@$Pt_{12}Ge_4$; Ra@$Pt_{12}Ge_4$; Sr@$Pt_{12}Ge_4$; Yb@$Pt_{12}Ge_4$; Ba@$Pt_{12}Sn_4$; Ra@$Pt_{12}Sn_4$; Sr@$Pt_{12}Sn_4$; Yb@$Pt_{12}Sn_4$;
Ba@$Pt_{12}Pb_4$; Ra@$Pt_{12}Pb_4$; Sr@$Pt_{12}Pb_4$; Yb@$Pt_{12}Pb_4$;

Table S21. The molecular formula of the stable structure of A$_2$ when the co-dual structure is embedded.

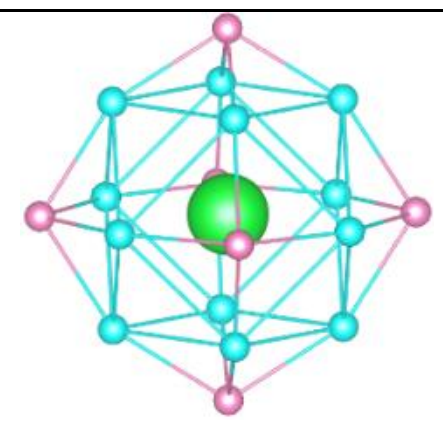

A$_2$ Co-dual structure

| **M@X$_{18}$ (65)** |
|---|
| Ba@Ag$_{18}$; Ra@Ag$_{18}$; Sr@Ag$_{18}$; Yb@Ag$_{18}$; Ba@Au$_{18}$; Ra@Au$_{18}$; Sr@Au$_{18}$; Yb@Au$_{18}$; Ba@Cs$_{18}$; Sr@Cs$_{18}$; Yb@Fr$_{18}$; Ba@K$_{18}$; Ra@K$_{18}$; Sr@K$_{18}$; Yb@K$_{18}$; Ba@Li$_{18}$; Ra@Li$_{18}$; Sr@Li$_{18}$; Yb@Li$_{18}$; Ba@Na$_{18}$; Ra@Na$_{18}$; Sr@Na$_{18}$; Yb@Na$_{18}$; Ba@Rb$_{18}$; Ra@Rb$_{18}$; Yb@Rb$_{18}$; Ra@Cs$_{18}$; Yb@Cs$_{18}$; Sr@Rb$_{18}$; Ba@Fr$_{18}$; Sr@Fr$_{18}$; Ra@Fr$_{18}$; Ba@As$_{18}$; Ra@As$_{18}$; Sr@As$_{18}$; Ba@Bi$_{18}$; Ra@Bi$_{18}$; Sr@Bi$_{18}$; Yb@Bi$_{18}$; Ba@Sb$_{18}$; Ra@Sb$_{18}$; Sr@Sb$_{18}$; Yb@Sb$_{18}$; Sr@Cu$_{18}$; Ra@Cu$_{18}$; Sr@Ag$_{18}$; Ba@Ag$_{18}$; Ra@Ag$_{18}$; Yb@Ag$_{18}$; Sr@Au$_{18}$; Ba@Au$_{18}$; Ra@Au$_{18}$; Yb@Au$_{18}$; Sr@V$_{18}$; Ba@V$_{18}$; Ra@V$_{18}$; Yb@V$_{18}$; Sr@Nb$_{18}$; Ba@Nb$_{18}$; Ra@Nb$_{18}$; Yb@Nb$_{18}$; Sr@Ta$_{18}$; Ba@Ta$_{18}$; Ra@Ta$_{18}$; Yb@Ta$_{18}$ |
| **M@X$_{12}$Y$_6$ (732)** |
| Ba@Al$_{12}$As$_6$; Ra@Al$_{12}$As$_6$; Sr@Al$_{12}$As$_6$; Yb@Al$_{12}$As$_6$; Ba@Ga$_{12}$As$_6$; Ra@Ga$_{12}$As$_6$; Sr@Ga$_{12}$As$_6$; Yb@Ga$_{12}$As$_6$; Ba@Ga$_{12}$Sb$_6$; Ra@Ga$_{12}$Sb$_6$; Sr@Ga$_{12}$Sb$_6$; Yb@Ga$_{12}$Sb$_6$; Ba@In$_{12}$Bi$_6$; Ra@In$_{12}$Bi$_6$; Sr@In$_{12}$Bi$_6$; Yb@In$_{12}$Bi$_6$; Ba@In$_{12}$Sb$_6$; Ra@In$_{12}$Sb$_6$; Sr@In$_{12}$Sb$_6$; Yb@In$_{12}$Sb$_6$; Ba@Tl$_{12}$Bi$_6$; Ra@Tl$_{12}$Bi$_6$; Sr@Tl$_{12}$Bi$_6$; Yb@Tl$_{12}$Bi$_6$; Ba@Tl$_{12}$Sb$_6$; Ra@Tl$_{12}$Sb$_6$; Sr@Tl$_{12}$Sb$_6$; Yb@Tl$_{12}$Sb$_6$; Ba@Ge$_{12}$Al$_6$; Ra@Ge$_{12}$Al$_6$; Sr@Ge$_{12}$Al$_6$; Yb@Ge$_{12}$Al$_6$; Ba@Ge$_{12}$Ga$_6$; Ra@Ge$_{12}$Ga$_6$; Sr@Ge$_{12}$Ga$_6$; Yb@Ge$_{12}$Ga$_6$; Ba@Pb$_{12}$In$_6$; Ra@Pb$_{12}$In$_6$; Sr@Pb$_{12}$In$_6$; Yb@Pb$_{12}$In$_6$; Ba@Pb$_{12}$Tl$_6$; Ra@Pb$_{12}$Tl$_6$; Sr@Pb$_{12}$Tl$_6$; Yb@Pb$_{12}$Tl$_6$; Ba@Si$_{12}$Al$_6$; Ra@Si$_{12}$Al$_6$; Sr@Si$_{12}$Al$_6$; Yb@Si$_{12}$Al$_6$; Ra@Si$_{12}$Ga$_6$; Sr@Si$_{12}$Ga$_6$; Yb@Si$_{12}$Ga$_6$; Ba@Sn$_{12}$Ga$_6$; Ra@Sn$_{12}$Ga$_6$; Sr@Sn$_{12}$Ga$_6$; Yb@Sn$_{12}$Ga$_6$; Ba@Sn$_{12}$In$_6$; Ra@Sn$_{12}$In$_6$; Sr@Sn$_{12}$In$_6$; Yb@Sn$_{12}$In$_6$; Ba@Sn$_{12}$Tl$_6$; Ra@Sn$_{12}$Tl$_6$; Sr@Sn$_{12}$Tl$_6$; Yb@Sn$_{12}$Tl$_6$; Ba@As$_{12}$Au$_6$; Ra@As$_{12}$Au$_6$; Sr@As$_{12}$Au$_6$; Yb@As$_{12}$Au$_6$; Ba@As$_{12}$Cu$_6$; Ra@As$_{12}$Cu$_6$; Sr@As$_{12}$Cu$_6$; Yb@As$_{12}$Cu$_6$; Ba@As$_{12}$Li$_6$; Ra@As$_{12}$Li$_6$; Sr@As$_{12}$Li$_6$; Yb@As$_{12}$Li$_6$; Ba@Bi$_{12}$Ag$_6$; Ra@Bi$_{12}$Ag$_6$; Sr@Bi$_{12}$Ag$_6$; Yb@Bi$_{12}$Ag$_6$; Ba@Bi$_{12}$Au$_6$; Ra@Bi$_{12}$Au$_6$; Sr@Bi$_{12}$Au$_6$; Yb@Bi$_{12}$Au$_6$; Ba@Bi$_{12}$Na$_6$; Ra@Bi$_{12}$Na$_6$; Sr@Bi$_{12}$Na$_6$; Yb@Bi$_{12}$Na$_6$; Ba@Sb$_{12}$Ag$_6$; Ra@Sb$_{12}$Ag$_6$; Sr@Sb$_{12}$Ag$_6$; Yb@Sb$_{12}$Ag$_6$; Ba@Sb$_{12}$Au$_6$; Ra@Sb$_{12}$Au$_6$; Sr@Sb$_{12}$Au$_6$; Yb@Sb$_{12}$Au$_6$; Ba@Sb$_{12}$Li$_6$; Ra@Sb$_{12}$Li$_6$; Sr@Sb$_{12}$Li$_6$; Yb@Sb$_{12}$Li$_6$; Ba@Sb$_{12}$Na$_6$; Ra@Sb$_{12}$Na$_6$; Sr@Sb$_{12}$Na$_6$; Yb@Sb$_{12}$Na$_6$; Ba@As$_{12}$Ag$_6$; Ra@As$_{12}$Ag$_6$; Sr@As$_{12}$Ag$_6$; Yb@As$_{12}$Ag$_6$; Ba@Bi$_{12}$Sb$_6$; Ra@Bi$_{12}$Sb$_6$; Sr@Bi$_{12}$Sb$_6$; Yb@Bi$_{12}$Sb$_6$; Ba@Sb$_{12}$Bi$_6$; Ra@Sb$_{12}$Bi$_6$; Sr@Sb$_{12}$Bi$_6$; Yb@Sb$_{12}$Bi$_6$; Ba@Po$_{12}$In$_6$; Ra@Po$_{12}$In$_6$; Sr@Po$_{12}$In$_6$; Yb@Po$_{12}$In$_6$; Ba@Po$_{12}$Tl$_6$; Ra@Po$_{12}$Tl$_6$; Sr@Po$_{12}$Tl$_6$; Yb@Po$_{12}$Tl$_6$; Ba@Te$_{12}$Al$_6$; Ra@Te$_{12}$Al$_6$; Sr@Te$_{12}$Al$_6$; Yb@Te$_{12}$Al$_6$; Ba@Te$_{12}$Ga$_6$; Ra@Te$_{12}$Ga$_6$; Sr@Te$_{12}$Ga$_6$; Yb@Te$_{12}$Ga$_6$; Ba@Te$_{12}$In$_6$; Ra@Te$_{12}$In$_6$; Sr@Te$_{12}$In$_6$; Yb@Te$_{12}$In$_6$; Ba@Te$_{12}$Tl$_6$; Ra@Te$_{12}$Tl$_6$; Sr@Te$_{12}$Tl$_6$; Yb@Te$_{12}$Tl$_6$; Ba@Na$_{12}$Sc$_6$; Ra@Na$_{12}$Sc$_6$; Sr@Na$_{12}$Sc$_6$; Yb@Na$_{12}$Sc$_6$; Ba@Na$_{12}$Y$_6$; Ra@Na$_{12}$Y$_6$; Sr@Na$_{12}$Y$_6$; Yb@Na$_{12}$Y$_6$; Ba@Ag$_{12}$Sc$_6$; Ra@Ag$_{12}$Sc$_6$; Sr@Ag$_{12}$Sc$_6$; Yb@Ag$_{12}$Sc$_6$; Ba@Au$_{12}$Sc$_6$; Ra@Au$_{12}$Sc$_6$; Sr@Au$_{12}$Sc$_6$; Yb@Au$_{12}$Sc$_6$; Ba@Li$_{12}$Co$_6$; Ra@Li$_{12}$Co$_6$; Sr@Li$_{12}$Co$_6$; Ba@Li$_{12}$Rh$_6$; Ra@Li$_{12}$Rh$_6$; Sr@Li$_{12}$Rh$_6$; Ba@Li$_{12}$Ir$_6$; Ra@Li$_{12}$Ir$_6$; Sr@Li$_{12}$Ir$_6$; Yb@Li$_{12}$Ir$_6$; Ba@Cu$_{12}$Co$_6$; Ra@Cu$_{12}$Co$_6$; Sr@Cu$_{12}$Co$_6$; Yb@Cu$_{12}$Co$_6$; Ba@Cu$_{12}$Rh$_6$; Ra@Cu$_{12}$Rh$_6$; Sr@Cu$_{12}$Rh$_6$; Yb@Cu$_{12}$Rh$_6$; Ba@Cu$_{12}$Ir$_6$; Ra@Cu$_{12}$Ir$_6$; Sr@Cu$_{12}$Ir$_6$; Yb@Cu$_{12}$Ir$_6$; Ba@Ag$_{12}$Co$_6$; Ra@Ag$_{12}$Co$_6$; Sr@Ag$_{12}$Co$_6$; Yb@Ag$_{12}$Co$_6$; Ba@Ag$_{12}$Rh$_6$; Ra@Ag$_{12}$Rh$_6$; Sr@Ag$_{12}$Rh$_6$; Yb@Ag$_{12}$Rh$_6$; Ba@Ag$_{12}$Ir$_6$; Ra@Ag$_{12}$Ir$_6$; Sr@Ag$_{12}$Ir$_6$; Yb@Ag$_{12}$Ir$_6$; Ba@Au$_{12}$Co$_6$; Ra@Au$_{12}$Co$_6$; Sr@Au$_{12}$Co$_6$; Yb@Au$_{12}$Co$_6$; Ba@Au$_{12}$Rh$_6$; Ra@Au$_{12}$Rh$_6$; Sr@Au$_{12}$Rh$_6$; Yb@Au$_{12}$Rh$_6$; Ba@Au$_{12}$Ir$_6$; Ra@Au$_{12}$Ir$_6$; Sr@Au$_{12}$Ir$_6$; Yb@Au$_{12}$Ir$_6$; Ba@Mg$_{12}$Cu$_6$; Ra@Mg$_{12}$Cu$_6$; Sr@Mg$_{12}$Cu$_6$; Yb@Mg$_{12}$Cu$_6$; Ba@Mg$_{12}$Ag$_6$; Ra@Mg$_{12}$Ag$_6$; Sr@Mg$_{12}$Ag$_6$; Yb@Mg$_{12}$Ag$_6$; Ba@Mg$_{12}$Au$_6$; Ra@Mg$_{12}$Au$_6$; Sr@Mg$_{12}$Au$_6$; Yb@Mg$_{12}$Au$_6$; Ba@Zn$_{12}$Li$_6$; Ra@Zn$_{12}$Li$_6$; Sr@Zn$_{12}$Li$_6$; Yb@Zn$_{12}$Li$_6$; Ba@Zn$_{12}$Ag$_6$; Ra@Zn$_{12}$Ag$_6$; Sr@Zn$_{12}$Ag$_6$; Yb@Zn$_{12}$Ag$_6$; Ba@Zn$_{12}$Au$_6$; Ra@Zn$_{12}$Au$_6$; Sr@Zn$_{12}$Au$_6$; Yb@Zn$_{12}$Au$_6$; Ba@Cd$_{12}$Na$_6$; Ra@Cd$_{12}$Na$_6$; Sr@Cd$_{12}$Na$_6$; Yb@Cd$_{12}$Na$_6$; Ba@Cd$_{12}$Ag$_6$; Ra@Cd$_{12}$Ag$_6$; Sr@Cd$_{12}$Ag$_6$; Yb@Cd$_{12}$Ag$_6$; Ba@Cd$_{12}$Au$_6$; Ra@Cd$_{12}$Au$_6$; Sr@Cd$_{12}$Au$_6$; Yb@Cd$_{12}$Au$_6$; Ba@Hg$_{12}$Na$_6$; Ra@Hg$_{12}$Na$_6$; Sr@Hg$_{12}$Na$_6$; Yb@Hg$_{12}$Na$_6$; Ba@Hg$_{12}$Ag$_6$; Ra@Hg$_{12}$Ag$_6$; Sr@Hg$_{12}$Ag$_6$; Yb@Hg$_{12}$Ag$_6$; Ba@Hg$_{12}$Au$_6$; Ra@Hg$_{12}$Au$_6$; Sr@Hg$_{12}$Au$_6$; Yb@Hg$_{12}$Au$_6$; Ba@Mg$_{12}$Mn$_6$; Ra@Mg$_{12}$Mn$_6$; Sr@Mg$_{12}$Mn$_6$; Yb@Mg$_{12}$Mn$_6$; Ba@Mg$_{12}$Tc$_6$; Ra@Mg$_{12}$Tc$_6$; Sr@Mg$_{12}$Tc$_6$; Yb@Mg$_{12}$Tc$_6$; Ba@Mg$_{12}$Re$_6$; Ra@Mg$_{12}$Re$_6$; Sr@Mg$_{12}$Re$_6$; Yb@Mg$_{12}$Re$_6$; Ba@Zn$_{12}$Mn$_6$; Ra@Zn$_{12}$Mn$_6$; Sr@Zn$_{12}$Mn$_6$; Yb@Zn$_{12}$Mn$_6$; Ba@Zn$_{12}$Tc$_6$; Ra@Zn$_{12}$Tc$_6$; Sr@Zn$_{12}$Tc$_6$; Yb@Zn$_{12}$Tc$_6$; Ba@Zn$_{12}$Re$_6$; Ra@Zn$_{12}$Re$_6$; Sr@Zn$_{12}$Re$_6$; Yb@Zn$_{12}$Re$_6$; Ba@Hg$_{12}$Tc$_6$; Ra@Hg$_{12}$Tc$_6$; Sr@Hg$_{12}$Tc$_6$; Yb@Hg$_{12}$Tc$_6$; Ba@Hg$_{12}$Re$_6$; Ra@Hg$_{12}$Re$_6$; Sr@Hg$_{12}$Re$_6$; Yb@Hg$_{12}$Re$_6$; Ba@Sc$_{12}$Nb$_6$; Ra@Sc$_{12}$Nb$_6$; Sr@Sc$_{12}$Nb$_6$; Yb@Sc$_{12}$Nb$_6$; Ba@Sc$_{12}$Ta$_6$; Ra@Sc$_{12}$Ta$_6$; Sr@Sc$_{12}$Ta$_6$; Yb@Sc$_{12}$Ta$_6$; Ba@Sc$_{12}$Sb$_6$; Ra@Sc$_{12}$Sb$_6$; Sr@Sc$_{12}$Sb$_6$; Yb@Sc$_{12}$Sb$_6$; Ba@Sc$_{12}$Bi$_6$; Ra@Sc$_{12}$Bi$_6$; Sr@Sc$_{12}$Bi$_6$; Yb@Sc$_{12}$Bi$_6$; Ba@Y$_{12}$Bi$_6$; Ra@Y$_{12}$Bi$_6$; Sr@Y$_{12}$Bi$_6$; Yb@Y$_{12}$Bi$_6$; Ba@Al$_{12}$V$_6$; Ra@Al$_{12}$V$_6$; Sr@Al$_{12}$V$_6$; Yb@Al$_{12}$V$_6$; Ba@Al$_{12}$Nb$_6$; Ra@Al$_{12}$Nb$_6$; Sr@Al$_{12}$Nb$_6$; Yb@Al$_{12}$Nb$_6$; Ba@Al$_{12}$Ta$_6$; Ra@Al$_{12}$Ta$_6$; Sr@Al$_{12}$Ta$_6$; Yb@Al$_{12}$Ta$_6$; Ba@Ga$_{12}$V$_6$; Ra@Ga$_{12}$V$_6$; Sr@Ga$_{12}$V$_6$; Yb@Ga$_{12}$V$_6$; Ba@Ga$_{12}$Nb$_6$; Ra@Ga$_{12}$Nb$_6$; Sr@Ga$_{12}$Nb$_6$; Yb@Ga$_{12}$Nb$_6$; Ba@Ga$_{12}$Ta$_6$; Ra@Ga$_{12}$Ta$_6$; Sr@Ga$_{12}$Ta$_6$; Yb@Ga$_{12}$Ta$_6$; Ba@In$_{12}$Nb$_6$; Ra@In$_{12}$Nb$_6$; Sr@In$_{12}$Nb$_6$; Yb@In$_{12}$Nb$_6$; Ba@In$_{12}$Ta$_6$; Ra@In$_{12}$Ta$_6$; Sr@In$_{12}$Ta$_6$; Yb@In$_{12}$Ta$_6$; Ba@Tl$_{12}$Nb$_6$; Ra@Tl$_{12}$Nb$_6$; Sr@Tl$_{12}$Nb$_6$; Yb@Tl$_{12}$Nb$_6$; Ba@Tl$_{12}$Ta$_6$; Ra@Tl$_{12}$Ta$_6$; Sr@Tl$_{12}$Ta$_6$; Yb@Tl$_{12}$Ta$_6$; Ba@Sc$_{12}$Rh$_6$; Ra@Sc$_{12}$Rh$_6$; Sr@Sc$_{12}$Rh$_6$; Yb@Sc$_{12}$Rh$_6$; Ba@Sc$_{12}$Ir$_6$; Ra@Sc$_{12}$Ir$_6$; Sr@Sc$_{12}$Ir$_6$; |

Yb@$Sc_{12}Ir_6$; Ba@$Al_{12}Co_6$; Ra@$Al_{12}Co_6$; Sr@$Al_{12}Co_6$; Yb@$Al_{12}Co_6$; Ba@$Al_{12}Rh_6$; Ra@$Al_{12}Rh_6$; Sr@$Al_{12}Rh_6$; Yb@$Al_{12}Rh_6$; Ba@$Al_{12}Ir_6$; Ra@$Al_{12}Ir_6$; Sr@$Al_{12}Ir_6$; Yb@$Al_{12}Ir_6$; Ba@$Ga_{12}Co_6$; Ra@$Ga_{12}Co_6$; Sr@$Ga_{12}Co_6$; Yb@$Ga_{12}Co_6$; Ba@$Ga_{12}Rh_6$; Ra@$Ga_{12}Rh_6$; Sr@$Ga_{12}Rh_6$; Yb@$Ga_{12}Rh_6$; Ba@$Ga_{12}Ir_6$; Ra@$Ga_{12}Ir_6$; Sr@$Ga_{12}Ir_6$; Yb@$Ga_{12}Ir_6$; Ba@$In_{12}Rh_6$; Ra@$In_{12}Rh_6$; Sr@$In_{12}Rh_6$; Yb@$In_{12}Rh_6$; Ba@$In_{12}Ir_6$; Ra@$In_{12}Ir_6$; Sr@$In_{12}Ir_6$; Yb@$In_{12}Ir_6$; Ba@$Ti_{12}Sc_6$; Ra@$Ti_{12}Sc_6$; Sr@$Ti_{12}Sc_6$; Yb@$Ti_{12}Sc_6$; Ba@$Ti_{12}Al_6$; Ra@$Ti_{12}Al_6$; Sr@$Ti_{12}Al_6$; Yb@$Ti_{12}Al_6$; Ba@$Ti_{12}Ga_6$; Ra@$Ti_{12}Ga_6$; Sr@$Ti_{12}Ga_6$; Yb@$Ti_{12}Ga_6$; Ba@$Ti_{12}In_6$; Ra@$Ti_{12}In_6$; Sr@$Ti_{12}In_6$; Yb@$Ti_{12}In_6$; Ba@$Ti_{12}Tl_6$; Ra@$Ti_{12}Tl_6$; Sr@$Ti_{12}Tl_6$; Yb@$Ti_{12}Tl_6$; Ba@$Zr_{12}Sc_6$; Ra@$Zr_{12}Sc_6$; Sr@$Zr_{12}Sc_6$; Yb@$Zr_{12}Sc_6$; Ba@$Zr_{12}Y_6$; Ra@$Zr_{12}Y_6$; Sr@$Zr_{12}Y_6$; Yb@$Zr_{12}Y_6$; Ba@$Zr_{12}Ga_6$; Ra@$Zr_{12}Ga_6$; Sr@$Zr_{12}Ga_6$; Yb@$Zr_{12}Ga_6$; Ba@$Zr_{12}In_6$; Ra@$Zr_{12}In_6$; Sr@$Zr_{12}In_6$; Yb@$Zr_{12}In_6$; Ba@$Zr_{12}Tl_6$; Ra@$Zr_{12}Tl_6$; Sr@$Zr_{12}Tl_6$; Yb@$Zr_{12}Tl_6$; Ba@$Hf_{12}Sc_6$; Ra@$Hf_{12}Sc_6$; Sr@$Hf_{12}Sc_6$; Yb@$Hf_{12}Sc_6$; Ba@$Hf_{12}Y_6$; Ra@$Hf_{12}Y_6$; Sr@$Hf_{12}Y_6$; Yb@$Hf_{12}Y_6$; Ba@$Hf_{12}Ga_6$; Ra@$Hf_{12}Ga_6$; Sr@$Hf_{12}Ga_6$; Yb@$Hf_{12}Ga_6$; Ba@$Hf_{12}In_6$; Ra@$Hf_{12}In_6$; Sr@$Hf_{12}In_6$; Yb@$Hf_{12}In_6$; Ba@$Hf_{12}Tl_6$; Ra@$Hf_{12}Tl_6$; Sr@$Hf_{12}Tl_6$; Yb@$Hf_{12}Tl_6$; Ba@$Sn_{12}Sc_6$; Ra@$Sn_{12}Sc_6$; Sr@$Sn_{12}Sc_6$; Yb@$Sn_{12}Sc_6$; Ba@$Pb_{12}Sc_6$; Ra@$Pb_{12}Sc_6$; Sr@$Pb_{12}Sc_6$; Yb@$Pb_{12}Sc_6$; Ba@$Pb_{12}Y_6$; Ra@$Pb_{12}Y_6$; Sr@$Pb_{12}Y_6$; Yb@$Pb_{12}Y_6$; Ba@$Ti_{12}Mn_6$; Ra@$Ti_{12}Mn_6$; Sr@$Ti_{12}Mn_6$; Yb@$Ti_{12}Mn_6$; Ba@$Ti_{12}Tc_6$; Ra@$Ti_{12}Tc_6$; Sr@$Ti_{12}Tc_6$; Yb@$Ti_{12}Tc_6$; Ba@$Ti_{12}Re_6$; Ra@$Ti_{12}Re_6$; Sr@$Ti_{12}Re_6$; Yb@$Ti_{12}Re_6$; Ba@$Zr_{12}Tc_6$; Ra@$Zr_{12}Tc_6$; Sr@$Zr_{12}Tc_6$; Yb@$Zr_{12}Tc_6$; Ba@$Zr_{12}Re_6$; Ra@$Zr_{12}Re_6$; Sr@$Zr_{12}Re_6$; Yb@$Zr_{12}Re_6$; Ba@$Hf_{12}Tc_6$; Ra@$Hf_{12}Tc_6$; Sr@$Hf_{12}Tc_6$; Yb@$Hf_{12}Tc_6$; Ba@$Hf_{12}Re_6$; Ra@$Hf_{12}Re_6$; Sr@$Hf_{12}Re_6$; Yb@$Hf_{12}Re_6$; Ba@$Si_{12}Mn_6$; Ra@$Si_{12}Mn_6$; Sr@$Si_{12}Mn_6$; Yb@$Si_{12}Mn_6$; Ba@$Si_{12}Tc_6$; Ra@$Si_{12}Tc_6$; Sr@$Si_{12}Tc_6$; Yb@$Si_{12}Tc_6$; Ba@$Si_{12}Re_6$; Ra@$Si_{12}Re_6$; Sr@$Si_{12}Re_6$; Yb@$Si_{12}Re_6$; Ba@$Ge_{12}Mn_6$; Ra@$Ge_{12}Mn_6$; Sr@$Ge_{12}Mn_6$; Yb@$Ge_{12}Mn_6$; Ba@$Ge_{12}Tc_6$; Ra@$Ge_{12}Tc_6$; Sr@$Ge_{12}Tc_6$; Yb@$Ge_{12}Tc_6$; Ba@$Ge_{12}Re_6$; Ra@$Ge_{12}Re_6$; Sr@$Ge_{12}Re_6$; Yb@$Ge_{12}Re_6$; Ba@$Sn_{12}Tc_6$; Ra@$Sn_{12}Tc_6$; Sr@$Sn_{12}Tc_6$; Yb@$Sn_{12}Tc_6$; Ba@$Sn_{12}Re_6$; Ra@$Sn_{12}Re_6$; Sr@$Sn_{12}Re_6$; Yb@$Sn_{12}Re_6$; Ba@$Pb_{12}Re_6$; Ra@$Pb_{12}Re_6$; Sr@$Pb_{12}Re_6$; Yb@$Pb_{12}Re_6$; Ba@$V_{12}Li_6$; Sr@$V_{12}Li_6$; Yb@$V_{12}Li_6$; Ra@$V_{12}Cu_6$; Sr@$V_{12}Cu_6$; Yb@$V_{12}Cu_6$; Ba@$V_{12}Ag_6$; Yb@$V_{12}Ag_6$; Ba@$V_{12}Au_6$; Sr@$V_{12}Au_6$; Yb@$V_{12}Au_6$; Ba@$Nb_{12}Li_6$; Ra@$Nb_{12}Li_6$; Sr@$Nb_{12}Li_6$; Yb@$Nb_{12}Li_6$; Ba@$Nb_{12}Cu_6$; Ra@$Nb_{12}Cu_6$; Sr@$Nb_{12}Cu_6$; Yb@$Nb_{12}Cu_6$; Ba@$Nb_{12}Ag_6$; Ra@$Nb_{12}Ag_6$; Sr@$Nb_{12}Ag_6$; Yb@$Nb_{12}Ag_6$; Ba@$Nb_{12}Au_6$; Ra@$Nb_{12}Au_6$; Sr@$Nb_{12}Au_6$; Yb@$Nb_{12}Au_6$; Ba@$Ta_{12}Li_6$; Ra@$Ta_{12}Li_6$; Sr@$Ta_{12}Li_6$; Yb@$Ta_{12}Li_6$; Ba@$Ta_{12}Cu_6$; Ra@$Ta_{12}Cu_6$; Sr@$Ta_{12}Cu_6$; Yb@$Ta_{12}Cu_6$; Ba@$Ta_{12}Ag_6$; Ra@$Ta_{12}Ag_6$; Sr@$Ta_{12}Ag_6$; Yb@$Ta_{12}Ag_6$; Ba@$Ta_{12}Au_6$; Ra@$Ta_{12}Au_6$; Sr@$Ta_{12}Au_6$; Yb@$Ta_{12}Au_6$; Ba@$V_{12}Nb_6$; Ra@$V_{12}Nb_6$; Sr@$V_{12}Nb_6$; Yb@$V_{12}Nb_6$; Ba@$V_{12}Ta_6$; Ra@$V_{12}Ta_6$; Sr@$V_{12}Ta_6$; Yb@$V_{12}Ta_6$; Ba@$V_{12}As_6$; Ra@$V_{12}As_6$; Sr@$V_{12}As_6$; Yb@$V_{12}As_6$; Ba@$V_{12}Sb_6$; Ra@$V_{12}Sb_6$; Yb@$V_{12}Sb_6$; Ba@$Nb_{12}V_6$; Ra@$Nb_{12}V_6$; Sr@$Nb_{12}V_6$; Yb@$Nb_{12}V_6$; Ba@$Nb_{12}Ta_6$; Ra@$Nb_{12}Ta_6$; Sr@$Nb_{12}Ta_6$; Yb@$Nb_{12}Ta_6$; Ba@$Nb_{12}As_6$; Ra@$Nb_{12}As_6$; Sr@$Nb_{12}As_6$; Yb@$Nb_{12}As_6$; Ba@$Nb_{12}Sb_6$; Ra@$Nb_{12}Sb_6$; Sr@$Nb_{12}Sb_6$; Yb@$Nb_{12}Sb_6$; Ba@$Nb_{12}Bi_6$; Ra@$Nb_{12}Bi_6$; Sr@$Nb_{12}Bi_6$; Yb@$Nb_{12}Bi_6$; Ba@$Ta_{12}V_6$; Ra@$Ta_{12}V_6$; Sr@$Ta_{12}V_6$; Yb@$Ta_{12}V_6$; Ba@$Ta_{12}Nb_6$; Ra@$Ta_{12}Nb_6$; Sr@$Ta_{12}Nb_6$; Yb@$Ta_{12}Nb_6$; Ba@$Ta_{12}As_6$; Ra@$Ta_{12}As_6$; Sr@$Ta_{12}As_6$; Yb@$Ta_{12}As_6$; Ba@$Ta_{12}Sb_6$; Ra@$Ta_{12}Sb_6$; Sr@$Ta_{12}Sb_6$; Yb@$Ta_{12}Sb_6$; Ba@$Ta_{12}Bi_6$; Ra@$Ta_{12}Bi_6$; Sr@$Ta_{12}Bi_6$; Yb@$Ta_{12}Bi_6$; Ba@$As_{12}V_6$; Ra@$As_{12}V_6$; Sr@$As_{12}V_6$; Yb@$As_{12}V_6$; Ba@$As_{12}Nb_6$; Ra@$As_{12}Nb_6$; Sr@$As_{12}Nb_6$; Yb@$As_{12}Nb_6$; Ba@$As_{12}Ta_6$; Ra@$As_{12}Ta_6$; Sr@$As_{12}Ta_6$; Yb@$As_{12}Ta_6$; Ba@$Sb_{12}V_6$; Ra@$Sb_{12}V_6$; Sr@$Sb_{12}V_6$; Yb@$Sb_{12}V_6$; Ba@$Sb_{12}Nb_6$; Ra@$Sb_{12}Nb_6$; Sr@$Sb_{12}Nb_6$; Yb@$Sb_{12}Nb_6$; Ba@$Sb_{12}Ta_6$; Ra@$Sb_{12}Ta_6$; Sr@$Sb_{12}Ta_6$; Yb@$Sb_{12}Ta_6$; Ba@$Bi_{12}Nb_6$; Ra@$Bi_{12}Nb_6$; Sr@$Bi_{12}Nb_6$; Yb@$Bi_{12}Nb_6$; Ba@$Bi_{12}Ta_6$; Ra@$Bi_{12}Ta_6$; Sr@$Bi_{12}Ta_6$; Yb@$Bi_{12}Ta_6$; Ba@$Cr_{12}Al_6$; Ra@$Cr_{12}Al_6$; Sr@$Cr_{12}Al_6$; Yb@$Cr_{12}Al_6$; Ba@$Cr_{12}Ga_6$; Ra@$Cr_{12}Ga_6$; Sr@$Cr_{12}Ga_6$; Yb@$Cr_{12}Ga_6$; Ba@$Mo_{12}Sc_6$; Ra@$Mo_{12}Sc_6$; Sr@$Mo_{12}Sc_6$; Yb@$Mo_{12}Sc_6$; Ra@$Mo_{12}Al_6$; Ra@$Mo_{12}Ga_6$; Yb@$Mo_{12}In_6$; Ra@$Mo_{12}Tl_6$; Ba@$W_{12}Sc_6$; Ra@$W_{12}Sc_6$; Sr@$W_{12}Sc_6$; Yb@$W_{12}Sc_6$; Ba@$W_{12}Al_6$; Ra@$W_{12}Al_6$; Sr@$W_{12}Al_6$; Yb@$W_{12}Al_6$; Ba@$W_{12}Ga_6$; Ra@$W_{12}Ga_6$; Sr@$W_{12}Ga_6$; Yb@$W_{12}Ga_6$; Ba@$W_{12}In_6$; Ra@$W_{12}In_6$; Sr@$W_{12}In_6$; Yb@$W_{12}In_6$; Ra@$W_{12}Tl_6$; Yb@$W_{12}Tl_6$; Ba@$Te_{12}Sc_6$; Ra@$Te_{12}Sc_6$; Sr@$Te_{12}Sc_6$; Yb@$Te_{12}Sc_6$; Ba@$Po_{12}Sc_6$; Ra@$Po_{12}Sc_6$; Sr@$Po_{12}Sc_6$; Yb@$Po_{12}Sc_6$; Ba@$Po_{12}Y_6$; Ra@$Po_{12}Y_6$; Sr@$Po_{12}Y_6$; Yb@$Po_{12}Y_6$; Ba@$Mn_{12}Li_6$; Ra@$Mn_{12}Li_6$; Yb@$Mn_{12}Li_6$; Ba@$Mn_{12}Cu_6$; Ra@$Mn_{12}Cu_6$; Sr@$Mn_{12}Cu_6$; Yb@$Mn_{12}Cu_6$; Ba@$Mn_{12}Ag_6$; Ra@$Mn_{12}Ag_6$; Sr@$Mn_{12}Ag_6$; Ba@$Mn_{12}Au_6$; Yb@$Mn_{12}Au_6$; Sr@$Tc_{12}Li_6$; Yb@$Tc_{12}Li_6$; Ba@$Tc_{12}Cu_6$; Ra@$Tc_{12}Cu_6$; Sr@$Tc_{12}Cu_6$; Yb@$Tc_{12}Cu_6$; Ra@$Tc_{12}Ag_6$; Yb@$Tc_{12}Ag_6$; Ba@$Tc_{12}Au_6$; Ra@$Tc_{12}Au_6$; Sr@$Tc_{12}Au_6$; Yb@$Tc_{12}Au_6$; Ba@$Re_{12}Li_6$; Sr@$Re_{12}Li_6$; Yb@$Re_{12}Li_6$; Ba@$Re_{12}Cu_6$; Ra@$Re_{12}Cu_6$; Sr@$Re_{12}Cu_6$; Yb@$Re_{12}Cu_6$; Ba@$Re_{12}Ag_6$; Ra@$Re_{12}Ag_6$; Sr@$Re_{12}Ag_6$; Yb@$Re_{12}Ag_6$; Ba@$Re_{12}Au_6$; Ra@$Re_{12}Au_6$; Sr@$Re_{12}Au_6$; Yb@$Re_{12}Au_6$;

Table S22. The molecular formula of the stable structure of A$_3$ when the co-dual structure is embedded.

| |
|---|
| 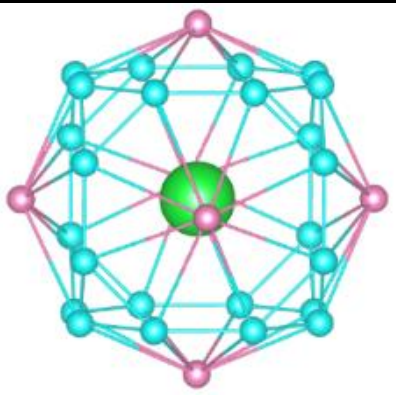 A$_3$ Co-dual structure |
| **M@X$_{30}$ (22)** |
| Ba@Al$_{30}$; Ra@Al$_{30}$; Sr@Al$_{30}$; Yb@Al$_{30}$; Ra@B$_{30}$; Sr@B$_{30}$; Yb@B$_{30}$; Ba@Ga$_{30}$; Ra@Ga$_{30}$; Sr@Ga$_{30}$; Yb@Ga$_{30}$; Ba@In$_{30}$; Ra@In$_{30}$; Sr@In$_{30}$; Yb@In$_{30}$; Ba@Tl$_{30}$; Ra@Tl$_{30}$; Sr@Tl$_{30}$; Yb@Tl$_{30}$; Ba@Y$_{30}$; Ra@Y$_{30}$; Yb@Y$_{30}$; |
| **M@X$_{24}$Y$_6$ (79)** |
| Ba@Be$_{24}$B$_6$. Ra@Be$_{24}$B$_6$; Sr@Be$_{24}$B$_6$; Yb@Be$_{24}$B$_6$; Ba@Cd$_{24}$Tl$_6$; Ra@Cd$_{24}$Tl$_6$; Sr@Cd$_{24}$Tl$_6$; Yb@Cd$_{24}$Tl$_6$; Ra@Hg$_{24}$Tl$_6$; Sr@Hg$_{24}$Tl$_6$; Yb@Hg$_{24}$Tl$_6$; Ba@Mg$_{24}$Al$_6$; Ra@Mg$_{24}$Al$_6$; Sr@Mg$_{24}$Al$_6$; Yb@Mg$_{24}$Al$_6$; Ba@Hg$_{24}$In$_6$; Ba@Mg$_{24}$Ga$_6$; Ra@Mg$_{24}$Ga$_6$; Sr@Mg$_{24}$Ga$_6$; Yb@Mg$_{24}$Ga$_6$; Ba@Mg$_{24}$In$_6$; Ra@Mg$_{24}$In$_6$; Sr@Mg$_{24}$In$_6$; Yb@Mg$_{24}$In$_6$; Ba@Mg$_{24}$Tl$_6$; Ra@Mg$_{24}$Tl$_6$; Sr@Mg$_{24}$Tl$_6$; Yb@Mg$_{24}$Tl$_6$; Ba@Zn$_{24}$Al$_6$; Ra@Zn$_{24}$Al$_6$; Sr@Zn$_{24}$Al$_6$; Yb@Zn$_{24}$Al$_6$; Ba@Zn$_{24}$Ga$_6$; Ra@Zn$_{24}$Ga$_6$; Sr@Zn$_{24}$Ga$_6$; Yb@Zn$_{24}$Ga$_6$; Ba@Zn$_{24}$In$_6$; Ra@Zn$_{24}$In$_6$; Sr@Zn$_{24}$In$_6$; Yb@Zn$_{24}$In$_6$; Ba@Al$_{24}$Ga$_6$; Ra@Al$_{24}$Ga$_6$; Sr@Al$_{24}$Ga$_6$; Yb@Al$_{24}$Ga$_6$; Ba@Ga$_{24}$Al$_6$; Ra@Ga$_{24}$Al$_6$; Sr@Ga$_{24}$Al$_6$; Yb@Ga$_{24}$Al$_6$; Ba@Ga$_{24}$In$_6$; Ra@Ga$_{24}$In$_6$; Sr@Ga$_{24}$In$_6$; Yb@Ga$_{24}$In$_6$; Ba@In$_{24}$Ga$_6$; Ra@In$_{24}$Ga$_6$; Sr@In$_{24}$Ga$_6$; Yb@In$_{24}$Ga$_6$; Ba@In$_{24}$Tl$_6$; Ra@In$_{24}$Tl$_6$; Sr@In$_{24}$Tl$_6$; Yb@In$_{24}$Tl$_6$; Ba@Tl$_{24}$In$_6$; Ra@Tl$_{24}$In$_6$; Sr@Tl$_{24}$In$_6$; Yb@Tl$_{24}$In$_6$; Sr@Zn$_{24}$Mn$_6$; Yb@Zn$_{24}$Re$_6$; Ba@Hg$_{24}$Re$_6$; Ra@Hg$_{24}$Re$_6$; Sr@Hg$_{24}$Re$_6$; Ba@Sc$_{24}$Y$_6$; Ra@Sc$_{24}$Y$_6$; Sr@Sc$_{24}$Y$_6$; Yb@Sc$_{24}$Y$_6$; Yb@Sc$_{24}$In$_6$; Yb@Sc$_{24}$Tl$_6$; Yb@Y$_{24}$Tl$_6$; Ba@In$_{24}$Y$_6$; Ra@In$_{24}$Y$_6$; Yb@In$_{24}$Y$_6$; |

Table S23. The molecular formula of the stable structure of A$_4$ when the co-dual structure is embedded.

| |
|---|
| 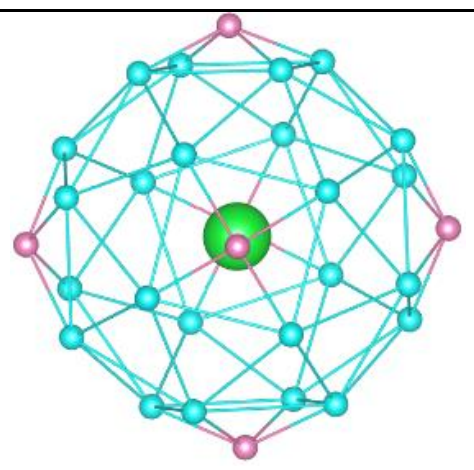 A$_4$ Co-dual structure |
| **M@X$_{30}$ (19)** |
| Ba@Al$_{30}$; Ra@Al$_{30}$; Sr@Al$_{30}$; Yb@Al$_{30}$; Ba@Ga$_{30}$; Ra@Ga$_{30}$; Sr@Ga$_{30}$; Yb@Ga$_{30}$; Ba@In$_{30}$; Ra@In$_{30}$; Sr@In$_{30}$; Yb@In$_{30}$; Sr@Sc$_{30}$; Ba@Sc$_{30}$; Ra@Sc$_{30}$; Yb@Sc$_{30}$; Sr@Y$_{30}$; Ba@Y$_{30}$; Ra@Y$_{30}$; |
| **M@X$_{24}$Y$_6$ (106)** |
| Ba@Be$_{24}$B$_6$; Ra@Be$_{24}$B$_6$; Sr@Be$_{24}$B$_6$; Yb@Be$_{24}$B$_6$; Ba@Mg$_{24}$Al$_6$; Ra@Mg$_{24}$Al$_6$; Sr@Mg$_{24}$Al$_6$; Yb@Mg$_{24}$Al$_6$; Ba@Mg$_{24}$Ga$_6$; Ra@Mg$_{24}$Ga$_6$; Sr@Mg$_{24}$Ga$_6$; Yb@Mg$_{24}$Ga$_6$; Ra@Zn$_{24}$Al$_6$; Ba@Ga$_{24}$Al$_6$; Ra@Ga$_{24}$Al$_6$; Sr@Ga$_{24}$Al$_6$; Yb@Ga$_{24}$Al$_6$; Ba@In$_{24}$Ga$_6$; Ra@In$_{24}$Ga$_6$; Sr@In$_{24}$Ga$_6$; Yb@In$_{24}$Ga$_6$; Ba@Tl$_{24}$In$_6$; Ra@Tl$_{24}$In$_6$; Sr@Tl$_{24}$In$_6$; Yb@Tl$_{24}$In$_6$; Ra@Sc$_{24}$Ga$_6$; Sr@Sc$_{24}$Ga$_6$; Ra@Y$_{24}$Sc$_6$; Ra@Y$_{24}$Tl$_6$; Ba@Ga$_{24}$Sc$_6$; Ra@Ga$_{24}$Sc$_6$; Sr@Ga$_{24}$Sc$_6$; Ba@In$_{24}$Sc$_6$; Ra@In$_{24}$Sc$_6$; Ba@Li$_{24}$Mn$_6$; Sr@Li$_{24}$Mn$_6$; Yb@Li$_{24}$Mn$_6$; Sr@Li$_{24}$Tc$_6$; Ba@Cu$_{24}$Mn$_6$; Ra@Cu$_{24}$Mn$_6$; Sr@Cu$_{24}$Mn$_6$; Yb@Cu$_{24}$Mn$_6$; Ba@Cu$_{24}$Tc$_6$; Ra@Cu$_{24}$Tc$_6$; Sr@Cu$_{24}$Tc$_6$; Yb@Cu$_{24}$Tc$_6$; Ba@Cu$_{24}$Re$_6$; Ra@Cu$_{24}$Re$_6$; Sr@Cu$_{24}$Re$_6$; Ba@Ag$_{24}$Mn$_6$; Ra@Ag$_{24}$Mn$_6$; Sr@Ag$_{24}$Mn$_6$; Yb@Ag$_{24}$Mn$_6$; Ba@Ag$_{24}$Tc$_6$; Ra@Ag$_{24}$Tc$_6$; Sr@Ag$_{24}$Tc$_6$; Yb@Ag$_{24}$Tc$_6$; Ba@Ag$_{24}$Re$_6$; Ra@Ag$_{24}$Re$_6$; Sr@Ag$_{24}$Re$_6$; Yb@Ag$_{24}$Re$_6$; Ba@Au$_{24}$Mn$_6$; Ra@Au$_{24}$Mn$_6$; Sr@Au$_{24}$Mn$_6$; Yb@Au$_{24}$Mn$_6$; Ba@Au$_{24}$Tc$_6$; Ra@Au$_{24}$Tc$_6$; Sr@Au$_{24}$Tc$_6$; Yb@Au$_{24}$Tc$_6$; Ba@Au$_{24}$Re$_6$; Ra@Au$_{24}$Re$_6$; Sr@Au$_{24}$Re$_6$; Yb@Au$_{24}$Re$_6$; Ba@Mg$_{24}$Sc$_6$; Ra@Mg$_{24}$Sc$_6$; Yb@Mg$_{24}$Sc$_6$; Ba@Ca$_{24}$Y$_6$; Ba@Cd$_{24}$Sc$_6$; Ra@Cd$_{24}$Sc$_6$; Sr@Cd$_{24}$Sc$_6$; Yb@Cd$_{24}$Sc$_6$; Ra@Cd$_{24}$Y$_6$; Sr@Cd$_{24}$Y$_6$; Yb@Cd$_{24}$Y$_6$; Ba@Hg$_{24}$Sc$_6$; Ra@Hg$_{24}$Sc$_6$; Sr@Hg$_{24}$Sc$_6$; Yb@Hg$_{24}$Sc$_6$; Ba@Mg$_{24}$Mn$_6$; Ra@Mg$_{24}$Mn$_6$; Sr@Mg$_{24}$Mn$_6$; Ba@Mg$_{24}$Tc$_6$; Ra@Mg$_{24}$Tc$_6$; Sr@Mg$_{24}$Tc$_6$; Yb@Mg$_{24}$Tc$_6$; Ra@Mg$_{24}$Re$_6$; Sr@Mg$_{24}$Re$_6$; Ba@Zn$_{24}$Mn$_6$; Ra@Zn$_{24}$Mn$_6$; Sr@Zn$_{24}$Mn$_6$; Yb@Zn$_{24}$Mn$_6$; Ba@Zn$_{24}$Re$_6$; Sr@Zn$_{24}$Re$_6$; Ba@Hg$_{24}$Tc$_6$; Ra@Hg$_{24}$Tc$_6$; Yb@Hg$_{24}$Re$_6$ |

Table S24. The molecular formula of the stable structure of $A_5$ when the co-dual structure is embedded.

| |
|---|
| 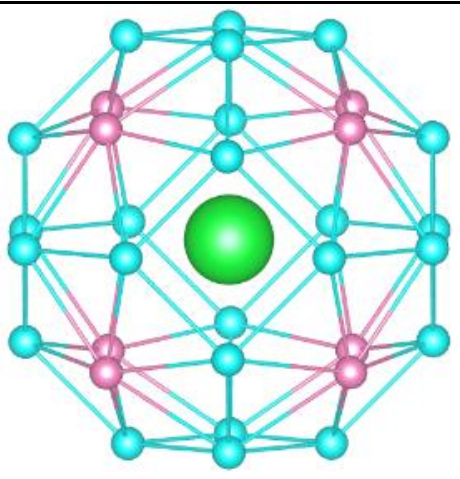<br>$A_5$ Co-dual structure |
| **$M@X_{32}$ (28)** |
| $Ba@Ag_{32}$; $Ra@Ag_{32}$; $Sr@Ag_{32}$; $Ba@Au_{32}$; $Ra@Au_{32}$; $Sr@Au_{32}$; $Yb@Au_{32}$; $Ba@Li_{32}$; $Ra@Li_{32}$; $Sr@Li_{32}$; $Yb@Li_{32}$; $Ba@Na_{32}$; $Ra@Na_{32}$; $Sr@Na_{32}$; $Yb@Na_{32}$; $Yb@Ag_{32}$; $Sr@Cu_{32}$; $Ba@Cu_{32}$; $Ra@Cu_{32}$; $Yb@Cu_{32}$; $Sr@K_{32}$; $Ba@K_{32}$; $Ra@K_{32}$; $Yb@K_{32}$; $Sr@Cs_{32}$; $Ba@Cs_{32}$; $Ra@Cs_{32}$; $Yb@Cs_{32}$; |
| **$M@X_{24}Y_8$ (470)** |
| $Ba@Ag_{24}Ge_8$; $Ra@Ag_{24}Ge_8$; $Sr@Ag_{24}Ge_8$; $Yb@Ag_{24}Ge_8$; $Ra@Ag_{24}Pb_8$; $Sr@Ag_{24}Pb_8$; $Yb@Ag_{24}Pb_8$; $Ba@Ag_{24}Si_8$; $Yb@Ag_{24}Si_8$; $Ba@Ag_{24}Sn_8$; $Ra@Ag_{24}Sn_8$; $Sr@Ag_{24}Sn_8$; $Yb@Ag_{24}Sn_8$; $Ba@Au_{24}Ge_8$; $Ra@Au_{24}Ge_8$; $Sr@Au_{24}Ge_8$; $Yb@Au_{24}Ge_8$; $Ba@Au_{24}Pb_8$; $Ra@Au_{24}Pb_8$; $Sr@Au_{24}Pb_8$; $Yb@Au_{24}Pb_8$; $Ba@Au_{24}Si_8$; $Ra@Au_{24}Si_8$; $Sr@Au_{24}Si_8$; $Yb@Au_{24}Si_8$; $Ba@Au_{24}Sn_8$; $Ra@Au_{24}Sn_8$; $Sr@Au_{24}Sn_8$; $Yb@Au_{24}Sn_8$; $Ba@Li_{24}Ge_8$; $Ra@Li_{24}Ge_8$; $Sr@Li_{24}Ge_8$; $Yb@Li_{24}Ge_8$; $Ba@Li_{24}Si_8$; $Ra@Li_{24}Si_8$; $Sr@Li_{24}Si_8$; $Yb@Li_{24}Si_8$; $Ba@Li_{24}Sn_8$; $Ra@Li_{24}Sn_8$; $Sr@Li_{24}Sn_8$; $Yb@Li_{24}Sn_8$; $Ba@Be_{24}Li_8$; $Ra@Be_{24}Li_8$; $Sr@Be_{24}Li_8$; $Yb@Be_{24}Li_8$; $Ba@Cd_{24}Ag_8$; $Sr@Cd_{24}Ag_8$; $Yb@Cd_{24}Ag_8$; $Ba@Cd_{24}Au_8$; $Ra@Cd_{24}Au_8$; $Sr@Cd_{24}Au_8$; $Yb@Cd_{24}Au_8$; $Ra@Hg_{24}Ag_8$; $Sr@Hg_{24}Na_8$; $Ba@Ra_{24}Cs_8$; $Ra@Ra_{24}Cs_8$; $Ba@Mg_{24}Ag_8$; $Ra@Mg_{24}Ag_8$; $Sr@Mg_{24}Ag_8$; $Yb@Mg_{24}Ag_8$; $Ba@Mg_{24}Au_8$; $Ra@Mg_{24}Au_8$; $Sr@Mg_{24}Au_8$; $Yb@Mg_{24}Au_8$; $Ba@Mg_{24}Cu_8$; $Ra@Mg_{24}Cu_8$; $Sr@Mg_{24}Cu_8$; $Yb@Mg_{24}Cu_8$; $Ba@Mg_{24}Li_8$; $Ra@Mg_{24}Li_8$; $Sr@Mg_{24}Li_8$; $Yb@Mg_{24}Li_8$; $Ba@Mg_{24}Na_8$; $Ra@Mg_{24}Na_8$; $Sr@Mg_{24}Na_8$; $Yb@Mg_{24}Na_8$; $Yb@Zn_{24}Ag_8$; $Ba@Zn_{24}Au_8$; $Ra@Zn_{24}Au_8$; $Sr@Zn_{24}Au_8$; $Yb@Zn_{24}Au_8$; $Ba@Ra_{24}Rb_8$; $Sr@Ra_{24}Rb_8$; $Ra@Sr_{24}K_8$; $Sr@Sr_{24}K_8$; $Yb@Sr_{24}K_8$; $Sr@Zn_{24}Li_8$; $Ba@Al_{24}Ge_8$; $Ra@Al_{24}Ge_8$; $Sr@Al_{24}Ge_8$; $Yb@Al_{24}Ge_8$; $Ba@Al_{24}Si_8$; $Ra@Al_{24}Si_8$; $Sr@Al_{24}Si_8$; $Yb@Al_{24}Si_8$; $Ba@Ga_{24}Ge_8$; $Ra@Ga_{24}Ge_8$; $Sr@Ga_{24}Ge_8$; $Yb@Ga_{24}Ge_8$; $Ba@Ga_{24}Si_8$; $Ra@Ga_{24}Si_8$; $Sr@Ga_{24}Si_8$; $Yb@Ga_{24}Si_8$; $Yb@Ga_{24}Sn_8$; $Ba@In_{24}Pb_8$; $Ra@In_{24}Pb_8$; $Sr@In_{24}Pb_8$; $Yb@In_{24}Pb_8$; $Ba@In_{24}Sn_8$; $Ra@In_{24}Sn_8$; $Sr@In_{24}Sn_8$; $Yb@In_{24}Sn_8$; $Ba@Tl_{24}Pb_8$; $Ra@Tl_{24}Pb_8$; $Sr@Tl_{24}Pb_8$; $Yb@Tl_{24}Pb_8$; $Ba@Tl_{24}Sn_8$; $Ra@Tl_{24}Sn_8$; $Ba@Ge_{24}Ag_8$; $Ra@Ge_{24}Ag_8$; $Sr@Ge_{24}Ag_8$; $Yb@Ge_{24}Ag_8$; $Ba@Ge_{24}Li_8$; $Ra@Ge_{24}Li_8$; $Sr@Ge_{24}Li_8$; $Yb@Ge_{24}Li_8$; $Ba@Si_{24}Li_8$; $Ra@Si_{24}Li_8$; $Ba@Pb_{24}Na_8$; $Ra@Pb_{24}Na_8$; $Sr@Pb_{24}Na_8$; $Yb@Pb_{24}Na_8$; $Ba@Si_{24}Ag_8$; $Ba@Sn_{24}Li_8$; $Ra@Sn_{24}Li_8$; $Sr@Sn_{24}Li_8$; $Yb@Sn_{24}Li_8$; $Ba@Sn_{24}Na_8$; $Ra@Sn_{24}Na_8$; $Sr@Sn_{24}Na_8$; $Ba@Ge_{24}Au_8$; $Ra@Ge_{24}Au_8$; $Sr@Ge_{24}Au_8$; $Yb@Ge_{24}Au_8$; $Ba@Ge_{24}Cu_8$; $Ra@Ge_{24}Cu_8$; $Yb@Ge_{24}Cu_8$; $Ba@Si_{24}Au_8$; $Ra@Si_{24}Au_8$; $Sr@Si_{24}Au_8$; $Yb@Si_{24}Au_8$; $Ba@Si_{24}Cu_8$; $Ra@Si_{24}Cu_8$; $Sr@Si_{24}Cu_8$; $Yb@Si_{24}Cu_8$; $Ba@Ge_{24}As_8$; $Ra@Ge_{24}As_8$; $Sr@Ge_{24}As_8$; $Yb@Ge_{24}As_8$; $Ba@Ge_{24}Sb_8$; $Ra@Ge_{24}Sb_8$; $Sr@Ge_{24}Sb_8$; $Yb@Ge_{24}Sb_8$; $Ba@Pb_{24}Bi_8$; $Ra@Pb_{24}Bi_8$; $Sr@Pb_{24}Bi_8$; $Yb@Pb_{24}Bi_8$; $Ba@Pb_{24}Sb_8$; $Ra@Pb_{24}Sb_8$; $Sr@Pb_{24}Sb_8$; $Yb@Pb_{24}Sb_8$; $Ba@Si_{24}As_8$; $Ra@Si_{24}As_8$; $Sr@Si_{24}As_8$; $Ba@Sn_{24}Bi_8$; $Ra@Sn_{24}Bi_8$; $Sr@Sn_{24}Bi_8$; $Yb@Sn_{24}Bi_8$; $Ba@Sn_{24}Sb_8$; $Ra@Sn_{24}Sb_8$; $Sr@Sn_{24}Sb_8$; $Yb@Sn_{24}Sb_8$; $Ba@As_{24}Mg_8$; $Ra@As_{24}Mg_8$; $Sr@As_{24}Mg_8$; $Yb@As_{24}Mg_8$; $Ba@Bi_{24}Mg_8$; $Ra@Bi_{24}Mg_8$; $Sr@Bi_{24}Mg_8$; $Yb@Bi_{24}Mg_8$; $Ba@Sb_{24}Cd_8$; $Ra@Sb_{24}Cd_8$; $Sr@Sb_{24}Cd_8$; $Yb@Sb_{24}Cd_8$; $Ba@Sb_{24}Mg_8$; $Ra@Sb_{24}Mg_8$; $Sr@Sb_{24}Mg_8$; $Ba@Sb_{24}Zn_8$; $Ra@Sb_{24}Zn_8$; $Sr@Sb_{24}Zn_8$; $Yb@Sb_{24}Zn_8$; $Ba@As_{24}Zn_8$; $Ra@As_{24}Zn_8$; $Sr@As_{24}Zn_8$; $Yb@As_{24}Zn_8$; $Ba@Bi_{24}Cd_8$; $Ra@Bi_{24}Cd_8$; $Sr@Bi_{24}Cd_8$; $Yb@Bi_{24}Cd_8$; $Ba@Bi_{24}Hg_8$; $Ra@Bi_{24}Hg_8$; $Sr@Bi_{24}Hg_8$; $Yb@Bi_{24}Hg_8$; $Ba@Sb_{24}Hg_8$; $Ra@Sb_{24}Hg_8$; $Sr@Sb_{24}Hg_8$; $Yb@Sb_{24}Hg_8$; $Ba@Cu_{24}Ti_8$; $Ra@Cu_{24}Ti_8$; $Ba@Au_{24}Ti_8$; $Ra@Au_{24}Ti_8$; $Sr@Au_{24}Ti_8$; $Yb@Au_{24}Ti_8$; $Ba@Li_{24}Pd_8$; $Yb@Li_{24}Pd_8$; $Ba@Cu_{24}Ni_8$; $Ra@Cu_{24}Ni_8$; $Yb@Cu_{24}Ni_8$; $Ba@Cu_{24}Pt_8$; $Ba@Ag_{24}Ni_8$; $Ra@Ag_{24}Ni_8$; $Sr@Ag_{24}Ni_8$; $Ba@Ag_{24}Pd_8$; $Ra@Ag_{24}Pd_8$; $Sr@Ag_{24}Pd_8$; $Ba@Ag_{24}Pt_8$; $Ra@Ag_{24}Pt_8$; $Sr@Ag_{24}Pt_8$; $Ba@Au_{24}Ni_8$; $Ra@Au_{24}Ni_8$; $Sr@Au_{24}Ni_8$; $Yb@Au_{24}Ni_8$; $Ba@Au_{24}Pd_8$; $Ra@Au_{24}Pd_8$; $Sr@Au_{24}Pd_8$; $Yb@Au_{24}Pd_8$; $Ba@Au_{24}Pt_8$; $Ra@Au_{24}Pt_8$; $Sr@Au_{24}Pt_8$; $Yb@Au_{24}Pt_8$; $Ba@Mg_{24}Cu_8$; $Ra@Mg_{24}Cu_8$; $Sr@Mg_{24}Cu_8$; $Yb@Mg_{24}Cu_8$; $Ba@Mg_{24}Ag_8$; $Ra@Mg_{24}Ag_8$; $Sr@Mg_{24}Ag_8$; $Yb@Mg_{24}Ag_8$; $Ba@Mg_{24}Au_8$; $Ra@Mg_{24}Au_8$; $Sr@Mg_{24}Au_8$; $Yb@Mg_{24}Au_8$; $Ra@Cd_{24}Na_8$; $Ba@Cd_{24}Ag_8$; $Ra@Cd_{24}Ag_8$; $Sr@Cd_{24}Ag_8$; $Yb@Cd_{24}Ag_8$; $Ba@Cd_{24}Au_8$; $Ra@Cd_{24}Au_8$; $Sr@Cd_{24}Au_8$; $Yb@Cd_{24}Au_8$; $Ba@Hg_{24}Na_8$; $Ra@Hg_{24}Na_8$; $Sr@Hg_{24}Na_8$; $Yb@Hg_{24}Na_8$; $Ba@Hg_{24}Ag_8$; $Ra@Hg_{24}Ag_8$; $Sr@Hg_{24}Ag_8$; $Yb@Hg_{24}Ag_8$; $Ra@Mg_{24}Mn_8$; $Sr@Mg_{24}Mn_8$; $Ba@Mg_{24}Tc_8$; $Ra@Zn_{24}Mn_8$; $Ra@Zn_{24}Re_8$; $Sr@Zn_{24}Re_8$; $Yb@Zn_{24}Re_8$; $Ba@Hg_{24}Tc_8$; $Ra@Hg_{24}Tc_8$; $Yb@Hg_{24}Tc_8$; $Ba@Hg_{24}Re_8$; $Ra@Hg_{24}Re_8$; $Sr@Hg_{24}Re_8$; $Yb@Hg_{24}Re_8$; $Ba@Sc_{24}Ti_8$; $Ba@Sc_{24}Sn_8$; $Ra@Sc_{24}Sn_8$; $Sr@Sc_{24}Sn_8$; $Yb@Sc_{24}Sn_8$; $Ba@Sc_{24}Pb_8$; $Ra@Sc_{24}Pb_8$; $Sr@Sc_{24}Pb_8$; $Ba@Y_{24}Pb_8$; $Ra@Y_{24}Pb_8$; $Sr@Y_{24}Pb_8$; $Yb@Y_{24}Pb_8$; $Ba@Al_{24}Ti_8$; $Ra@Al_{24}Ti_8$; $Sr@Al_{24}Ti_8$; $Yb@Al_{24}Ti_8$; $Ba@Ga_{24}Ti_8$; $Ra@Ga_{24}Ti_8$; $Sr@Ga_{24}Ti_8$; $Yb@Ga_{24}Ti_8$; $Ba@Ga_{24}Zr_8$; $Ra@Ga_{24}Zr_8$; $Sr@Ga_{24}Zr_8$; $Yb@Ga_{24}Zr_8$; $Ba@Ga_{24}Hf_8$; $Ra@Ga_{24}Hf_8$; $Sr@Ga_{24}Hf_8$; $Yb@Ga_{24}Hf_8$; $Ba@In_{24}Ti_8$; $Ra@In_{24}Ti_8$; $Sr@In_{24}Ti_8$; $Ba@In_{24}Zr_8$; $Ra@In_{24}Zr_8$; $Sr@In_{24}Zr_8$; $Yb@In_{24}Zr_8$; $Ba@In_{24}Hf_8$; $Ra@In_{24}Hf_8$; $Sr@In_{24}Hf_8$; $Yb@In_{24}Hf_8$; $Ba@Tl_{24}Zr_8$; $Ra@Tl_{24}Zr_8$; $Sr@Tl_{24}Zr_8$; $Ba@Tl_{24}Hf_8$; $Ra@Tl_{24}Hf_8$; $Sr@Tl_{24}Hf_8$; $Yb@Tl_{24}Hf_8$; $Ba@Sc_{24}Ru_8$; $Ra@Sc_{24}Ru_8$; $Sr@Sc_{24}Ru_8$; $Yb@Sc_{24}Ru_8$; $Ba@Sc_{24}Os_8$; $Ra@Sc_{24}Os_8$; $Ba@Al_{24}Fe_8$; $Ba@Al_{24}Ru_8$; $Ra@Al_{24}Ru_8$; $Sr@Al_{24}Ru_8$; $Yb@Al_{24}Ru_8$; $Ba@Al_{24}Os_8$; $Ra@Al_{24}Os_8$; $Sr@Al_{24}Os_8$; $Yb@Al_{24}Os_8$; $Ra@Ga_{24}Fe_8$; $Sr@Ga_{24}Fe_8$; $Yb@Ga_{24}Fe_8$; $Ba@Ga_{24}Ru_8$; $Ra@Ga_{24}Ru_8$; $Sr@Ga_{24}Ru_8$; $Yb@Ga_{24}Ru_8$; $Ba@Ga_{24}Os_8$; $Ra@Ga_{24}Os_8$; $Sr@Ga_{24}Os_8$; $Yb@Ga_{24}Os_8$; $Ba@In_{24}Ru_8$; $Ra@In_{24}Ru_8$; $Sr@In_{24}Ru_8$; $Ba@In_{24}Os_8$; |
| Ra@In$_{24}$Os$_{8}$; Sr@In$_{24}$Os$_{8}$; Ra@Ti$_{24}$Cu$_{8}$; Ba@Ti$_{24}$Ag$_{8}$; Sr@Ti$_{24}$Ag$_{8}$; Yb@Ti$_{24}$Ag$_{8}$; Ba@Zr$_{24}$Ag$_{8}$; Ra@Zr$_{24}$Ag$_{8}$; Sr@Zr$_{24}$Ag$_{8}$; Ba@Zr$_{24}$Au$_{8}$; Ra@Zr$_{24}$Au$_{8}$; Sr@Zr$_{24}$Au$_{8}$; Yb@Zr$_{24}$Au$_{8}$; Ba@Hf$_{24}$Ag$_{8}$; Ra@Hf$_{24}$Ag$_{8}$; Sr@Hf$_{24}$Ag$_{8}$; Yb@Hf$_{24}$Ag$_{8}$; Ba@Hf$_{24}$Au$_{8}$; Ra@Hf$_{24}$Au$_{8}$; Sr@Hf$_{24}$Au$_{8}$; Yb@Hf$_{24}$Au$_{8}$; Ba@Ti$_{24}$V$_{8}$; Ra@Ti$_{24}$V$_{8}$; Sr@Ti$_{24}$V$_{8}$; Yb@Ti$_{24}$V$_{8}$; Ba@Ti$_{24}$Nb$_{8}$; Ra@Ti$_{24}$Nb$_{8}$; Sr@Ti$_{24}$Nb$_{8}$; Yb@Ti$_{24}$Nb$_{8}$; Ba@Ti$_{24}$Ta$_{8}$; Ra@Ti$_{24}$Ta$_{8}$; Sr@Ti$_{24}$Ta$_{8}$; Yb@Ti$_{24}$Ta$_{8}$; Ba@Ti$_{24}$As$_{8}$; Ra@Ti$_{24}$As$_{8}$; Sr@Ti$_{24}$As$_{8}$; Yb@Ti$_{24}$As$_{8}$; Ba@Ti$_{24}$Sb$_{8}$; Ra@Ti$_{24}$Sb$_{8}$; Yb@Ti$_{24}$Sb$_{8}$; Ba@Ti$_{24}$Bi$_{8}$; Ra@Ti$_{24}$Bi$_{8}$; Sr@Ti$_{24}$Bi$_{8}$; Ba@Zr$_{24}$Nb$_{8}$; Ra@Zr$_{24}$Nb$_{8}$; Sr@Zr$_{24}$Nb$_{8}$; Yb@Zr$_{24}$Nb$_{8}$; Ba@Zr$_{24}$Ta$_{8}$; Ra@Zr$_{24}$Ta$_{8}$; Sr@Zr$_{24}$Ta$_{8}$; Yb@Zr$_{24}$Ta$_{8}$; Ba@Zr$_{24}$Sb$_{8}$; Ra@Zr$_{24}$Sb$_{8}$; Sr@Zr$_{24}$Sb$_{8}$; Yb@Zr$_{24}$Sb$_{8}$; Ba@Zr$_{24}$Bi$_{8}$; Ra@Zr$_{24}$Bi$_{8}$; Sr@Zr$_{24}$Bi$_{8}$; Yb@Zr$_{24}$Bi$_{8}$; Ba@Hf$_{24}$Nb$_{8}$; Ra@Hf$_{24}$Nb$_{8}$; Sr@Hf$_{24}$Nb$_{8}$; Yb@Hf$_{24}$Nb$_{8}$; Ba@Hf$_{24}$Ta$_{8}$; Ra@Hf$_{24}$Ta$_{8}$; Sr@Hf$_{24}$Ta$_{8}$; Yb@Hf$_{24}$Ta$_{8}$; Ba@Hf$_{24}$Sb$_{8}$; Ra@Hf$_{24}$Sb$_{8}$; Sr@Hf$_{24}$Sb$_{8}$; Yb@Hf$_{24}$Sb$_{8}$; Ba@Hf$_{24}$Bi$_{8}$; Ra@Hf$_{24}$Bi$_{8}$; Sr@Hf$_{24}$Bi$_{8}$; Yb@Hf$_{24}$Bi$_{8}$; Ba@Si$_{24}$V$_{8}$; Ra@Si$_{24}$V$_{8}$; Sr@Si$_{24}$V$_{8}$; Ba@Si$_{24}$Nb$_{8}$; Ra@Si$_{24}$Nb$_{8}$; Sr@Si$_{24}$Nb$_{8}$; Ba@Si$_{24}$Ta$_{8}$; Ra@Si$_{24}$Ta$_{8}$; Sr@Si$_{24}$Ta$_{8}$; Yb@Si$_{24}$Ta$_{8}$; Ba@Ge$_{24}$V$_{8}$; Sr@Ge$_{24}$V$_{8}$; Ra@Ge$_{24}$Nb$_{8}$; Sr@Ge$_{24}$Nb$_{8}$; Yb@Ge$_{24}$Nb$_{8}$; Ba@Ge$_{24}$Ta$_{8}$; Ra@Ge$_{24}$Ta$_{8}$; Sr@Ge$_{24}$Ta$_{8}$; Ba@Sn$_{24}$Nb$_{8}$; Ra@Sn$_{24}$Nb$_{8}$; Sr@Sn$_{24}$Nb$_{8}$; Ba@V$_{24}$Zn$_{8}$; Yb@V$_{24}$Zn$_{8}$; Ra@Nb$_{24}$Zn$_{8}$; Sr@Nb$_{24}$Zn$_{8}$; Ra@Nb$_{24}$Cd$_{8}$; |

Table S25. The molecular formula of the stable structure of $A_6$ when the co-dual structure is embedded.

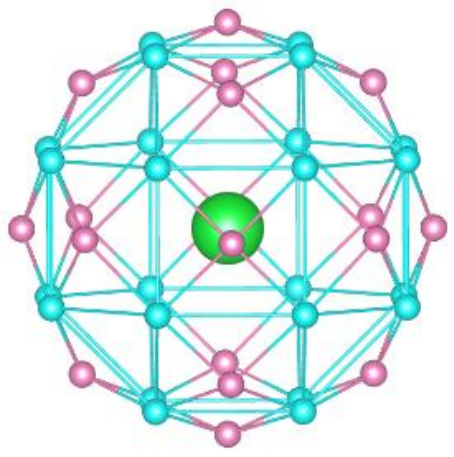

$A_6$ Co-dual structure

| |
|---|
| **M@$X_{42}$ (0)** |
| --- |
| **M@$X_{24}Y_{18}$ (45)** |
| Ra@Cd$_{24}$Au$_{18}$; Yb@Cd$_{24}$Au$_{18}$; Ba@Hg$_{24}$Ag$_{18}$; Ra@Hg$_{24}$Ag$_{18}$; Sr@Hg$_{24}$Ag$_{18}$; Yb@Hg$_{24}$Ag$_{18}$; Ba@Hg$_{24}$Au$_{18}$; Ra@Hg$_{24}$Au$_{18}$; Sr@Hg$_{24}$Au$_{18}$; Yb@Hg$_{24}$Au$_{18}$; Ba@Cd$_{24}$Ag$_{18}$; Ra@Cd$_{24}$Ag$_{18}$; Sr@Cd$_{24}$Ag$_{18}$; Yb@Cd$_{24}$Ag$_{18}$; Ba@Zn$_{24}$Cu$_{18}$; Ra@Zn$_{24}$Cu$_{18}$; Sr@Zn$_{24}$Cu$_{18}$; Yb@Zn$_{24}$Cu$_{18}$; Ra@Ga$_{24}$Cu$_{18}$; Ba@In$_{24}$Au$_{18}$; Ra@In$_{24}$Au$_{18}$; Sr@In$_{24}$Au$_{18}$; Yb@In$_{24}$Au$_{18}$; Ra@Al$_{24}$Zn$_{18}$; Sr@Al$_{24}$Zn$_{18}$; Yb@Al$_{24}$Zn$_{18}$; Yb@In$_{24}$Ag$_{18}$; Ra@Mg$_{24}$Cu$_{18}$; Yb@Mg$_{24}$Cu$_{18}$; Ba@Mg$_{24}$Ag$_{18}$; Yb@Mg$_{24}$Ag$_{18}$; Ba@Mg$_{24}$Au$_{18}$; Ra@Mg$_{24}$Au$_{18}$; Sr@Mg$_{24}$Au$_{18}$; Yb@Mg$_{24}$Au$_{18}$; Ra@Cd$_{24}$Ag$_{18}$; Yb@Cd$_{24}$Ag$_{18}$; Ra@Cd$_{24}$Au$_{18}$; Sr@Cd$_{24}$Au$_{18}$; Yb@Cd$_{24}$Au$_{18}$; Ba@Hg$_{24}$Na$_{18}$; Ra@Hg$_{24}$Na$_{18}$; Sr@Hg$_{24}$Na$_{18}$; Ba@Hg$_{24}$Ag$_{18}$; Ra@Hg$_{24}$Ag$_{18}$; |

Table S26. The molecular formula of the stable structure of $A_7$ when the co-dual structure is embedded.

| |
|---|
| 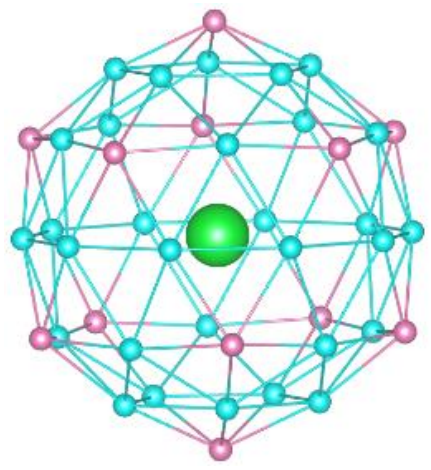 $A_7$ Co-dual structure |
| **$M@X_{42}$ (0)** |
| --- |
| **$M@X_{30}Y_{12}$ (85)** |
| $Sr@Ag_{30}Al_{12}$; $Yb@Ag_{30}Al_{12}$; $Ba@Ag_{30}In_{12}$; $Ra@Ag_{30}In_{12}$; $Sr@Ag_{30}In_{12}$; $Yb@Au_{30}Al_{12}$; $Ra@Na_{30}In_{12}$; $Yb@Na_{30}In_{12}$; $Ba@Au_{30}Bi_{12}$; $Ra@Au_{30}Bi_{12}$; $Sr@Au_{30}Bi_{12}$; $Yb@Au_{30}Bi_{12}$; $Ba@Li_{30}As_{12}$; $Ra@Li_{30}As_{12}$; $Sr@Li_{30}As_{12}$; $Yb@Li_{30}As_{12}$; $Ba@Li_{30}Sb_{12}$; $Ra@Li_{30}Sb_{12}$; $Sr@Li_{30}Sb_{12}$; $Yb@Li_{30}Sb_{12}$; $Ba@Na_{30}Bi_{12}$; $Ra@Na_{30}Bi_{12}$; $Sr@Na_{30}Bi_{12}$; $Yb@Na_{30}Bi_{12}$; $Ba@Na_{30}Sb_{12}$; $Ra@Na_{30}Sb_{12}$; $Sr@Na_{30}Sb_{12}$; $Yb@Na_{30}Sb_{12}$; $Yb@Na_{30}Sc_{12}$; $Ba@Na_{30}Y_{12}$; $Ra@Na_{30}Y_{12}$; $Sr@Na_{30}Y_{12}$; $Ba@Ag_{30}Sc_{12}$; $Ra@Ag_{30}Sc_{12}$; $Sr@Ag_{30}Sc_{12}$; $Yb@Ag_{30}Sc_{12}$; $Ba@Au_{30}Sc_{12}$; $Ra@Au_{30}Sc_{12}$; $Sr@Au_{30}Sc_{12}$; $Yb@Au_{30}Sc_{12}$; $Sr@Li_{30}V_{12}$; $Ba@Li_{30}Nb_{12}$; $Ra@Li_{30}Nb_{12}$; $Sr@Li_{30}Nb_{12}$; $Yb@Li_{30}Nb_{12}$; $Ba@Li_{30}Ta_{12}$; $Ra@Li_{30}Ta_{12}$; $Sr@Li_{30}Ta_{12}$; $Yb@Li_{30}Ta_{12}$; $Ba@Cu_{30}V_{12}$; $Ra@Cu_{30}V_{12}$; $Sr@Cu_{30}V_{12}$; $Yb@Cu_{30}V_{12}$; $Ba@Cu_{30}Nb_{12}$; $Ra@Cu_{30}Nb_{12}$; $Sr@Cu_{30}Nb_{12}$; $Yb@Cu_{30}Nb_{12}$; $Ba@Cu_{30}Ta_{12}$; $Ra@Cu_{30}Ta_{12}$; $Sr@Cu_{30}Ta_{12}$; $Yb@Cu_{30}Ta_{12}$; $Ba@Ag_{30}V_{12}$; $Ra@Ag_{30}V_{12}$; $Sr@Ag_{30}V_{12}$; $Yb@Ag_{30}V_{12}$; $Ba@Ag_{30}Nb_{12}$; $Ra@Ag_{30}Nb_{12}$; $Sr@Ag_{30}Nb_{12}$; $Yb@Ag_{30}Nb_{12}$; $Ba@Ag_{30}Ta_{12}$; $Ra@Ag_{30}Ta_{12}$; $Sr@Ag_{30}Ta_{12}$; $Yb@Ag_{30}Ta_{12}$; $Ba@Au_{30}V_{12}$; $Ra@Au_{30}V_{12}$; $Sr@Au_{30}V_{12}$; $Yb@Au_{30}V_{12}$; $Ba@Au_{30}Nb_{12}$; $Ra@Au_{30}Nb_{12}$; $Sr@Au_{30}Nb_{12}$; $Yb@Au_{30}Nb_{12}$; $Ba@Au_{30}Ta_{12}$; $Ra@Au_{30}Ta_{12}$; $Sr@Au_{30}Ta_{12}$; $Yb@Au_{30}Ta_{12}$; |

Table S27. The molecular formula of the stable structure of $A_8$ when the co-dual structure is embedded.

| |
|---|
| 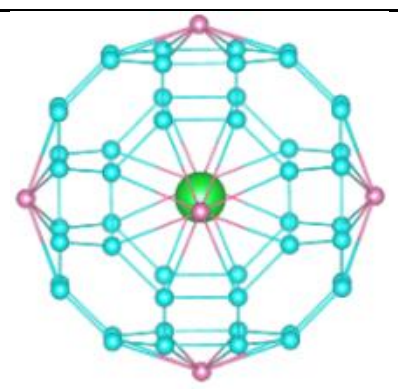 $A_8$ Co-dual structure |
| **$M@X_{54}$ (0)** |
| - - - |
| **$M@X_{48}Y_6$ (0)** |
| - - - |

Table S28. The molecular formula of the stable structure of $A_9$ when the co-dual structure is embedded.

| |
|---|
| 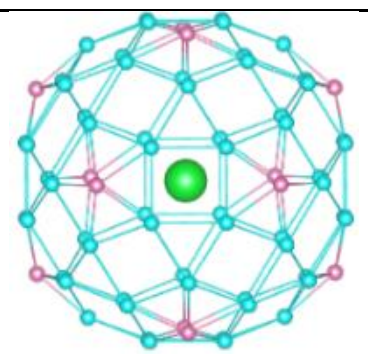 $A_9$ Co-dual structure |
| **$M@X_{72}$ (0)** |
| --- |
| **$M@X_{60}Y_{12}$ (0)** |
| --- |

Table S29. The molecular formula of the stable structure of $A_{10}$ when the co-dual structure is embedded.

| |
|---|
| 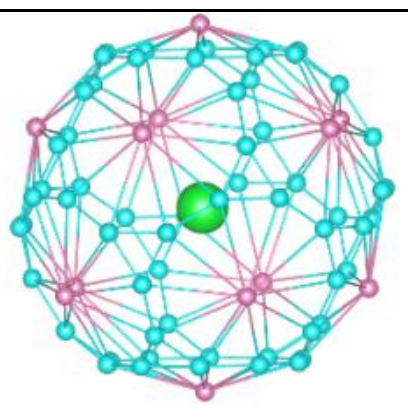<br>$A_{10}$ Co-dual structure |
| **$M@X_{72}$ (0)** |
| --- |
| **$M@X_{60}Y_{12}$ (0)** |
| --- |

Table S30. The molecular formula of the stable structure of $A_{11}$ when the co-dual structure is embedded.

| |
|---|
| 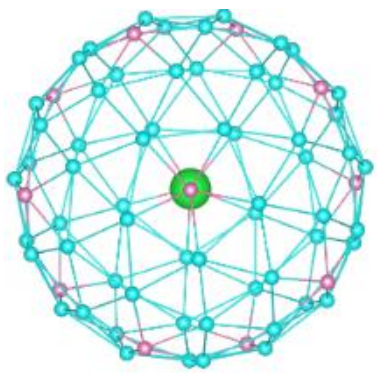<br>$A_{11}$ Co-dual structure |
| **$M@X_{72}$ (0)** |
| --- |
| **$M@X_{60}Y_{12}$ (0)** |
| --- |

Table S31. The molecular formula of the stable structure of $A_{12}$ when the co-dual structure is embedded.

| |
|---|
| 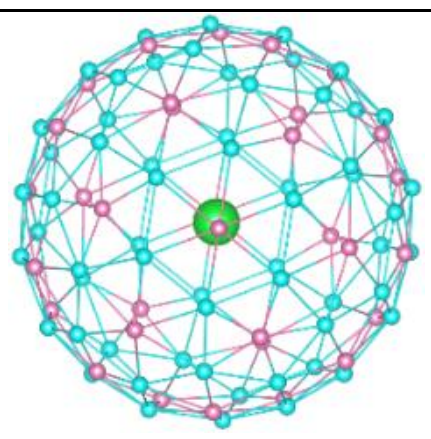<br>$A_{12}$ Co-dual structure |
| **$M@X_{80}$ (0)** |
| --- |
| **$M@X_{60}Y_{20}$ (0)** |
| --- |

Table S32. The molecular formula of the stable structure of $A_{13}$ when the co-dual structure is embedded.

| |
|---|
| 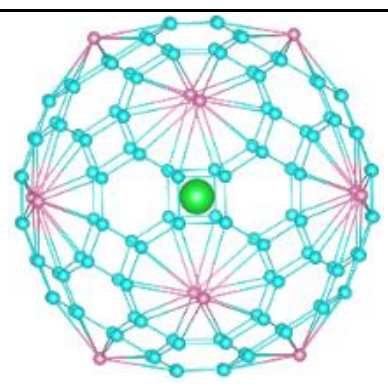 $A_{13}$ Co-dual structure |
| **M@$X_{132}$ (0)** |
| --- |
| **M@$X_{120}Y_{12}$ (0)** |
| --- |